\documentclass[twocolumn]{aastex631}
\usepackage{natbib}
\usepackage{url}
\usepackage{color}
\usepackage{amsmath}
\definecolor{red}{rgb}{1,0,0}
\definecolor{orange}{rgb}{1,0.5,0}
\definecolor{green}{rgb}{0,127,0}
\definecolor{grey}{rgb}{0.6627,0.6627,0.6627}
\definecolor{skyblue}{rgb}{0.53,0.808,0.98}

\newcommand{\lya}{Ly$\alpha$}

\newcommand{\ha}{H$\alpha$}

\newcommand{\oii}{[\ion{O}{2}]}

\newcommand{\oiii}{[\ion{O}{3}]}

\newcommand{\eazy}{\texttt{\textsc{EAZY}}}
\newcommand{\bpass}{\texttt{\textsc{BPASS}}}
\newcommand{\cloudy}{\texttt{\textsc{Cloudy}}}
\newcommand{\rband}{$r_{062}$}
\newcommand{\zband}{$z_{087}$}
\newcommand{\yband}{$Y_{106}$}
\newcommand{\jband}{$J_{129}$}
\newcommand{\hband}{$H_{158}$}
\newcommand{\fband}{$F_{184}$}

\newcommand{\hst}{\textit{HST}}
\newcommand{\rst}{\textit{Roman}}
\newcommand{\jwst}{\textit{JWST}}

\newcommand{\zphot}{$z_{\mathrm{phot}}$}

\newcommand\myvdots{\vbox{\baselineskip=1pt \lineskiplimit=0pt \kern10pt \hbox{.}\hbox{.}\hbox{.}}} 

\shorttitle{Going Wide and Deep with Roman}
\shortauthors{Bagley et al.}

\begin{document}

\title{Going Wide and Deep with Roman: The $z\sim6-9$ UV luminosity function 
in a Roman Deep Field}

\correspondingauthor{Micaela B. Bagley}
\email{mbagley@utexas.edu}

\author[0000-0002-9921-9218]{Micaela B. Bagley}
\affil{Department of Astronomy, The University of Texas at Austin, Austin, TX, USA}

\author[0000-0001-8519-1130]{Steven L. Finkelstein}
\affil{Department of Astronomy, The University of Texas at Austin, Austin, TX, USA}
\affiliation{Cosmic Frontier Center, The University of Texas at Austin, Austin, TX 78712}

\author{James Rhoads}
\affil{Astrophysics Science Division, NASA Goddard Space Flight Center, Greenbelt, MD, USA}

\author{Sangeeta Malhotra}
\affil{Astrophysics Science Division, NASA Goddard Space Flight Center, Greenbelt, MD, USA}

\author[0000-0003-3466-035X]{{L. Y. Aaron} {Yung}}
\affil{Space Telescope Science Institute, Baltimore, MD, USA}

\author{Rachel S. Somerville}
\affil{Center for Computational Astrophysics, Flatiron Institute, NY, USA}

\author[0000-0001-7503-8482]{Casey Papovich}
\affil{Department of Physics and Astronomy, Texas A\&M University, College Station, TX, USA}
\affil{George P. and Cynthia Woods Mitchell Institute for Fundamental Physics and Astronomy, Texas A\&M University, College Station, TX, USA}

\begin{abstract}
We present a trade study of possible ultra-deep surveys with the \textit{Nancy Grace Roman Space Telescope}, optimizing the depth-area-filter parameter space for high-redshift galaxy science. Using a mock galaxy catalog derived from a 2 degree$^2$ lightcone created using the Santa Cruz semi-analytic model and populated with over $7.6$ million galaxies at $0 < z < 10$ with $M_\mathrm{UV} \lesssim -15$, with realistic clustering and synthetic photometry, we evaluate sixteen 500-hour survey configurations spanning $0.28–2$ degree$^2$ and four filter combinations.
We demonstrate that even a single \rst\ pointing dramatically reduces cosmic variance compared to \hst-like observations, more faithfully recovering the true UV luminosity function. For each survey configuration, we explore photometric redshift recovery, sample contamination, and measurements of the rest-UV luminosity function and non-ionizing UV luminosity density ($\rho_{\mathrm{UV}}$) across four redshift bins at $z\sim6–9$. We find that inclusion of the \rband\ filter is essential for studies at $z\sim5–6$, reducing sample contamination from nearly 100\% to negligible levels and recovering the bright end of the luminosity function. The \fband\ filter improves galaxy recovery at $z > 9$ and is critical for stellar contamination removal at all redshifts. Based on these results, we recommend that a \rst\ ultra-deep survey cover at least two \rst\ pointings (0.56 degree$^2$) with all six filters (\rband, \zband, \yband, \jband, \hband, \fband), reducing uncertainties on the rest-UV luminosity density by factors of 2–4 relative to the deepest existing JWST programs. Building off of the Deep Tier of the High Latitude Time Domain Survey to add depth and filter coverage to existing (or planned) observations is an excellent option.
\end{abstract}

\section{Introduction} \label{sec:intro}

Understanding the growth of galaxies at early times requires precise measurements of the ultraviolet (UV) luminosity function.  This is most critical at the faint end, where the faint-end slope, $\alpha$, directly determines the abundance of low-luminosity galaxies that are thought to dominate the ionizing photon budget of the early universe \citep{robertson2015,finkelstein2019a, yung2020b}. The integrated non-ionizing UV luminosity density, $\rho_{UV}$, is derived by integrating the luminosity function down to a limiting magnitude, often $M_{1500} = -13$ mag, below which observational constraints become increasingly uncertain \citep[e.g.][]{bouwens2015a,finkelstein2015a}.

Under assumptions about the ionizing photon production efficiency and the escape fraction of Lyman continuum radiation, $\rho_{UV}$ can be converted to the galaxy ionizing emissivity and compared with the emissivity required to initiate and maintain cosmic reionization \citep{madau1999,robertson2015,finkelstein2019a, yung2020b, yung2020a}. However, uncertainties in the luminosity function—particularly in the normalization ($\phi^{\ast}$), characteristic magnitude ($M^{\ast}$), and faint-end slope ($\alpha$)—propagate directly into measurements of $\rho_{UV}$ and hence our understanding of galaxies' role in reionization. Small systematic errors in $\alpha$ can lead to order-of-magnitude differences in the predicted number density of faint galaxies.

Measurements of the high-redshift UV luminosity function have historically been limited by the trade-off between survey depth and area. The Hubble Ultra Deep Field \citep[HUDF;][]{beckwith2006} reached depths of m$_{F160W} \sim$ 29.5 over a single {\it Hubble} field of view, while the Hubble Frontier Fields \citep[HFF;][]{lotz2017} and their parallel observations extended coverage at slightly shallower depths, with the benefit of lensing magnification. These small fields, however, suffer from significant cosmic variance, leading to field-to-field variations that can exceed the measurement uncertainties \citep[e.g.][]{trenti2008,robertson2014}. More recently, {\it JWST} has begun to probe even fainter magnitudes in programs such as JADES \citep{eisenstein2023a}, CEERS \citep{finkelstein2025}, and NGDEEP \citep{bagley2024b}, but these too cover relatively modest areas.

The {\it Nancy Grace Roman Space Telescope} \citep[$Roman$][]{spergel2015b} will transform our ability to measure the UV luminosity function at high redshifts ($z >$ 6). Roman's Wide-Field Instrument \citep[WFI;][]{akeson2019} offers a field of view more than 100 times larger than {\it HST}/WFC3, enabling surveys that simultaneously achieve HUDF-like depths while covering areas sufficient to overcome cosmic variance. \citet{wang2022b} demonstrated the potential of Roman for time-domain science in deep fields, but a comprehensive assessment of survey strategy optimization for high-redshift galaxy science has been lacking.

In this paper, we present a trade study examining potential Roman ultra-deep field surveys, evaluating how different combinations of survey depth, area, and filter coverage affect the recovery of high-redshift galaxies and measurements of the UV luminosity function.  In this analysis we focus on filters bluer than 2$\mu$m, as the passively cooled nature of {\it Roman} results in prohibitively large integration times in the F213W filter to reach comparable depths as the bluer filters.  
This paper is organized as follows. In Section~\ref{sec:sam}, we briefly 
describe the lightcone, semi-analytic model, and mock catalog. 
We perform an initial analysis in Section~\ref{sec:noiseless}, comparing
measurements of \hst\ deep fields with single \rst/WFI pointings.
In Section~\ref{sec:tradestudy}, we quantify the source recovery and contamination in each survey and redshift 
sample in Sections~\ref{sec:recov} and \ref{sec:contam},
respectively. 

We assume a $\Lambda$CDM cosmology consistent with the latest 
\textit{Planck} results with $H_0 = 67.74$ km s$^{-1}$ Mpc$^{-1}$,
$\Omega_m = 0.309$ and $\Omega_{\Lambda}=0.691$
\citep{planck2018}. All magnitudes are presented in the AB system
\citep{oke1983} unless otherwise noted.

\begin{figure*}
\plotone{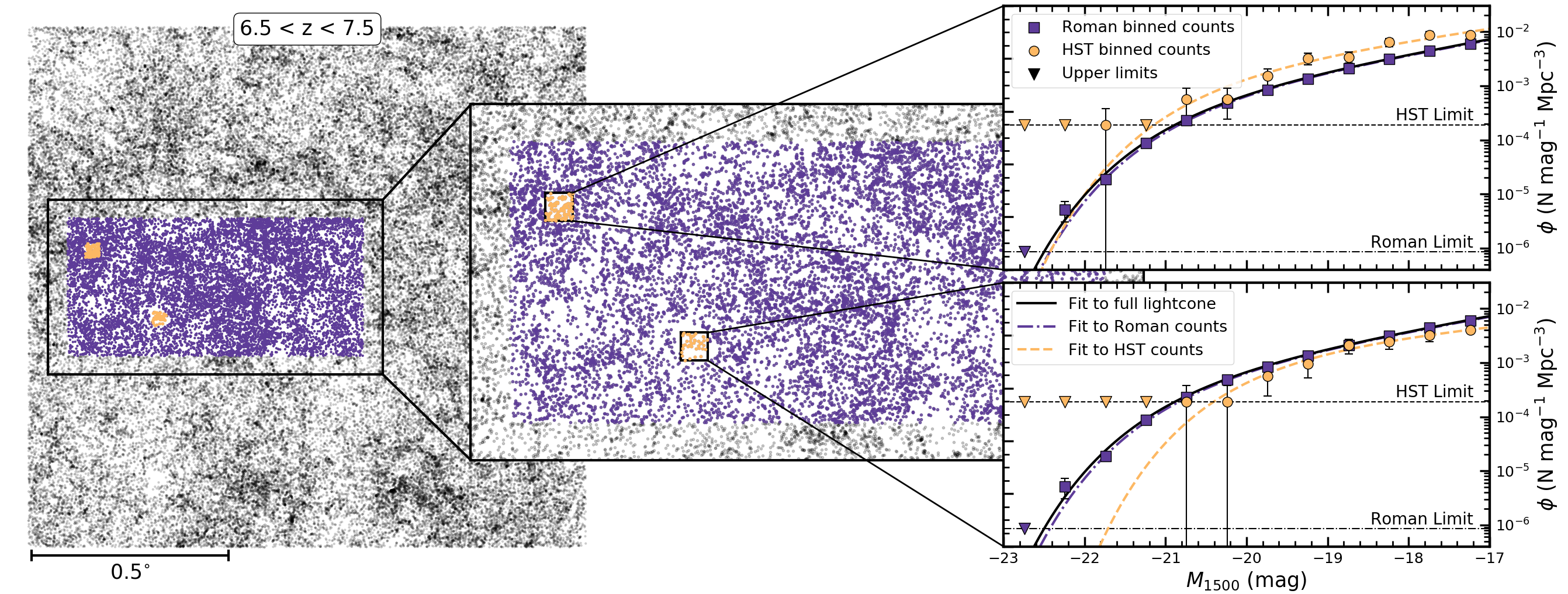}
\caption{An illustration of the constraining power of a \rst\ Ultra Deep 
Field. In the left panel, we show the positions of sources in the full
2 degree$^2$ catalog down to the depth of the HUDF 
($m_{\mathrm{F160W}} \leq 29.5$) and in the redshift range $6.5 < z < 7.5$. 
The middle panel zooms in on a region that is $\sim0\fdg85 \times 0\fdg475$.
The sources that would be observed in two \hst-sized pointings are shaded 
orange, while the sources observed by a single \rst\ pointing are in purple. 
In the right panels, we show the luminosity functions as measured in each 
of the pointings. The two \hst/WFC3 pointings cover regions of relative 
over- and underdensity, and as a result over- and underestimate the underlying
luminosity function. However, the \rst-sized pointing is large enough to 
marginalize over the local density variations, and the resulting measurement
of the luminosity function recovers the true value.
\label{fig:sam}}
\end{figure*}

\section{Santa Cruz Semi-Analytic Model}\label{sec:sam}
For this project, we are using a mock catalog generated using the Santa Cruz 
SAM \citep{somerville1999b,somerville2008,somerville2015b}. 
The base of the mock catalog is a 2 degree$^2$ lightcone constructed with dark matter halos extracted from the 
Small MultiDark Planck (SMDPL) $N$-body simulation from the MultiDark suite \citep{klypin2016}. 
The simulation resolves halos down to $M_{\rm halo} \sim10^{10}$ M$_{\odot}$ in a cube 
400 Mpc/$h$ on a side.
A mock lightcone of dark matter halos is created 
by passing through the simulation volume along many 
different slightlines to provide mock positions in the sky for the synthetic 
sources, as described by \citet{somerville2021} and \citet{yung2022}. The 
SAM uses an extended Press-Schechter formalism-based algorithm to construct merger 
histories for individual halos \citep{somerville1999a}, and track the evolution of galaxy properties within these halos using a set of carefully curated analytic and empirical recipes, including baryonic processes such as 
star formation, chemical evolution, black hole growth, and stellar and AGN feedback. We refer the reader to the schematic flowchart in \citet{yung2022} for an overview of the internal workflow of the Santa Cruz SAM. The predicted galaxy populations have been extensively tested and shown to well-reproduce observed 
galaxy properties and population statistics out to $z\sim10$ 
\citep{somerville2015b,yung2019a,yung2019b}, and the spatial distribution of sources in these lightcones are in good agreements with observed two-point correlation functions \citep{yung2022, yung2023}. 

The star formation and chemical evolution histories are then forward modeled to create synthetic 
spectral energy distributions based on the stellar population synthesis model by \citet{bc03}. The photometry in the Roman filter set are computed by red-shifting the rest-frame spectra to their redshift in the lightcone and accounting for attenuation by the intervening neutral hydrogen along the line of sight \citep{madau1996} and ISM dust attenuation assuming the dust attenuation curve of \citet{calzetti2000}.

We refer readers to \citet{somerville2008, somerville2015b, somerville2021} for details on the 
SAM framework and model parameters, and \citet{yung2019a, yung2021} for model calibration. The simulated data products and specifications of the 2 degree$^2$ mock 
catalog we use in this paper is presented and described in detail by 
\citet{yung2023}.

\section{The Need for A \rst\ Ultra Deep Field}\label{sec:noiseless}
Deep \hst\ (and now \jwst) observations provide important insight into the population of 
faint galaxies, yet only probe small areas. For example, the Hubble Ultra
Deep Field \citep[HUDF;][]{beckwith2006} reaches a depth of $\sim29.6$ in 
WFC3/IR filter F160W \citep[e.g.,][]{bouwens2008,oesch2009,finkelstein2015a} in a total of $\sim$46.5 hours. However, this depth is achieved over an area of 
only 5.1 arcmin$^2$. The \rst\ WFI offers the opportunity to achieve a 
similar depth in a given time, but over an area more than 100$\times$ wider.

In this section, we explore the improvements made possible with \rst\ imaging
by simulating ``perfect'' \hst\ and \rst\ observations of the mock lightcone, 
i.e., noiseless, fully complete observations with no contamination. 
We consider three ``surveys'': a single, deep HUDF-like pointing with sources 
selected down to $m_{\mathrm{F160W}} \leq 29.5$ (Section~\ref{sec:1hudf});
an approximation of all \hst\ deep fields, comprised of a single HUDF-like pointing combined with a set of eight \hst-sized pointings 
with $m_{\mathrm{F160W}} \leq 29.0$ (to represent the HFF parallel fields, Section~\ref{sec:hffs}); and a single 
\rst/WFI pointing reaching $m_{\mathrm{F160W}} \leq 29.0$. 
We also define the ``truth'' by fitting sources in the lightcone down to a 
magnitude of $m_{\mathrm{F160W}} \leq 30$. Here we adopt an \hst\ filter (F160W) for direct comparison with the \hst\ deep surveys mentioned above.
In all cases, we select sources from the SAM mock catalog directly using their
absolute UV magnitudes, $M_{1500}$, as measured from their modelled spectral 
energy distributions in a 100\AA-wide tophat filter centered at 1500\AA. 
For each ``survey'', we convert between the given apparent magnitude limit in 
\hst/WFC3 F160W using the midpoint of each redshift range for the conversion, 
and select all sources in the relevant footprints that are brighter than that
limit in $M_{1500}$.

From the SAM mock catalog, we extract pointings that cover the same areas as 
the \hst\ and \rst\ fields-of-view.
For the \hst\ fields, we adopt the WFC3/IR footprint, which covers
$123\arcsec \times 134\arcsec$ and has an area of 4.58 arcmin$^2$.
The \rst\ WFI is comprised of 18 detectors arranged in a $6\times3$ 
array, covering a field-of-view that is $0\fdg8 \times 0\fdg4$ including the 
gaps between each detector. In practice, a \rst\ deep field would likely 
employ dithers large enough to fill in the detector gaps, thereby covering 
a larger area but with some variation in depth across the field. For our 
analysis, we consider a simplified \rst/WFI
field-of-view, adopting a single $0\fdg75 \times 0\fdg375$ rectangle that is
equivalent to the \rst\ field area excluding detector gaps. In this way 
we neglect the need for dithers and extract \rst\ fields from the SAM 
catalog that cover 0.281 degree$^2$ each but with the same axis ratio as 
the real detector.
We extract 1517 unique (non-overlapping) \hst\ fields from the 2 degree$^2$ 
catalog footprint, while only 4 unique \rst\ fields fit in the same 
area.

\subsection{A Single HUDF}\label{sec:1hudf}

\begin{figure*}
\epsscale{1.1}
\plotone{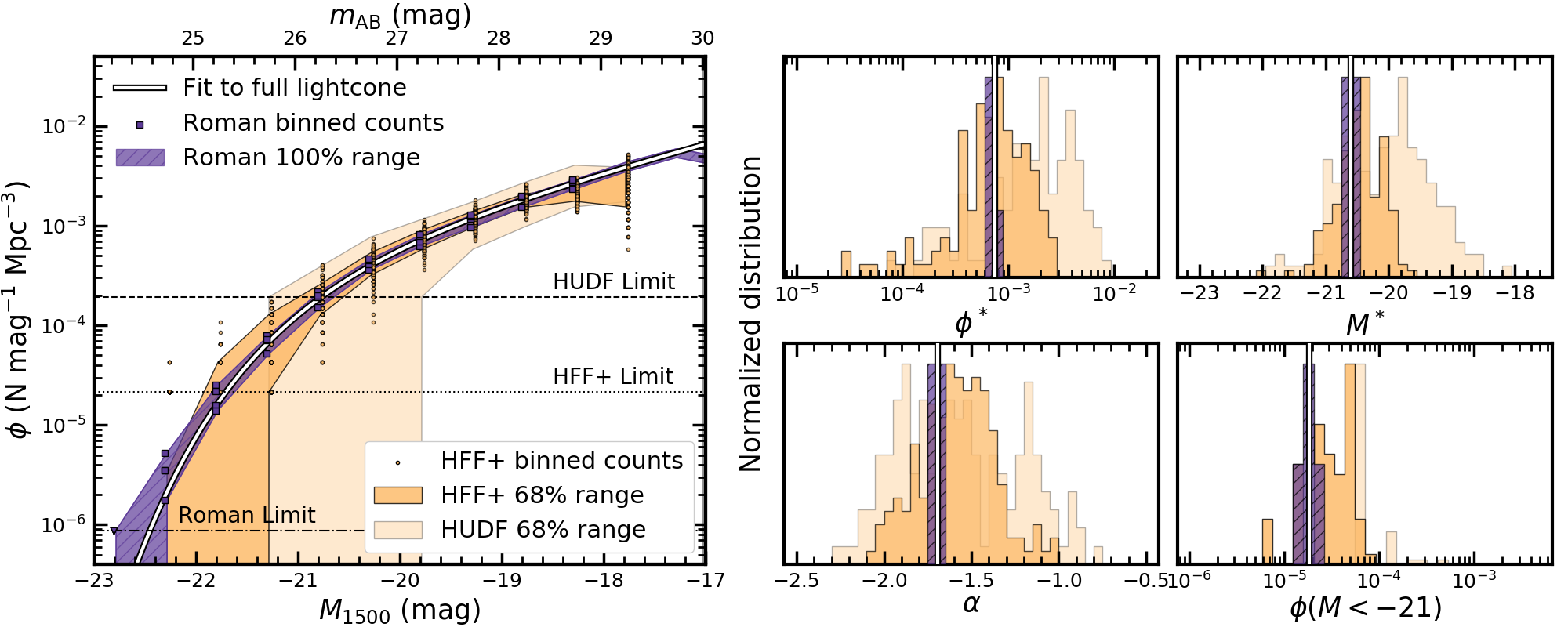}
\caption{The unscattered UV luminosity function measured in the redshift 
range $6.7 < z < 7.7$ for 150 randomly-selected single HUDFs 
($m_{\mathrm{F160W}} \leq 29.5$, light orange), 150 \hst\ deep field surveys 
(`HFF+', eight pointings with $m_{\mathrm{F160W}} \leq 29.0$ plus one
with $m_{\mathrm{F160W}} \leq 29.5$, darker orange), and four \rst\ fields 
($m_{\mathrm{F160W}} \leq 29.0$ purple). 
\textit{Left:} The shaded regions show the 68\% range of the binned number densities 
for both \hst\ surveys and the 100\% range for the four \rst\ fields.
The binned number densities measured in the HFF (orange circles) and \rst\ 
field (purple squares) are plotted slightly offset from their magnitude bin 
centers for clarity. 
The fit to the full 2 degree$^2$ lightcone ($m_{\mathrm{F160W}} \leq 30.0$)
is shown as the white curve and is taken as ``truth.''
\textit{Right}: We show the distribution of Schechter function parameters
fit to each set of binned number densities, including $\phi^*$ (top left), 
$M^*$ (top right) and $\alpha$ (bottom left). Finally, we show the cumulative 
number of bright ($M_{1500} < -21$) galaxies selected in each pointing or
set of pointings. In these panels, the \rst, 150 single HUDFs, and 150 sets 
of HFF+ pointings are shown as the filled purple hatched, light orange, and
dark orange histograms, respectively. The value from the fit to the full 
lightcone is plotted as a vertical white line. The distributions are 
normalized by their maximum value for easy comparison.
The rest-UV luminosity function measured in either a single HUDF-sized survey
or a survey consisting of nine \hst-sized pointings can vary significantly
from the ``truth,'' resulting in a range of Schechter function fits (including 
faint end slope measurements spanning $>1$) and calculated number of galaxies 
at the bright end. 
In a similar amount of time, a \rst\ survey could reach the depth of the HUDF 
while covering an area large enough to marginalize over density variations 
and recover the ``true'' luminosity function.
\label{fig:unscattered}}
\end{figure*}

In Figure~\ref{fig:sam} we show an illustration of the constraining 
power \rst\ brings to the study of high-redshift galaxy evolution.
The positions of all $6.5 < z < 7.5$ sources across the full 2 degree$^2$ 
catalog down to a depth of WFC3/IR $m_{\mathrm{F160W}} \leq 29.5$ (to
approximate the HUDF)
are shown as black points in the left panel. 
The orange points indicate sources covered by two \hst/WFC3 pointings, one
of which is `observing' an underdense region in the catalog and the other 
an area of overdensity.
The sources covered by a single \rst\ pointing are shown in purple. 

In the right panels, we show the rest-UV luminosity function calculated using
the galaxy counts in each pointing. 
The binned luminosity densities measured in the \hst\ and 
\rst\ fields are shown as orange circles and purple squares, respectively. 
Here we use the full volume probed by each field-of-view, i.e., assuming 
100\% completeness, and with magnitude bins of width $\Delta m=0.5$ mag. 
We fit a \citet{schechter1976} function to the binned densities, 
using a Markov Chain Monte Carlo (MCMC) method that is described in detail in 
Section~\ref{sec:lffits}. 
We show the Schechter function fits to the \hst\ 
binned densities (orange dashed curve), the \rst\ binned densities (purple
dash-dotted curve), and the binned densities in the full 2 degree$^2$
catalog (black solid curve). 
The luminosity functions measured in the \hst\ fields over- and under-predict 
the density of sources\footnote{The two luminosity function fits have similar characteristic densities ($\phi^* =2.6\times10^{-3}$ and $2.3\times10^{-3}$ Mpc$^{-3}$, respectively), but significantly different characteristic magnitudes ($M^* = -20.3$ and $-19.6$), resulting in cumulative number densities that are 73.9\% larger and 28.9\% smaller than truth, respectively. For comparison, the \rst\ Schechter function results in a cumulative number density that is only 2.9\% smaller than truth.} in the full 2 degree$^2$. Additionally, the 
luminosity function fits deviate significantly from the fit to the full 
catalog.
On the other hand, the \rst-measured luminosity function almost exactly 
traces that of the full catalog. A single \rst\ pointing covers a large enough 
field-of-view to smooth over most of the density variations in the lightcone at this redshift. 

Next, we consider the rest-UV luminosity function as calculated in 150 
randomly-selected single WFC3 pointings down to $m_{\mathrm{F160W}}=29.5$. 
Here we are exploring the variation in the UV luminosity function as measured
in 150 HUDF-type pointings placed randomly throughout the lightcone. We 
follow the same procedure described above, using 0.5 mag wide bins and fitting
Schechter functions to the binned densities using an MCMC analysis. We show 
the 68\% range of the binned densities in the 150 fields as the light orange 
shaded region in the left panel of Figure~\ref{fig:unscattered}. 
For comparison, the fit to all galaxies in the lightcone with
$m_{\mathrm{F160W}} \leq 30.0$ is shown as the white curve. The purple squares
show the binned densities measured in the four \rst\ pointings (with sources selected at $m_{\mathrm{F160W}} \leq 29.0$), with the full
(100\%) range of these densities plotted as the purple shaded hatched region.

There is significant field-to-field variation in the densities measured in 
\hst-sized pointings. The 68\% range in measured densities spans more than 
0.5 dex at the faint end, and increases to $>$1 dex at the bright end. 
Additionally, the bright limit for the HUDF-like pointings (i.e., the 
magnitude at which there is $\leq1$ galaxy in a given pointing) spans from 
$M_{1500} \sim -21.3$ to $M_{1500} \sim -19.8$. 
As illustrated in Figure~\ref{fig:sam}, it is unsurprising that the number 
densities and luminosity functions measured in a single \hst-sized pointing 
vary widely across the full SAM. Larger areas are needed to marginalize
over the variations in the density of dark matter halos. 
A single pointing is also insensitive to the rarer, brighter ($\lesssim M^*$)
galaxies, and larger areas are again required to sample this population. 
Studies aiming to develop a complete understanding of the rest-UV luminosity
function therefore often include multiple \hst\ surveys of varying depths
\citep[e.g.,][]{mcleod2015,mcleod2016,oesch2018,bouwens2019a,bouwens2021a,finkelstein2015a,finkelstein2022a,ishigaki2018}. 
In Section~\ref{sec:hffs}, we explore the extent of the improvement in the 
measurement of the UV luminosity function that is obtained by including eight 
additional moderately deep fields.

\subsection{Combination of Deep \hst\ Observations}\label{sec:hffs}
We next repeat the above procedure using 150 randomly-selected combinations 
of nine unique and non-contiguous \hst-sized fields. These nine pointings are 
comprised of one field reaching a depth of $m_{\mathrm{F160W}}=29.5$ to 
represent the HUDF and eight pointings reaching $m_{\mathrm{F160W}} = 29.0$ to 
represent the two HUDF parallel fields and the six HFF parallel fields. 
In Figure~\ref{fig:unscattered}, we refer to these sets of nine pointings 
collectively as HFF$+$ and plot the 68\% range of binned densities measured in 
the 150 sets of HFF$+$ pointings in dark orange. We also show the binned 
densities measued in each set of HFF$+$ pointings as small orange squares. The 
combined pointings probe a volume $9\times$ larger ($4.66\times10^4$ Mpc$^3$ 
versus $5.18\times10^3$ Mpc$^3$ for a single pointing) at all magnitudes 
brighter than $m_{\mathrm{F160W}}=29.0$ and therefore help constrain the 
luminosity function at magnitudes $\sim1-2$ brighter than the single HUDF-like 
pointing alone. 

Yet even when combining nine \hst-sized pointings, there is still significant
variation at both the bright and faint ends of the measured luminosity 
functions. The 68\% range at the bright end is still $\sim$0.5 dex wide, and 
the bright limit spans $\sim$0.7 magnitudes. The range of measured densities 
is relatively constrained from $M_{1500} \sim -21$ to $-18.5$, but is less
constrained in the faintest magnitude bin where the measured density is 
determined by only the single HUDF pointing in each HFF$+$ set. The combination of nine \hst-sized pointings is not enough to overcome the 
field-to-field density variation in the SAM lightcone.
Deeper, gravitationally lensed observations such as those from the primary Hubble Frontier Field pointings can offer further constraints at the faint end, but with larger cosmic variance uncertainties due to the smaller survey volumes \citep[e.g., a fractional uncertainty of $\sim35-65\%$ at $z\sim7-10$;][]{robertson2014}.
Studies measuring the luminosity function therefore often include several additional surveys
and a ``wedding cake'' approach, combining shallow wide-area surveys \citep[often ground-based, e.g.,][]{bowler2015} with a few deeper pointings. However, this approach requires combining different filter sets, imaging depths, resolutions, selection functions, and sample completenesses (and therefore effective volumes). Consistently-selected high-redshift galaxy samples can help mitigate some of the systematic uncertainties resulting from these heterogeneous studies. 

In the four right panels in Figure~\ref{fig:unscattered}, we plot histograms 
of the best-fitting Schechter function parameters for each observation set
as well as the cumulative number of sources detected with $M_{1500} \leq -21$.
The histograms are color-coded as in the left panel: light orange for the 
HUDF-like survey, darker orange for the HFF$+$ survey, purple hatched for 
\rst. The ``true'' values, obtained by fitting the full lightcone down to 
$m_{\mathrm{F160W}} \leq 30.0$, are plotted as vertical white lines.
The distributions of Schechter function parameters recovered by the HUDF-like 
and HFF$+$ surveys are very broad, spanning more than two orders of magnitude 
in $\phi^*$, three magnitudes in $M^*$, and with faint-end slopes ranging from
$\alpha < -2$ to $\alpha > -1$. While the HFF$+$ parameter distributions do 
improve on those of the HUDF surveys, the field-to-field variations in the 
lightcone are still dominant. The combination of nine independent \hst-size 
fields-of-view is not sufficient to fully constrain the luminosity function. 

However, the parameters measured from the four \rst\ pointings closely recover
the true parameter values. The \rst\ pointings each individually cover an area
wide enough to capture the average source density in the lightcone and 
constrain $\phi^*$ and $M^*$. Additionally, the \rst\ fields probe a much 
larger volume and therefore more fully sample the population of bright 
galaxies, constraining the bright end of the luminosity function. As the 
Schechter function parameters are degenerate, the strong constraints of the 
bright end and knee in the \rst\ observations allow for good constraints on 
the faint-end slope, even though the \rst\ fields considered here (down to $m_{\mathrm{F160W}}=29.0$) are 
shallower than the HUDF-like (and one ninth of the HFF$+$) surveys.
A survey that simultaneously reaches or exceeds the depth of the HUDF and
covers an area wide enough to overcome cosmic variance will be a game changer 
in measuring and constraining the rest-UV luminosity function in the early 
universe.

\section{A Trade Study of \rst\ Deep Fields}\label{sec:tradestudy}
\begin{deluxetable*}{lcccccccc}
\centering
\tablecaption{Depth/Area Trade Study -- PSFs \label{tab:grid}}
\tablehead{
\colhead{ID} & \colhead{Area} & \colhead{5$\sigma$ Limit} & 
\colhead{\rband} & \colhead{\zband} & \colhead{\yband} & \colhead{\jband} & 
\colhead{\hband} & \colhead{\fband} \\
\colhead{} & \colhead{(degree$^2$)} & \colhead{(AB mag)} &
\colhead{(hours)} & \colhead{(hours)} & \colhead{(hours)} & \colhead{(hours)} & 
\colhead{(hours)} & \colhead{(hours)}}
\decimalcolnumbers
\startdata
Base1 & 0.28 & 30.05 & \nodata & 224.3 & 86.9 & 93.5 & 102.8 & \nodata \\
\hspace{4mm}+\rband & 0.28 & 29.89 & 97.2 & 176.7 & 69.7 & 75.0 & 82.4 & \nodata \\
\hspace{4mm}+\fband & 0.28 & 29.87 & \nodata & 158.2 & 62.4 & 67.1 & 73.8 & 141.9 \\
\hspace{4mm}+\rband+\fband & 0.28 & 29.76 & 73.7 & 134.0 & 52.8 & 56.9 & 62.5 & 120.2 \\
Base2 & 0.56 & 29.67 & \nodata & 109.5 & 43.2 & 46.5 & 51.0 & \nodata \\
\hspace{4mm}+\rband & 0.56 & 29.52 & 48.3 & 87.8 & 34.6 & 37.2 & 40.9 & \nodata \\
\hspace{4mm}+\fband & 0.56 & 29.50 & \nodata & 78.6 & 31.0 & 33.3 & 36.6 & 70.4 \\
\hspace{4mm}+\rband+\fband & 0.56 & 29.38 & 36.6 & 66.6 & 26.2 & 28.3 & 31.0 & 59.7 \\
Base4 & 1.12 & 29.30 & \nodata & 54.4 & 21.4 & 23.1 & 25.3 & \nodata \\
\hspace{4mm}+\rband & 1.12 & 29.14 & 24.4 & 44.4 & 17.5 & 18.8 & 20.7 & \nodata \\
\hspace{4mm}+\fband & 1.12 & 29.12 & \nodata & 39.0 & 15.4 & 16.6 & 18.2 & 35.0 \\
\hspace{4mm}+\rband+\fband & 1.12 & 29.00 & 18.5 & 33.7 & 13.3 & 14.3 & 15.7 & 30.2 \\
Base7 & 2.0 & 28.99 & \nodata & 27.5 & 10.8 & 11.7 & 12.8 & \nodata \\
\hspace{4mm}+\rband & 2.0 & 28.84 & 12.1 & 22.0 & 8.7 & 9.4 & 10.3 & \nodata \\
\hspace{4mm}+\fband & 2.0 & 28.82 & \nodata & 19.7 & 7.8 & 8.4 & 9.2 & 17.7 \\
\hspace{4mm}+\rband+\fband & 2.0 & 28.70 & 9.2 & 16.7 & 6.6 & 7.1 & 7.8 & 15.0 \\
\enddata
\tablecomments{
Columns are:
(1) Survey identification, where `Base' indicates \textit{zYJH} photometry, 
to which \rband, \fband, or both (+\rband+\fband) are added, and 
the number indicates how many \rst\ pointings are included;
(2) Survey area; 
(3) $5\sigma$ limiting magnitude in the detection filters (\yband, \jband, 
\hband, \fband),
while the dropout bands (\rband, \zband) are always 0.5 magnitude deeper;
(4-9) Exposure time in each filter for a single \rst\ pointing. 
For example, `Base4' indicates 4 \rst\ pointings of \textit{zYJH}
photometry, and so the exposure time in each filter should be multiplied 
by 4 to obtain the total program time.
Each survey takes approximately 500 hours.
}
\end{deluxetable*}

While Section 3 established the potential of a Roman deep field to uncover the shape of the true UV luminosity function, a deep field observation strategy requires finding the optimal compromise between depth, survey area, and filter coverage.
We therefore consider a grid of possible 
deep field survey strategies, each of which comes with trade-offs related to 
the recovery of high-redshift galaxies, contamination from low-redshift 
galaxies, and the measured number densities as a function of redshift and 
magnitude. 
We adopt four survey areas: 0.28 degree$^2$ (a single \rst\ pointing), 0.56 degree$^2$ (two pointings), 1.12 degree$^2$ (four pointings, and the full 2 degree$^2$ of the mock catalog (an area equivalent to that covered by approximately seven \rst\ pointings). As in Section~\ref{sec:noiseless}, we consider a simplified 
WFI detector layout and neglect dithers, assuming that the full area 
is covered to a uniform depth. For each survey area, we explore four 
different filter combinations: a `base' survey consisting of \zband, \yband, 
\jband\ and \hband; the base configuration with the \rband\ filter added; 
the base configuration with the \fband\ filter added; and finally, the base 
configuration with both the \rband\ and \fband\ filters added.
We therefore consider a total of sixteen possible \rst\ surveys, each
totalling $\sim$500 hours of exposure time, where the combination of 
area covered and maximum time determines the depths attained in each filter.
We note that we do not consider the $K_{213}$ filter in this analysis, as it was not included in the SAM mock catalog (the $K_{213}$ band was added to the WFI after the SAM inputs were constructed). With a maximum redshift of $z=10$ in the SAM mock catalog, we limit our samples to $z\leq9$ where galaxies are detectable in two or more filters even without including $K_{213}$. Additionally, due to \rst's passive cooling, the $K_{213}$ filter achieves a much shallower flux limit ($\sim1.5$ mag brighter) than the other filters in comparable time. 

In the following sections, we describe how we determine the filter exposure 
times and depths for each survey (Section~\ref{sec:exptimes}) and how we 
perturb the SAM mock catalog fluxes to mimic those depths 
(Section~\ref{sec:perturbation}).

\subsection{Survey Exposure Times and Filter Depths}\label{sec:exptimes}
We estimate exposure times in each filter using two methods, the 
Pandeia\footnote{\url{https://outerspace.stsci.edu/display/PEN}}
exposure time calculator engine \citep[version 1.6.2;][]{pandeia}, and the 
anticipated performance tables released by NASA Goddard Space Flight 
Center\footnote{\url{https://roman.gsfc.nasa.gov/science/apttables2021/table-exposuretimes.html}} for the WFI. 
In determining the estimated depths with each tool, we assume a point source 
morphology. This choice is motivated because 
galaxies at $z\sim6-9$ are expected to have effective radii of $\sim$0.25$-$0.3
kpc \citep{kawamata2018}, corresponding to $\sim$0\farcs053$-$0\farcs058.
The full width at half maximum of the $r$ filter is expected to be 0\farcs058, 
with all other filters larger by a factor of two or more. Even in the $r$ band,
high-redshift galaxies will be barely resolved. 

For the Pandeia calculations, we consider a flat spectrum in $f_{\nu}$ and 
adopt the 1024$\times$1024 detector subarray. As Pandeia was written for 
\jwst, exposures are defined in terms of \jwst\ readout modes, the number 
of frames averaged in each group, the number of groups per integration, and 
the number of integrations per exposure. We use the MEDIUM8 readout mode 
(eight frames per group followed by two skipped frames) and ten groups, as 
recommended to match the expected performance of the WFI imaging. 
We then increase the number of integrations per exposure as needed to achieve
a signal-to-noise (S/N) of 5 for a given magnitude in each filter. 
For the calculations, we assume a low background level, which corresponds to 
10\% of the maximum background as measured at the ``Minzodi' location 
(ecliptic longitude, latitude = 266.3\degr, -50.0\degr).
The Pandeia backgrounds are calculated by the \jwst\ Backgrounds Tool\footnote{\url{github.com/spacetelescope/jwst_backgrounds}}.
Finally, source photometry is calculated in a circular aperture of radius 
$r=0\farcs22$ (2 pixels) and background subtracted using the sky count rate
measured in an annulus with inner and outer radii 0\farcs4 and 0\farcs6.

The Goddard anticipated performance tables provide estimates for the exposure
time required to reach a given signal-to-noise in all filters. The exposure 
times are provided in groups of six readout frames. As with the Pandeia 
calculations, we adopt the table with S/N computed in a circular aperture with 
radius $r=0\farcs22$. The assumed zodiacal light level is set to 1.2$\times$
the minimum value, corresponding to the sky background at an approximate 
location of ecliptic longitude, latitude = 90\degr, 60\degr, and 
representing the background in the majority of the continuous viewing zone.

The depths predicted by these two tools differ by approximately 0.6 magnitudes
for a given exposure time, with the Goddard tables estimating the deeper depth. 
We believe the main differences between the two exposure time calculators are 
due to the different background models adopted, and the differing techniques 
used to extract photometry. For this paper, we therefore adopt an average of
the exposure times estimated by the two tools. Our adopted exposure times and
corresponding depths may not reflect the exact sensitivity of the WFI once it 
is on sky, but they are sufficient for our goal of exploring a grid of 
potential \rst\ deep field survey strategies.
However, while we have assumed that the true WFI performance will lie somewhere 
between the estimates from the two tools, this difference in depths should 
also be taken as a source of uncertainty in our results. 

\begin{figure}
\epsscale{1.1}
\plotone{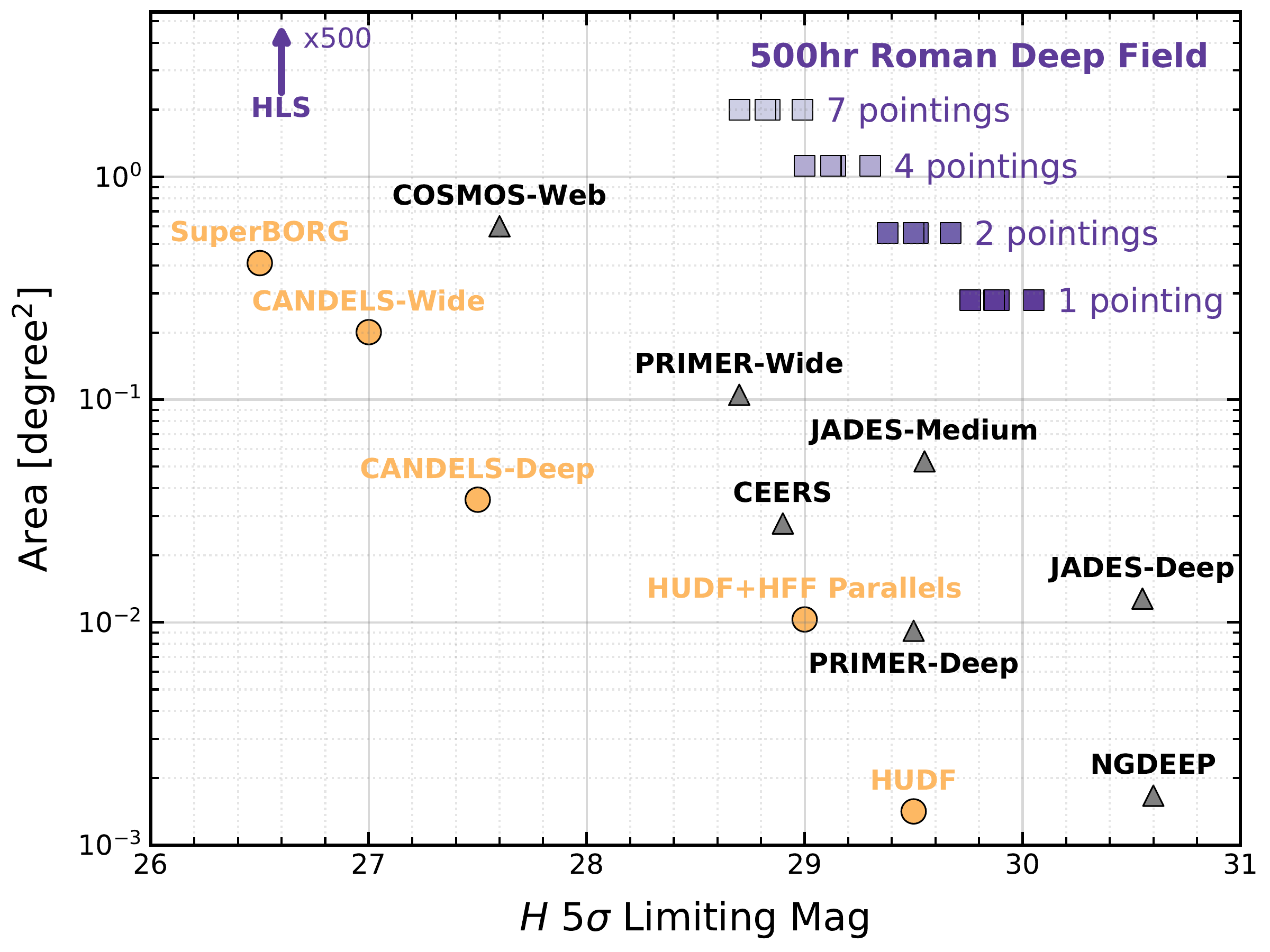}
\caption{
The area as a function of $H$-band depth for a variety of extragalactic 
surveys.
We show \hst\ surveys as orange circles and current and scheduled \jwst\ 
surveys as grey triangles. The grid of 16 \rst\ deep field surveys we 
consider in this paper are plotted as purple squares. The \rst\ deep fields
all occupy a unique part of the parameter space, covering large areas at the
depth of all but the deepest extragalactic surveys. The \rst\ High Latitude 
Wide Area Survey depth is indicated by the purple arrow, while the planned 
area coverage lies far above the limits of this figure.
\label{fig:surveys}}
\end{figure}

We present the exposure times and depths per filter in Table~\ref{tab:grid}. 
Each row represents one of the 16 surveys, where `Base' indicates imaging in 
\zband, \yband, \jband\ and \hband, and the number following the underscore 
indicates how many \rst\ pointings are included. To each of the four base 
surveys, we add versions that include \rband, \fband, and both \rband\ and 
\fband. 
Each survey consists of at least one filter that 
will be used to image blueward of the \lya\ break for redshifts $z=6-9$. These 
dropout filters (\rband\ and \zband) go 0.5 magnitudes deeper in order to 
enable the detection of \lya\ breaks at the 5$\sigma$ depth of the detection 
filters. Table~\ref{tab:grid} lists the 5$\sigma$ limiting magnitude in the detection filters and the exposure time in hours per pointing
required to reach the required depths, with the total exposure time per 
survey adding to approximately 500 hours.

\begin{figure*}
\epsscale{1.15}
\plotone{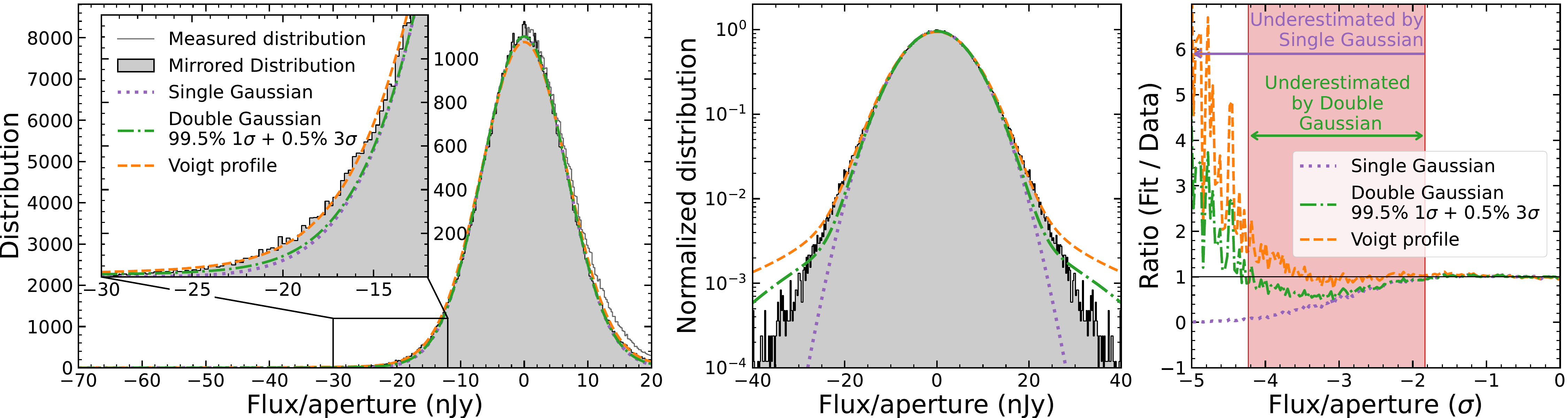}
\caption{Noise as measured in sky apertures with $r=0\farcs2$ across the 
F160W image (0\farcs06/pixel) in the EGS field, avoiding source flux and 
bad pixels.
\textit{Left:} The measured distribution of fluxes in the sky apertures is
shown as a grey curve. We fit the distribution mirrored about the peak
(filled grey histogram) to characterize the noise in the image while avoiding
all source flux. 
We fit the mirrored distribution with a single Gaussian (purple dotted curve),
a composite of two Gaussians (green dash-dotted curve), and a Voigt profile
(orange dashed curve). The inset zooms in on the wings of the distribution,
where the Gaussian fits underestimate the measured noise. 
\textit{Middle}: The mirrored distribution is normalized to a peak of 1 and 
plotted on a logarithmic scale. 
\textit{Right:} We plot the ratio of each profile fit to the measured data as a function of Gaussian $\sigma$. The 
single Gaussian significantly underestimates the noise in the wings of the distribution. The vertical red shaded region indicates the regime in which 
even the double Gaussian underestimates the noise. The Voigt profile 
traces the distribution out to $\sim$3.8$\sigma$ ($\sim$25 nJy) and 
overestimates the noise further in the wings. 
Noise in images is not completely normally-distributed. Fully reproducing
the photometric scatter of an imaging survey requires a noise distribution 
that scatters additional sources further out in the wings. We adopt the 
Voigt profile as the conservative option that does not underestimate this 
scatter.
\label{fig:voigt}}
\end{figure*}

We plot the 16 survey areas as a function of depth in Figure~\ref{fig:surveys},
comparing them to several existing and scheduled \hst\ and \jwst\ surveys. 
The \rst\ deep fields proposed here occupy an important part of the parameter
space, simultaneously reaching the depths of several of the deepest existing
surveys while covering areas approaching or exceeding the widest existing
programs. For reference, we also show the depth of the \rst\ High Latitude 
Wide Area Survey, which will cover at least 1700 degree$^2$ and so lies far
above the figure shown here.

\subsection{Perturbing Source Fluxes} \label{sec:perturbation}
We next approximate ``observing'' the SAM mock catalog by applying 
photometric scatter to the catalog fluxes according to the depths 
presented in Table~\ref{tab:grid}.
This step requires adopting a distribution for the flux perturbations that 
is appropriate for the noise properties in WFI images. 
It is often assumed that image noise is normally distributed about zero, 
such that the significance of a detection above the level of noise 
corresponds to the probability that the detection is real and not a 
noise fluctuation. 
However, as demonstrated by \citet{schmidt2014}, the 
noise distribution in images can have highly non-Gaussian tails, and
fully reproducing the photometric scatter of an imaging survey requires 
scattering additional sources further out in the wings. 

We quantify this departure from Gaussianity using an approach similar to that
of \citet{schmidt2014} and \citet{finkelstein2015a}. For this analysis, we 
explored the noise in CANDELS images of the EGS field drizzled on a 
0\farcs06/pixel scale. We placed a grid of over 2.6 million 
$r=0\farcs2$ apertures across the images in each filter, using the 
segmentation and weight maps to avoid 
source flux and bad pixels, respectively. The distribution of negative fluxes 
measured in these sky apertures is shown in grey in Figure~\ref{fig:voigt},
where we have mirrored the distribution about the peak at zero flux. In this 
way, we exclude from our fitting any positive flux contributions from 
unresolved sources. We focus here on the F160W image, as the wings of the 
noise distribution are largest in this filter.
A single Gaussian profile can fit the core of the distribution but 
underestimates the wings, dropping below the observed distribution at
$\sim$1.5$-$2.3$\sigma$ ($\sim$10$-$15 nJy from the profile center). 
We also show the result of fitting 
the distribution with a composite of two Gaussian profiles, where the 
second Gaussian contributes 5\% to the total profile and has a standard 
deviation $3\times$ that of the first. 
This composite profile exceeds the amplitude of the tails
but still underestimates the distribution in the range 
$\sim$1.8$-$2.6$\sigma$ ($\sim$12$-$17 nJy). 

We therefore adopt a Voigt profile, which combines a Gaussian core with 
the wings of a Lorentzian distribution. The Voigt function depends on 
two parameters, the standard deviation $\sigma$ of the Gaussian component 
and the half-width at half-maximum $\lambda$ of the Lorentzian. 
We set $\lambda = 0.05 \sigma$, determined by fitting the noise measured 
in the EGS F160W image. The Voigt profile (orange dashed line in 
Figure~\ref{fig:voigt}) traces the observed distribution out to 
$\sim$3.8$\sigma$ ($\sim$25 nJy) but overestimates the strength of the 
wings further from the profile center. However, overestimation in the wings is 
preferable to underestimation closer to the distribution center where we 
would be under-scattering the fluxes of more sources.
We note that with the two-component Gaussian fit, increasing the standard 
deviation (i.e., from $3\times$ to $5\times$) and/or the percentage 
(i.e., from 5\% to 10-15\%) of the second component does as fine a job 
at approximating the measured distribution as the Voigt profile at 
$\sim$2$\sigma$, but at the expense of much broader wings at $\geq4\sigma$. 
The Voigt profile therefore provides a conservative approach that allows us to 
reproduce the photometric scatter present in \hst/WFC3 images out to 
$>$3.5$\sigma$ and to err on the side of larger perturbations for a small
fraction of sources in the wings. 

For each of the sixteen \rst\ deep surveys, we create a catalog of 
photometrically scattered fluxes. The flux of each source is perturbed 
by an amount pulled randomly from a Voigt profile centered at the true 
value. The $\sigma$ parameter for the Voigt profile is set at the $1\sigma$ 
flux that corresponds to each filter's $5\sigma$ magnitude limit in 
Table~\ref{tab:grid}, with the dropout filters ($z$ and $r$ where applicable) 
going 0.5 magnitudes deeper. 
We adopt the $1\sigma$ fluxes as the photometric uncertainties for each 
filter, and use these sixteen catalogs in our sample selection described 
in the following sections.

\subsection{Photometric Redshift Determination} \label{sec:photoz}
Our high-redshift sample selection is based on a photometric redshift 
analysis. 
Such an approach allows for a more flexible and complete selection of 
galaxies than strict color cuts \citep[e.g.,][]{mclure2010,finkelstein2015a,finkelstein2022a,finkelstein2022c,finkelstein2023a,finkelstein2023b,bouwens2019a,bouwens2021a,bowler2020,rojas-ruiz2020,bagley2024a}.
Photometric redshift fitting involves all available colors, and therefore 
depends not only on the detection of the \lya\ break but also on the slope of
the galaxy spectrum redward of the break. Selecting galaxies based on their 
redshift probability distributions may include sources that would be excluded 
from a color-color selection window. However, such galaxy samples may also 
be more highly contaminated by lower-redshift sources.

We begin by requiring a signal-to-noise (S/N) of 5 in the \hband\ filter
to select strongly-detected sources. We next calculate photometric redshifts 
for all detected sources using the Easy and Accurate \zphot\ from Yale 
\citep[\eazy\ version 2015-05-08;][]{brammer2008} software, which fits 
photometric measurements to synthetic spectral templates through chi-squared 
minimization. 
We allow \eazy\ to consider linear combinations of the input templates.
The redshift fitting is therefore less dependent on the choice of spectral 
templates, but does require that the colors of the templates fully span the 
expected colors of the sources to be fit.

As demonstrated by \citet{larson2022b}, the standard template 
set available with \eazy\ does not include templates blue enough to bracket 
the colors of galaxies that are observed at high redshifts with \hst\
\citep[e.g.,][]{bouwens2009,finkelstein2012a} and in the early results 
from \jwst \citep{arellano-cordova2022,rhoads2022}.
The authors therefore propose the inclusion of a new set of templates created
with \bpass\ spectral energy distribution templates \citep{eldridge2017,stanway2018}.
These new templates have low metallicities (0.05 $Z_{\odot}$) and stellar 
populations with ages of $10^6$, $10^{6.5}$ and $10^7$ Myr. 
\citet{larson2022b} also provide templates that include nebular emission 
lines modelled with \cloudy\ \citep{ferland2017}. However, the broadband 
magnitudes of galaxies in the SAM mock catalog do not include the 
contributions of emission lines, and so we only use the emission-line-free
templates from \citet{larson2022b}.
In total we use 15 templates, the three from \citet{larson2022b} and the 
12 \texttt{tweak\_fsps\_QSF\_12\_v3} \eazy\ templates that are based on the 
Flexible Stellar Population Sysnthsis (FSPS) models 
\citep{conroy2009,conroy2010}. 

We run \eazy\ in the redshift range $0.01 \leq z \leq 12$ in steps of 
$\Delta z =0.01$. The fluxes in each filter are perturbed as described in 
Section~\ref{sec:perturbation}, and the flux uncertainties are the $1\sigma$
fluxes corresponding to the filter's $5\sigma$ magnitude limit. We also 
add an additional minimum fractional error of 0.05 to the flux uncertainties 
in all filters (\eazy\ parameter \texttt{SYS\_ERR}) to mimic systematic 
uncertainties in flux measurements that may be expected with real data.
Redshift-dependent attenuation from the intergalactic medium is applied 
following \citet{madau1995} with \lya\ forest and damped \lya\ system 
absorption following \citet{inoue2014}. We adopt a flat luminosity prior 
because the luminosity density of galaxies at high redshift is not yet 
fully understood. A flat prior will not downselect true high-redshift sources,
yet considers high redshift and lower-redshift galaxies as equally likely.
This choice therefore results in more complete samples that are also 
more contaminated by low-redshift galaxies.

\begin{figure}
\epsscale{1.15}
\plotone{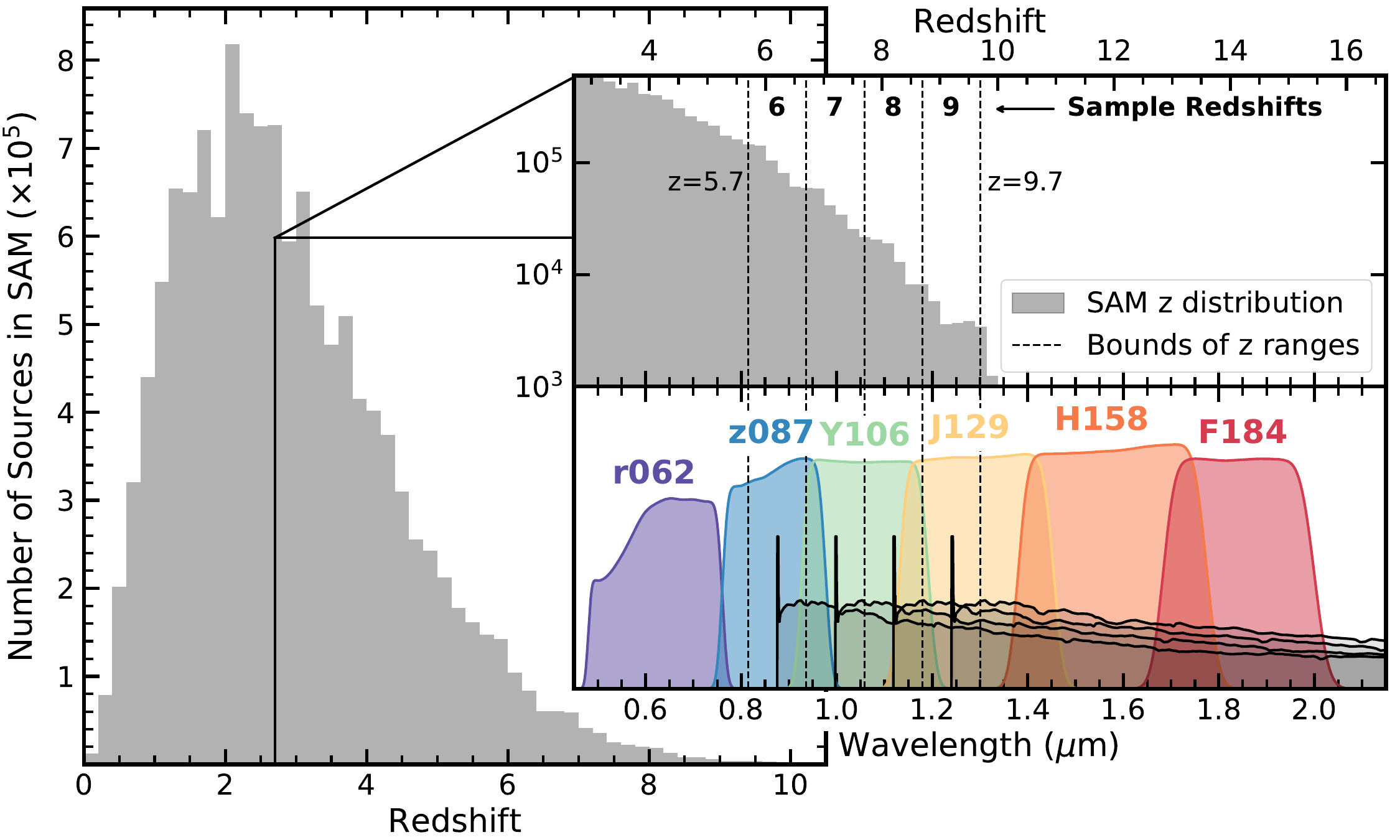}
\caption{
The redshift distribution of sources in the SAM mock catalog. 
The top inset shows the distribution of sources at $z>2.7$ on a log scale to 
highlight the distribution at high redshift. 
In the bottom inset, we show the \rst/WFI filter profiles and a model galaxy
spectrum plotted at each of the sample redshifts we consider: $z=6$, 7, 8 and 9.
The dotted vertical lines indicate the bounds of the $\Delta z = 1$ redshift 
ranges considered for each redshift sample. The lowest bound ($z=5.7$) is 
determined by the wavelength at which the transmission in the \rband\ filter 
drops to effectively zero such that a \lya\ break will completely drop out 
of this filter.
\label{fig:samfilts}}
\end{figure}

\subsection{Sample Selection} \label{sec:sample}
We select galaxies from the perturbed SAM mock catalogs in four 
$\Delta z=1$ redshift ranges centered at $z=6.2$, 7.2, 8.2 and 9.2. 
The ranges are chosen such that the Lyman-$\alpha$ break for sources at 
the low redshift edge of the $z\sim6$ bin has fully redshifted out of the 
\rband\ filter. Sources in all redshift ranges of interest therefore have 
the potential to be ``observed'' with a dropout filter, assuming 
the \rband\ filter is part of the survey in question.
Additionally, we select sources in the range $9.7 < z < 10$ in order to 
explore recovered number densities out to the edge of the SAM at $z=10$, 
though we do not calculate a luminosity function for sources in this 
reduced bin. 

\begin{figure*}
\epsscale{1.1}
\plotone{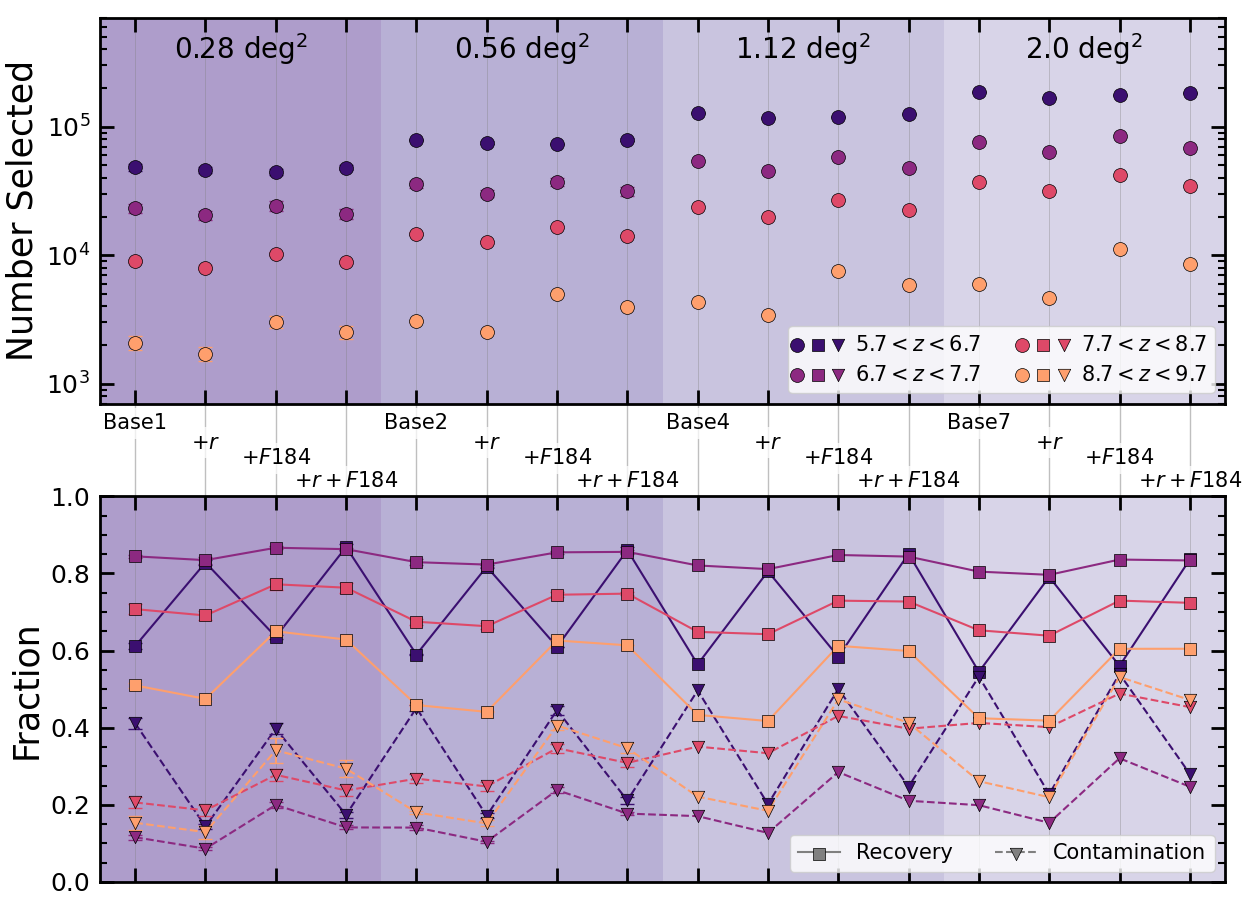}
\caption{
The results of sample selection for each survey. (\textit{Top:}) The 
number of sources selected in each redshift bin. 
(\textit{Bottom:}) The fraction of recovered sources (squares and solid lines)
and low-redshift contaminants (triangles and dashed lines) in each sample.
For surveys covering 1 and 2 \rst\ pointings (0.28 deg$^2$ and 0.56 deg$^2$), 
we extract the corresponding survey area multiple times (four and two times, 
respectively). Here we plot the median sample numbers and fractions for these
two survey areas with error bars (often smaller than the plot symbols)
indicating the standard deviations. 
Notably, the sample contamination increases in all samples with increasing 
survey area as the depth in each filter decreases. The large variation in 
the $z=6$ sample fractions highlights the need for the \rband\ filter to
improve both recovery and contamination.
The addition of \fband\ also improves recovery for the $z=9$ sample, but at 
the expense of increased contamination.
The table associated with this figure is available in the online journal.
\label{fig:samples}}
\end{figure*}

Our selection criteria are similar to those used to select candidates from 
real imaging catalogs \citep[e.g.,][]{finkelstein2015a,finkelstein2022a,rojas-ruiz2020,bouwens2019a,bouwens2021a,bagley2024a}.
In addition to the (S/N)$_H>5$ requirement, we consider the best-fit 
redshift value as well as the full photometric redshift probability 
distribution function (PDF) calculated for each source.
We select sources that satisfy the following criteria:
\begin{enumerate}
\item A detection in \hband\ with (S/N)$_H \geq 5$; 
\item At least 70\% of the integrated redshift PDF lies under the primary 
redshift probability peak, leaving less than 30\% of the redshift PDF for any 
additional redshift solutions; 
\item At least 35\% of the integrated redshift PDF lies in one of the sample 
redshift ranges (i.e., $6.7 < z < 7.7$ for the $z=7$ sample), ensuring 
that the redshift PDF is not too broad;
\item The integrated redshift PDF in the sample redshift range of interest 
is greater than the integral in all other integer redshift ranges, so that 
we do not select the same source multiple times in different redshift samples;
\item At least 60\% of the integrated redshift PDF lies above
$z\_{\mathrm{sample}} - 1$ (i.e., $\int P(z>6) > 0.6$ for the $z=7$ sample),
again selecting sources with a majority of their redshift PDF at 
high redshift;
\item The best-fitting redshift is above $z_{\mathrm{sample}} - 2$, 
indicating that \eazy\ securely selected the source at high redshift
\end{enumerate}

We perform this sample selection in subsets of the SAM mock catalog that 
match the areas of each survey. As described in Section~\ref{sec:noiseless}, 
we consider a simplified \rst/WFI footprint that is $0\fdg75 \times 0\fdg375$
and can fit four such unique rectangles in the 2 degree$^2$ mock catalog.
For the two smallest survey areas (``Base1'' covering 0.28 degree$^2$ and 
``Base2'' covering 0.56 degree$^2$), we extract the relevant \rst/WFI 
area multiple times, creating multiple realizations of each survey. 
Although the number of realizations is small, they provide an empirical 
estimate of the effect of cosmic variance on these single and double 
\rst\ pointings (see Section~\ref{sec:cv}).
Figure~\ref{fig:samples} provides a summary of the samples selected in each 
redshift range and survey. We show the total number of sources selected in 
the top panel, color-coded by redshift. In the bottom panel, we plot the 
recovery (squares connected by solid lines) and contamination fractions
(triangles with dashed lines). In both panels, error bars on the measurements
for the 0.28 and 0.56 degree$^2$ surveys indicate the standard deviation of 
these quantities in the available realizations.

Here, we define the recovery fraction as the number of selected sources  
with true redshifts in the sample range (e.g., $5.7 \leq z < 6.7$ for 
the $z=6$ sample) divided by the total number of sources with 
$\mathrm{S/N}_H \geq5$ in the SAM lightcone that are in the same redshift 
range. We distinguish this definition of recovery from the completeness 
needed to measure effective volumes for the luminosity function, because the 
SAM does not contain enough $z>9$ galaxies for adequate completeness 
statistics at all magnitudes. We therefore calculate sample completeness using
a separate set of simulated sources as discussed in Section~\ref{sec:veff}.
The contamination fraction is calculated as the number of selected sources 
with true redshifts that are $\Delta z > 1$ from the midpoint of the sample
redshift range (e.g., $z<6.2$ or $z>8.2$ for the $z=7$ sample) divided 
by the total number of sources in the sample.
Figure~\ref{fig:samples} provides a single set of recovery and contamination
fractions per redshift sample and survey. However, within a single sample, 
the recovery and contamination depend on both magnitude and redshift. 
We explore these dependencies in Section~\ref{sec:recov} and \ref{sec:contam}.

We note that in this work we are not promoting a specific selection criteria
for \rst\ deep observations. Instead, we wish to explore the benefits and 
drawbacks associated with different survey depths, areas and filter coverages.
While we aim to create as clean a sample as possible for our analysis, our 
focus is therefore on consistent sample selection across all surveys and a 
careful quantification of the resulting sample recovery and contamination
(see Sections~\ref{sec:recov} and \ref{sec:contam}).

\section{Recovery of High-Redshift Sources}\label{sec:recov}
We use the recovery fraction -- the fraction of real high-redshift galaxies that
have been recovered in each sample -- as a way to help evaluate the performance of 
the selection criteria in each survey.
We again note that this recovery fraction is separate from the 
survey completeness that is discussed in Section~\ref{sec:veff}.

In the bottom panel of Figure~\ref{fig:samples}, the fraction of recovered 
sources can vary significantly by redshift range and survey. The variation is 
most striking at $z\sim6$, where the recovery drops by $\sim$20--30\% when the
\rband\ filter is not in use. In this redshift bin, \rband\ is the only \rst/WFI
filter blueward of the \lya\ break. Surveys that do not use this filter 
will be selecting $z\sim6$ galaxies based on their SED shape, but at the 
cost of lower overall recovery and higher contamination from lower-redshift
galaxies. The recovery also varies in the $z\sim9$ redshift range, where 
the \lya\ break is in the \jband\ filter and \hband\ is the first 
detection band redward of the break. At this redshift, including \fband\ 
increases source recovery by providing a second detection filter and therefore 
an additional measurement of the rest-UV colors of the galaxy.
Overall, the sample recovery for a given filter set decreases slightly as the
survey area increases. This lower recovery is caused by the decreased depth in 
all filters. However, as mentioned above, Figure~\ref{fig:samples} provides 
a single snapshot of the recovery for each survey, which over-simplifies the 
performance of the surveys.

\begin{figure*}
\gridline{\fig{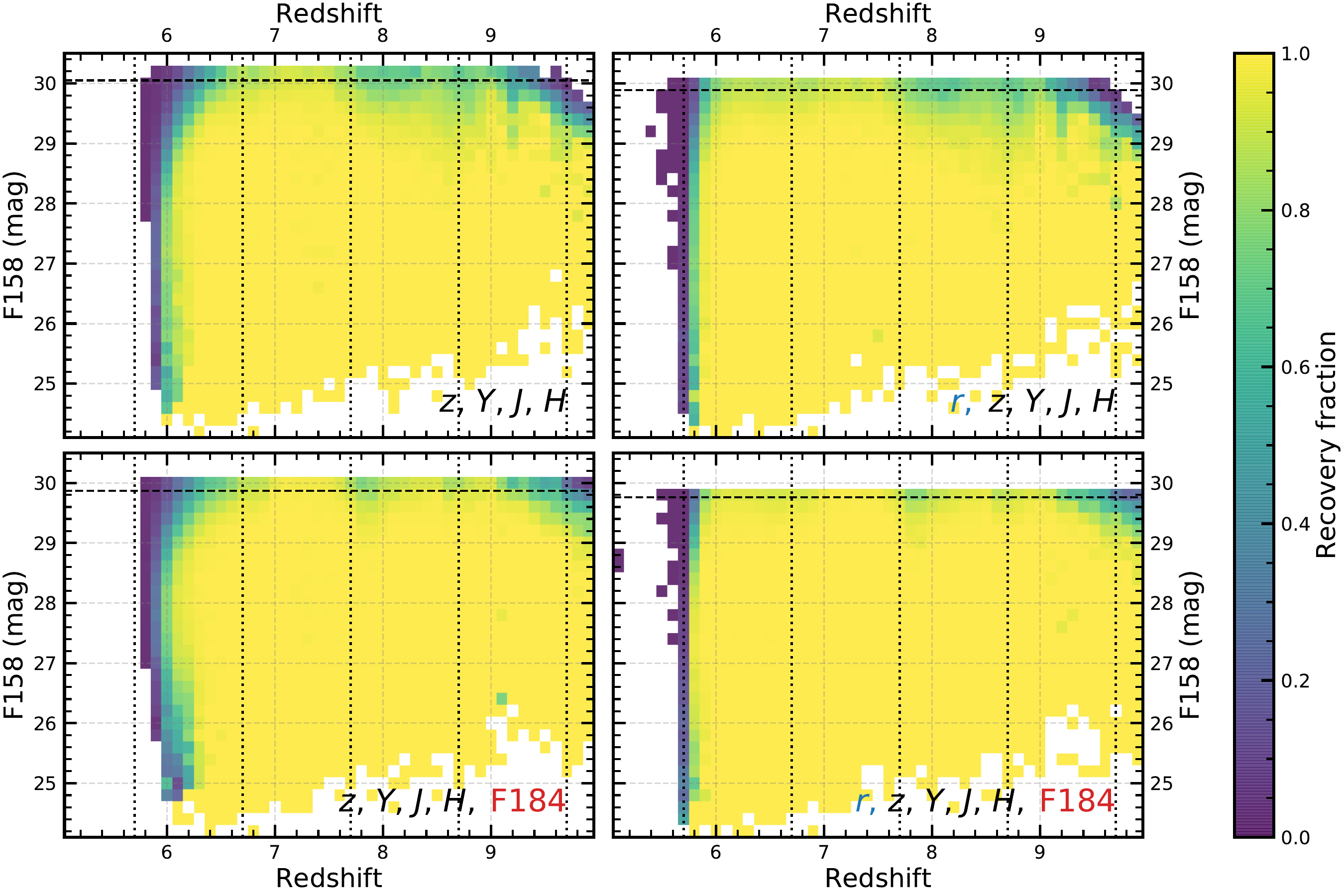}{0.49\textwidth}{}
          \fig{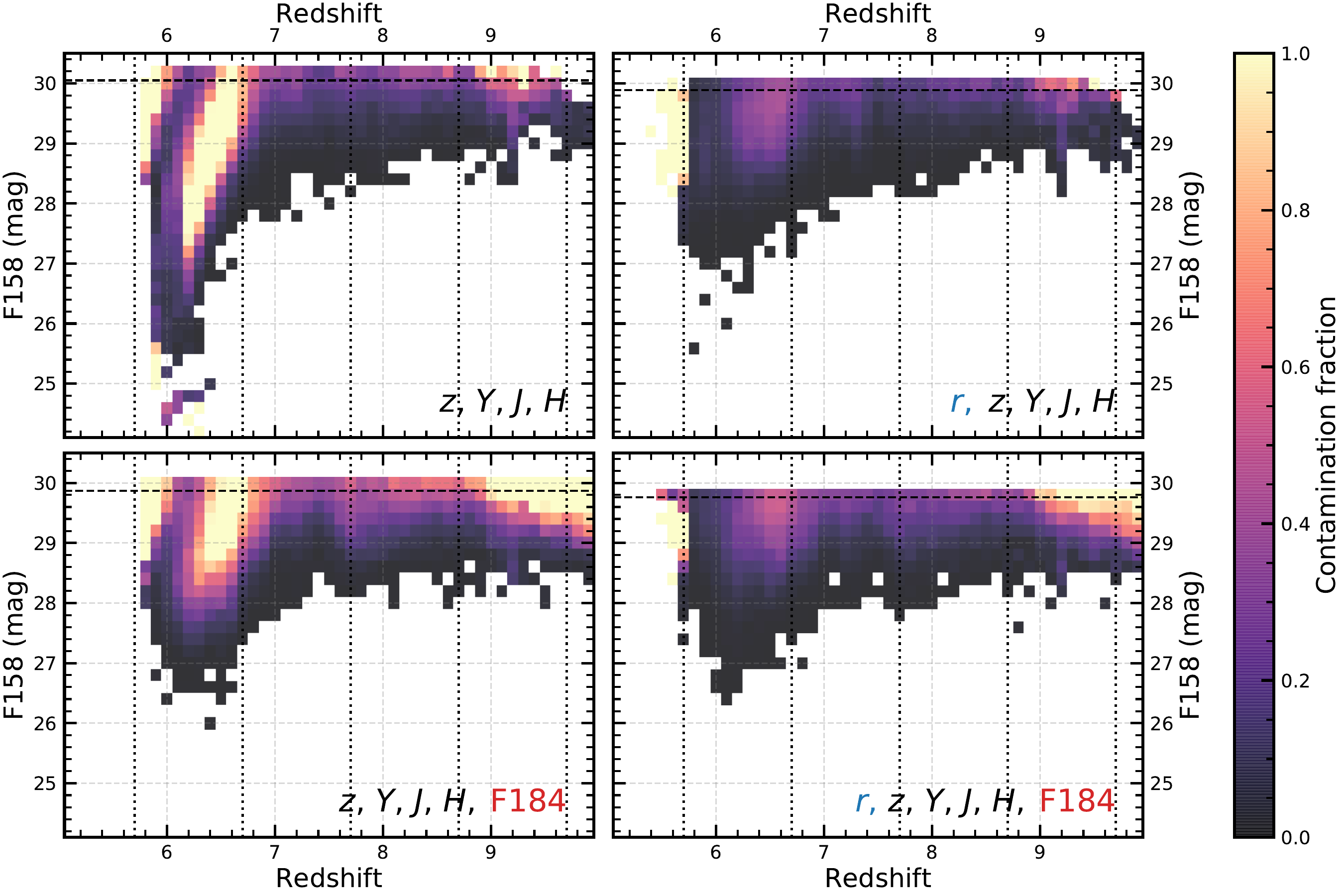}{0.49\textwidth}{}}
\caption{
The recovery and contamination of the four surveys covering a given survey 
area. Each set of four panels shows results for the Base 
(top left), Base+F602 (top right), Base+\fband\ (bottom left), and
Base+\rband+\fband\ (bottom right) surveys. The vertical dotted lines show the edges 
of each redshift sample, and the horizontal dashed lines indicate the 
5$\sigma$ magnitude limits for each survey.
\textit{Left:} The recovery fraction in bins of input redshift and observed 
(scattered) \hband\ magnitude. 
\textit{Right:} The contamination fraction in bins of input redshift and 
observed \hband\ magnitude.
The base survey has the lowest recovery and highest contamination 
(especially in the $z=6$ sample), whereas including \rband\ significantly 
improves both fractions. Overall, the Base+\rband+\fband\ survey has the
highest recovery, while the Base+\rband\ has the lowest contamination. 
These filter set trends are similar across all survey areas, with
the main difference being a decrease in overall depth with increasing area.
Results for all surveys are included in the online version as a figure set. 
We show the results for the surveys covering 0.28 degree$^2$ (1 \rst\
pointing) as an example in the PDF version. (The remaining figures are presented in the Appendix as Figure~\ref{fig:contam_recov_temp} for the arXiv version.)
\label{fig:contam_recov}}
\end{figure*}

We therefore next explore the recovery fraction in each survey as a function 
of redshift and magnitude. In the left panels of Figure~\ref{fig:contam_recov}, 
we show the recovery of each survey binned by observed (or photometrically
scattered) \hband\ magnitude and redshift. Each of the four panels shows the 
results for one filter set (the base set of \zband, \yband, \jband\ and 
\hband; the base set plus \rband; plus \fband; and plus \rband\ and \fband)
for the given survey area.

The recovery is lowest at the low-redshift end, where the \lya\ break occurs 
in the middle of the \zband\ filter. For surveys that do not include
bluer \rband\ imaging (left column in Figure~\ref{fig:contam_recov}), 
\eazy\ prefers a solution at $z\la2$ for many sources in the range 
$5.7 \leq z \lesssim 6.2$ because the decreased flux in \zband\ is often 
attributed to the shallower slope of a red or dusty galaxy at lower-redshift.
The recovery is the highest for the $z=7$ sample. Sources in this redshift 
range are detected in \yband, and \zband\ is always a dropout filter.
There are also at least two filters redward of the break (but somtimes three 
when \fband\ is included) to obtain a strong constraint on the shape of the 
SED. 
The recovery remains high for almost all magnitudes for the $z=8$ and 9 
samples, but drops slightly at the faintest magnitudes. 

The Base survey results in the lowest overall source recovery. 
Adding \fband\ helps at the higher redshift end by adding an additional 
detection filter redward of the break, but does not help with recovery at the 
low redshift end. The addition of \rband, however, significantly improves 
the recovery right down to $z=5.7$, the edge of the $z=6$ sample window.
The survey with the best recovery is Base+\rband+\fband\ (bottom right), as the 
benefits of both extra filters are present while still maintaining a 
significant survey depth. 
The recovery for each survey area (1, 2, 4 or 7 \rst\ pointings) follows 
this same pattern as a function of filter set. 
However, with each increase in survey area the achieved
depth decreases. This affects the $z=9$ sample the most, where the recovery 
drops fastest at faint magnitudes close to the survey detection limit.

\section{Contamination}\label{sec:contam}

There are multiple sources of contamination for a photometrically-selected
sample of high-redshift galaxy candidates. The dominant source of 
contamination is lower-redshift galaxies, as the red colors of passive 
or dusty galaxies can appear to be dropout galaxies in a broadband selection. 
Lower-redshift contaminants are of particular concern for surveys with 
only a single imaging filter blueward of the Lyman break, or for galaxies 
close to the magnitude limit of the survey. 
Galaxies with strong emission lines such as \oii, \oiii, or \ha\ could 
also boost the flux in the detection filters 
\citep[e.g.,][]{atek2011,pirzkal2015,faisst2016a}.
Low-mass stars and brown dwarfs can also be important sources of contamination
in a galaxy sample. These cooler stars and dwarfs can have similar 
near-infrared colors as high-redshift galaxies. They will also be unresolved 
in images, and are therefore difficult to separate from small, compact 
galaxies that will remain unresolved with the 0\farcs11 pixels of the \rst/WFI
images. In the following sections, we consider both types of contamination.

\subsection{Lower-redshift galaxies}

We begin by exploring the contamination in each sample from lower-redshift 
galaxies. The SAM mock catalog includes physically-motivated densities of 
galaxies at lower redshift. The number of low-redshift galaxies selected in 
our high-redshift samples provides a realistic estimate of the contamination
fractions than can be expected for these selection criteria. 

The dashed lines in the bottom panel of Figure~\ref{fig:samples} show the
contamination fraction in each redshift range and survey from sources that are 
more than $\Delta z = 1$ from the midpoint of the sample redshift range. 
The contamination in all samples is almost exclusively
from sources at $z<5$, but $\lesssim$6\% of each sample is comprised of 
galaxies at $z>5$ that have been selected in the wrong redshift bin (i.e., a 
galaxy at $5 < z<6.2$ or $z>8.2$ selected in the $z=7$ sample). This contamination
from $z>5$ galaxies is $<$1\% in the $z=6$ and 7 samples for all surveys. 
In the higher redshift samples ($z=8$ and 9), the fractions increase from $\sim$3\% to $\sim$6\% with increasing survey area, as the decreasing depths 
in all filters make it progressively harder to constrain the shapes of the SEDs.

Overall, the contamination fractions mirror many of the trends of the recovery
fractions. The contamination increases with increasing survey area as the depth
and recovery fraction decrease. The contamination at $z\sim6$ exhibits similar
but opposite behavior as the recovery, with contamination increasing by 
$\sim$20-30\% in surveys that do not include \rband\ imaging and are therefore 
not sensitive to the presence (or lack) of the \lya-break at these redshifts. 
The increased recovery generally offered by the \fband\ filter comes at the 
expense of higher contamination, especially at $z\sim9$. 
As with the sample recovery, the contamination fractions presented in
Figure~\ref{fig:samples} represent the total contamination in each sample
across all magnitudes, and in many cases are worse then the sample contamination 
for all but the faintest sources.

In the right panels of Figure~\ref{fig:contam_recov}, we show the contamination
fraction from galaxies with true redshifts of $z<5$ as a function of observed
\hband\ magnitude and redshift. 
Unsurprisingly, the Base survey has the highest levels of contamination 
from lower-redshift galaxies. The contamination is highest in the 
$z=6$ sample, reaching $80-90$\% in some regions and extending down to 
almost all magnitudes, because there is no filter blueward of the \lya\ 
break in this redshift range. Sources at $z\sim6$ and red galaxies at 
$z\lesssim2$ will both be detected in \zband, but will have a lower flux than 
in \yband. In these cases, \eazy\ cannot distinguish between a sharp spectral 
break at high redshift and the shallower slope of a redder SED. 
Therefore, for the surveys that lack \rband\ imaging (left column in 
Figure~\ref{fig:contam_recov}), the recovery is low and the contamination is 
high in this redshift range. 
The contamination drops significantly for $z>6.7$, reaching only $\sim$10\% 
down to within $\sim$0.5 magnitudes of the survey limit and $\sim$20$-$25\%
at fainter magnitudes. The contamination increases slightly at the faint end
of the $z=9$ sample, where faint low-redshift galaxies may fall below the 
detection limit in \zband\ and \yband\ while being partially detected in 
\jband. 

Adding either filter (\rband\ or \fband) to the base filter set helps to 
reduce the sample contamination.
As can be seen in the right column of Figure~\ref{fig:contam_recov}, 
including \rband\ imaging significantly reduces the contamination from 
low-redshift galaxies. The dropout filter reduces the contamination in the
$z=6$ sample from 80$-$90\% to 10$-$25\%.  
The addition of the \fband\ filter (bottom row) also decreases the contamination 
in the $z=6$ sample at bright magnitudes. This additional red filter probes the 
shape of the SED out to longer wavelengths, helping to distinguish a 
continually-rising red slope from a flat or bluer spectrum at high redshift.
However, the \fband\ lowers the depth in all filters, leading to an increased
contamination for faint magnitudes at $z\sim9$. 

\begin{figure*}[t!]
\epsscale{1.1}
\plotone{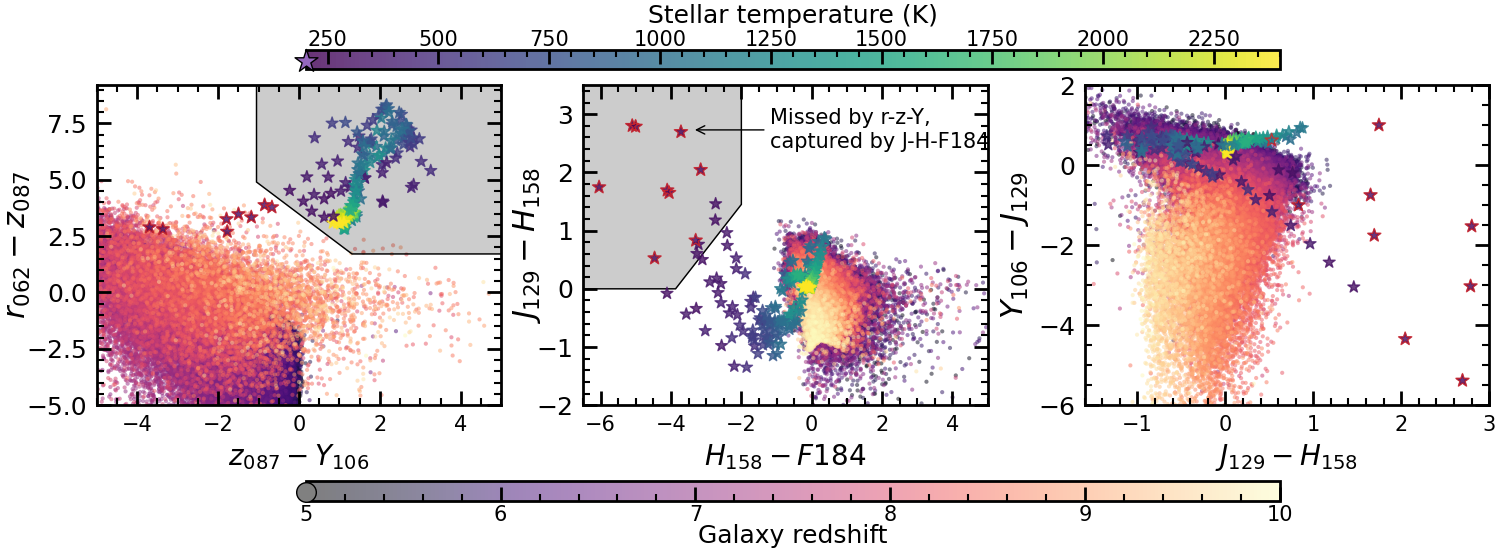}
\caption{
Separating stars and galaxies based on their colors in the \rst/WFI filters. We show three color-color plots, with the 180 substellar atmosphere models described in Section~\ref{sec:stars} plotted as stars that are color-coded by temperature. The small circles show the colors of every source from the SAM mock catalog that is selected in a redshift sample at the depth of the Base1+\rband+\fband\ survey ($m_{H158}=29.76$), color-coded by redshift. The grey regions in the left and middle panels show possible selection regions that together can cleanly separate the galaxies from the brown dwarf models. Models that overlap with galaxies in the left panel, thus requiring a second color selection for a clean galaxy-stellar separation, are plotted with red outlines.  
While there are only a few brown dwarfs with red outlines, they have similar \rband-\zband\ and \zband-\yband\ colors as galaxies at $z\gtrsim9$. At these redshifts, galaxies are more likely to be unresolved in \rst\ images and a clean identification of stars and brown dwarfs is therefore crucial. Such a clean separation requires the \rband\ and/or the \fband\ filters, as demonstrated in the right panel that does not use either filter.
\label{fig:stars}}
\end{figure*}

The survey with the lowest contamination at all redshifts and magnitudes
is Base+\rband\ (upper right corner of Figure~\ref{fig:contam_recov}). The 
survey that has the highest recovery, Base+\rband+\fband\ (lower right corner),
has slightly higher contamination at $z>9$. The addition of \fband\ reduces the 
overall depth, again leading to an increased number of faint low redshift 
galaxies being selected as $z\sim9$ galaxies. These contamination fraction trends 
are the same across all survey areas, where 
the Base+\rband\ surveys consistently have the lowest contamination.

The SAM mock catalog allows us to identify all contaminating sources in 
our samples and to correct the measured number densities accordingly. 
We discuss this correction in Section~\ref{sec:lffits}.

\subsection{Estimates of Stellar Contamination} \label{sec:stars}
The SAM mock catalog does not include stars, and so our samples do not 
suffer from stellar contamination. However, as contamination from cool 
stars and brown dwarfs is a common concern for photometrically-selected
high-redshift galaxy samples, we next explore the colors of the sources 
selected in our samples compared with stellar templates.

We calculate the \rst/WFI colors of a grid of 180
cool, substellar atmosphere models (Morley et al. 2024, in prep). These
models are compiled from the \texttt{Sonora-Bobcat} \citep{sonora_bobcat}, 
\texttt{Sonota-ElfOwl} \citep{sonora_elfowl}, 
\texttt{Sonora-Diamondback} \citep{morley2024a}, 
and \texttt{LOWZ} models \citep{meisner2021}. 
The grid includes both cloudy and cloud-free models, and chemical equilibrium models as well as those in chemical disequilibrium.
The grid covers temperatures in the range $T=200-2400$ K, surface 
gravities of $g=100$ and 3160, and metallicities of 
$[M/H]=0$, $-0.5$, and $-1$. We estimate the \rst\ colors of each 
stellar model by calculating the flux densities of the model SED in
the WFI filter curves.

In Figure~\ref{fig:stars}, we plot the colors of these templates 
color-coded by temperature. We also show the colors of all the SAM 
sources selected in redshift samples from the Base1+\rband+\fband\
survey (i.e., from the full lightcone area but at the depth for a 
0.28 deg$^2$ survey that includes all six filters). Cleanly separating 
all stellar models from the SAM galaxies requires two selections in 
color-color space, represented by the two left panels of 
Figure~\ref{fig:stars}, as a single selection region is not capable of 
capturing all stellar models. 
First, in the left panel, a color selection using \rband$-$\zband\ and 
\zband$-$\yband\ is able to identify all but a handful of the 
coolest stellar sources. The remaining sources (outlined in red in 
Figure~\ref{fig:stars}) are then easily identified by a selection using
\jband$-$\hband\ and \hband$-$\fband\ (middle panel). 

These color selections use both \rband\ and 
\fband, again highlighting the need for these two filters in 
identifying common contaminants of high-redshift samples. 
A color selection that depends on \zband, \yband\ and \jband\ (and not \rband, i.e., the right panel of Figure~\ref{fig:stars}) results in twice the number of stellar contaminants in the galaxy sample
as compared with the left panel of Figure~\ref{fig:stars}, additional
stellar contaminants that are not as cleanly separated using other 
available \rst\ colors.
Also, while there are relatively few stellar contaminants that cannot be 
isolated from the galaxies in the left panel (stars with red outlines), 
the few remaining have colors similar to those of $z\gtrsim9$ galaxies. 
Identifying stellar contaminants in these redshift ranges will be of 
critical importance, as galaxies at these redshifts are expected to be 
largely unresolved at the \rst/WFI resolution
\citep[with a 0.11\arcsec/pixel scale; see high-redshift galaxy samples from, e.g.,][]{finkelstein2023b}.
A second color selection is needed to separate these stellar models from the 
high-redshift galaxy sample, and as demonstrated in the middle panel, the 
\fband\ filter provides significant distinguishing power. 
The right panel of Figure~\ref{fig:stars} shows one of the available 
color-color plots that uses only the Base filter set.
The majority of the stellar colors are degenerate with those of the galaxies,
and the stars with red outlines are closer to the galaxy sample than they 
are in the middle panel. 
The cleanest galaxy sample is one selected with both the \rband\ and \fband.

\section{Rest-UV Luminosity Functions}\label{sec:lfs}

In this section, we calculate the rest-UV luminosity function from the observed (i.e. photometrically-scattered) mock catalog in our four redshift sample bins ($5.7<z<6.7$, $6.7<z<7.7$, $7.7<z<8.7$ and $8.7<z<9.7$) for each of the 16 surveys. As part of this calculation, we must convert the observed apparent magnitude into an absolute magnitude (Section~\ref{sec:m1500}), determine the completeness of each sample as a function of magnitude (Section~\ref{sec:veff}), and fit the volume number densities using an MCMC routine (Section~\ref{sec:lffits}).

\subsection{Absolute Magnitude at 1500\AA} \label{sec:m1500}
We begin by calculating the absolute magnitude at 1500\AA, $M_{1500}$, of 
each ``observed'' source. 
As we are considering a range of sample redshifts, there is no single 
filter that cleanly corresponds to 1500\AA\ in the rest frame for all 
sources. The same broadband filter will cover a different range of rest-frame
wavelengths at each redshift, and so the observed fluxes in the filter of 
sources at different redshifts are averaging over different regions of the 
rest-frame spectrum. The broadband fluxes cannot be directly converted to 
absolute magnitude in a consistent way without the use of a $K$ correction.

Instead, we use all available photometry redward of the \lya\ break to fit a 
slope to each source's SED and determine the luminosity of the fit at 1500\AA. 
For each source, we convert all ``observed'' fluxes and uncertainties to the 
rest frame using the \eazy\ best-fit redshift. We include in the fit 
measurements from all filters with at least 75\% of their full width at half 
maximum redward of \lya\ given the source redshift. 
In the case that more than two filters meet this requirement, we fit the 
broadband luminosities and uncertainties with a function of the form 
$L_{\lambda} \sim \lambda^{\beta}$ using the \texttt{\textsc{SciPy}}
\citep{virtanen2020} 
routine \texttt{curve\_fit} for non-linear least squares fitting.
In the event that the fitting fails to converge, which occurs in $<1$\% of 
cases, or for cases with exactly two filters redward of the break, we fit the 
photometry with a straight line. We then interpolate the fit function or line
and calculate the luminosity at $\lambda=1500$\AA. 
Finally, for sources with only a single filter redward of the break, we adopt
the luminosity measured in that filter for $M_{1500}$. 
This single filter case occurs primarily in the $z=9$ sample for surveys that 
do not include the \fband\ filter ($\sim$50\% of sources; $<1$\% for all 
other samples and surveys). 
The absolute magnitude conversion will therefore contribute to a higher 
uncertainty in the $z=9$ luminosity function measured for the Base and 
Base+\rband\ surveys than for the other surveys and redshift ranges.

\subsection{Effective Volumes} \label{sec:veff}
We next quantify the completeness of our sample selection in each redshift
bin. While the SAM mock catalog contains physically-motivated number densities 
of galaxies at all redshifts out to $z\sim10$, there are too few sources at
$z\gtrsim8$ for a statistically significant completeness analysis in multiple 
magnitude and redshift bins. We therefore distinguish between the survey 
recovery fractions presented in Section~\ref{sec:recov} and the survey 
completeness that is calculated here. In this section, we perform additional 
catalog-level simulations, applying the same selection criteria as those used 
for the SAM catalog sources, and determine the number of sources selected as a 
function of magnitude and redshift. 

Following procedures described in depth in 
\cite{rojas-ruiz2020}, \cite{finkelstein2022a}, and \citet{bagley2024a},
we create a catalog of synthetic sources with assigned redshifts, spectral 
templates, dust extinction, and observed fluxes. We tweak the input parameter 
distributions to ensure the resulting distributions of $M_{1500}$ and UV color 
([1500\AA]$-$[2800\AA]) match those of the sources in the SAM mock catalog at 
$z>5$. The input parameters are assigned as follows. 

The catalog contains $1.34\times10^6$ synthetic sources, distributed 
uniformly over the redshift range $4 \leq z \leq 10.7$ ($\sim$2$\times10^5$ 
per $\Delta z=1$). We assign to each source a \citet{bc03} spectral template 
with a \citet{chabrier2003} initial mass function and a constant star 
formation history, as we do not expect there to be large numbers of 
galaxies with declining star formation histories at these early times. 
The templates have metallicities of 0.02, 0.2, 0.4 or 1.0$Z_{\odot}$, pulled
from a lognormal distribution that peaks at 0.2$Z_{\odot}$. The stellar 
population ages are pulled from a lognormal distribution peaking at 
$\sim$100 Myr and are constrained to be less than the age of the Universe 
at each source's assigned redshift. 

%
%
%
%
%

\begin{figure*}
\epsscale{1.1}
\plotone{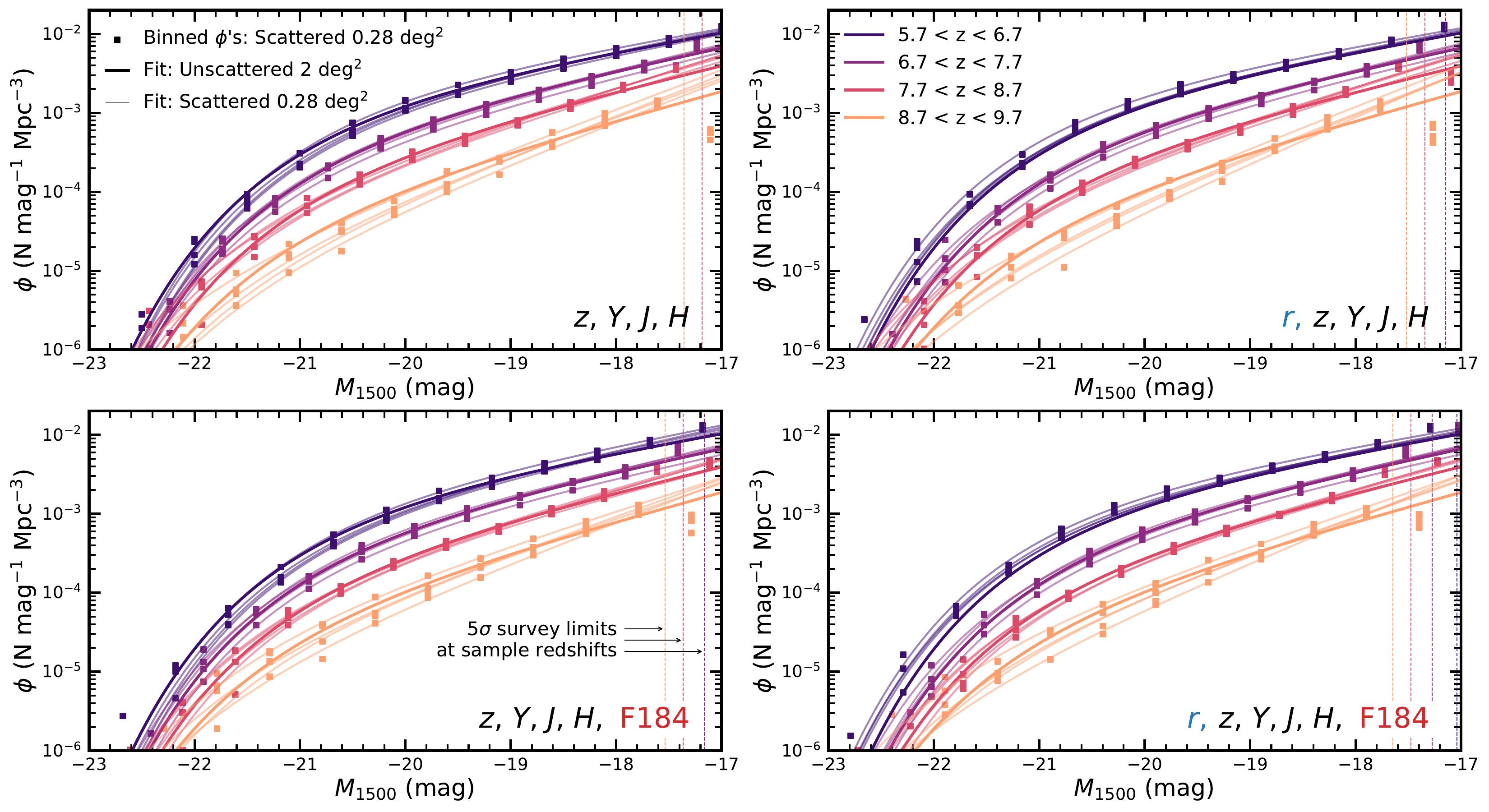}
\caption{The rest-UV luminosity function as measured in the four surveys covering a given area. Each set of four panels shows results for the Base (top left), Base+F602 
(top right), Base+\fband\ (bottom left), and Base+\rband+\fband\ (bottom right) surveys.
In each panel, we show the luminosity function in each sample redshift 
range. The ``true'' luminosity function, obtained by fitting the full 
2 degree$^2$ unscattered light cone down to a magnitude limit of $m_{\mathrm{F160W}} \leq 30.0$, is plotted as a thick curve color-coded by redshift range. 
The binned number densities are plotted as squares, and the fits to the 
``observed'' (photometrically scattered and identified via our selection criteria) pointings are plotted
as thin curves. The 5$\sigma$ survey limits at the redshift midpoint of each sample are plotted as vertical dotted lines. 
For surveys covering 0.28 and 0.56 degrees$^2$, we show the luminosity function calculated from each realization of the survey that we extract from the 2.0 degree$^2$ light cone. 
Results for all surveys are included in the online version as a figure set. We show the results for the surveys covering 0.28 degree$^2$ (1 \rst\ pointing) as an example in the PDF version.
(The remaining figures are presented in the Appendix as Figure~\ref{fig:lfs_temp} for the arXiv version.)
\label{fig:lfs}}
\end{figure*}

We apply interstellar medium dust extinction to each spectral template using 
a \citet{calzetti2000} extinction law. The color excess $E(B-V)$ for each 
source is pulled from a composite distribution of two truncated Gaussians, 
where three quarters of the sources are pulled from a narrow ($\sigma=0.05$) 
component with a peak at $\mu=0.01$, and the remaining quarter are pulled 
from a slightly wider and redder component ($\mu=0.25$, $\sigma=0.1$). 
We next apply redshift-dependent attenuation due to the intergalactic medium 
using the prescription of \citet{madau1995}, in keeping with the dust 
attenuation present in the SAM and assumed by \eazy. 
Finally, we assign each source an observed 
$H$-band magnitude (using \hst/WFC3 F160W as the reference in order to use one set of simulations for \hst, \rst\ and \jwst\ measurements), pulled from a composite of a truncated Gaussian 
distribution with $\mu=30.1$ mag and $\sigma=1.5$ mag and a uniform 
distribution in the range $22 \leq H \leq 34$ mag, which is used to boost 
the number of bright sources. 
We normalize each spectral template to the assigned $H$-band magnitude in 
the F160W filter, and calculate the flux density of the normalized
template in each \rst/WFI filter.

We then select synthetic sources using the same process as with the 
sources in the SAM mock catalog. First, we apply photometric scatter 
appropriate for the depth of each \rst\ survey as described in 
Section~\ref{sec:perturbation}. We run \eazy\ on all synthetic sources 
with the $1\sigma$ depth in each filter adopted as the photometric 
uncertainties. We select sources using the criteria described in 
Section~\ref{sec:sample}, and then calculate the fraction of recovered 
sources in bins of magnitude and redshift.  
In each magnitude and redshift bin, the completeness fraction is calculated 
as the number of recovered sources divided by the number of input sources. 
The completeness fractions are fully consistent with the recovery fractions 
displayed in Figure~\ref{fig:contam_recov}, yet 
are calculated with enough input sources in each redshift and magnitude 
bin to allow for robust estimates even at the highest redshifts and faintest
magnitudes.

The effective volume probed in each survey is then calculated according 
to:
\begin{equation}
V_{\mathrm{eff}}(M_{1500}) = \int P(M_{1500}, z) \frac{\mathrm{d}V}{\mathrm{d}z} \mathrm{d}z,
\label{eqn:veff}
\end{equation}
where $\mathrm{d}V/\mathrm{d}z$ is the differential comoving volume element
and $P(M_{1500}, z)$ is the probability that a source at a given redshift 
$z$ and with absolute magnitude $M_{1500}$ will be selected by our sample
criteria. This volume is used in Section~\ref{sec:lffits} in calculating 
the volume number densities of sources in each redshift range and survey.

\subsection{Survey Luminosity Functions} \label{sec:lffits}
We now measure the rest-UV luminosity function in the range $6 \leq z \leq 9$ for each of the 16 deep field surveys.  
For each survey, we extract the corresponding area from the 2 degree$^2$ mock catalog, creating multiple unique realizations of each redshift sample for the surveys covering 1 and 2 pointings (0.28 degree$^2$ and 0.56 degree$^2$, respectively).
We next bin the observed densities for a given survey and redshift sample realization by absolute magnitude. 
The densities are calculated as the number of sources selected in the sample, corrected for contamination, and divided by the effective volume in each bin, $V_{\mathrm{eff}}(M_{1500})$ from equation~\ref{eqn:veff}. 
As in Section~\ref{sec:contam}, we define contaminants as any galaxy with a true redshift that is $\Delta z>1$ from the sample midpoint. 
The magnitude bins have width $\Delta M_{1500} = 0.5$ mag cover 10 magnitudes, extending far brighter than the brightest observed candidates and one magnitude fainter than each survey's $5\sigma$ limit (see Table~\ref{tab:grid}) at the sample redshift midpoint. For example, for the $z=6$ sample (midpoint $z=6.2$) in the Base1 survey ($m_{5\sigma}=30.05$, $M_{1500} = -16.75$), we consider bins in the range $-25.75 < M_{1500} < -15.75$.  

We calculate the luminosity function using a Markov Chain Monte Carlo (MCMC) analysis, fitting the binned densities with a \citet{schechter1976} function. While many observations at $z\gtrsim 8$ are better fit with a double power law
\citep[e.g.,][]{bowler2015,bowler2020,finkelstein2022b}, we use a Schechter function form to match the SAM inputs. Our conclusions would not have changed if the SAM had exhibited a double power law behavior, because we are comparing each survey's recovery of the true distribution.
We perform the fitting with the \texttt{emcee} Python package \citep{emcee}.
We use 10 walkers with 10$^5$ steps as a burn-in, followed by 10$^4$ additional steps used to calculate the posterior distributions of each parameter.
A least squares fit to the binned densities serves as the initial guess for each luminosity function parameter, and the walkers are perturbed about these values when initializing the chains. At each step, we fit the binned densities down to the $5\sigma$ magnitude limit for the given survey at the sample redshift midpoint.
We use the Cash statistic \citep{cash1979,ryan2011} as the goodness-of-fit measure, which represents the likelihood that the observed number of galaxies matches the expected number under the assumption of Poisson probability distributions. This statistic is well-suited for cases where there are few sources per bin, such as at bright magnitudes and/or high redshifts, and approaches the behavior of a normally distributed statistic when there are many sources per bin. 
We use two proposals with differential evolution for updating the coordinates for the walkers, \texttt{DEMove} \citep{terbraak2006,nelson2014} and \texttt{DESnookerMove} \citep{terbraak2008}. We weight the moves such that at each step, the sampler will randomly select either \texttt{DEMove} with 80\% probability or \texttt{DESnookerMove} with 20\% probability.
Following burn-in, we thin the chains by half the autocorrelation time and take the median values of the remaining elements in the chains as the fits for the luminosity function parameters $\phi^*$, $M^*$, and $\alpha$.

The resulting luminosity functions are shown in Figure~\ref{fig:lfs}. As with Figure~\ref{fig:contam_recov}, each set of four panels shows the results calculated for the four filter sets in a given survey area. We show the binned densities as squares, the Schechter function fits as thin curves, and the ``true'' luminosity function fit to the full, unscattered light cone as thick curves. The luminosity functions are color-coded by sample redshift bin. We show the measured luminosity functions for each survey realization (e.g., there are four realizations of the surveys covering a single \rst\ pointing), and the range of densities reflect the level of variation imparted by probing different areas in the light cone (see Section~\ref{sec:cv}). We see that the surveys do best at recovering the true luminosity function at $z\sim7$ and 8. At $z\sim6$, surveys either under- or over-estimate the density at the bright end depending on whether the \rband\ band is included. The faint end of the measured $z\sim9$ is steeper than the true slope, an effect that gets worse with increasing area. This steepening has important implications for the recovery of the luminosity density at high redshift, and is discussed in more depth in the following section.

\section{Rest-UV Luminosity Density}\label{sec:rhouv}
Our final step is to calculate the evolution of the integrated non-ionizing UV
luminosity density as a function of redshift. This quantity, $\rho_{\mathrm{UV}}$,
is found by integrating the rest-UV luminosity function down to given magnitude 
limit and has important implications for the galactic contribution to the epoch of 
reionization. Given assumptions on the ratio of ionizing and non-ionizing UV 
luminosity emitted by galaxies and the escape of ionizing radiation, 
$\rho_{\mathrm{UV}}$ can be converted to an ionizing emissivity and compared against
that required to maintain an ionized IGM. 
Here, we calculate $\rho_{\mathrm{UV}}$ by integrating the luminosity functions
measured in Section~\ref{sec:lfs} down to a limit of $M_{1500} = -17$. 
This magnitude limit is on par with the observation limits at $z\sim6-9$ in 
lensed fields with \hst\ \citep[e.g.,][]{} or deep fields with \jwst\ 
\citep[e.g.,][]{leung2023}. While fainter limits are often adopted to include the 
contribution of the population of galaxies fainter than current observational 
limits, we are not attempting to evaluate the true value of $\rho_{\mathrm{UV}}$.
Instead, we are concerned with the relative comparison of $\rho_{\mathrm{UV}}$ 
as measured in each survey in order to evaluate the trade-offs of depth and area 
in recovering this quantity.

\begin{figure*}
\gridline{\fig{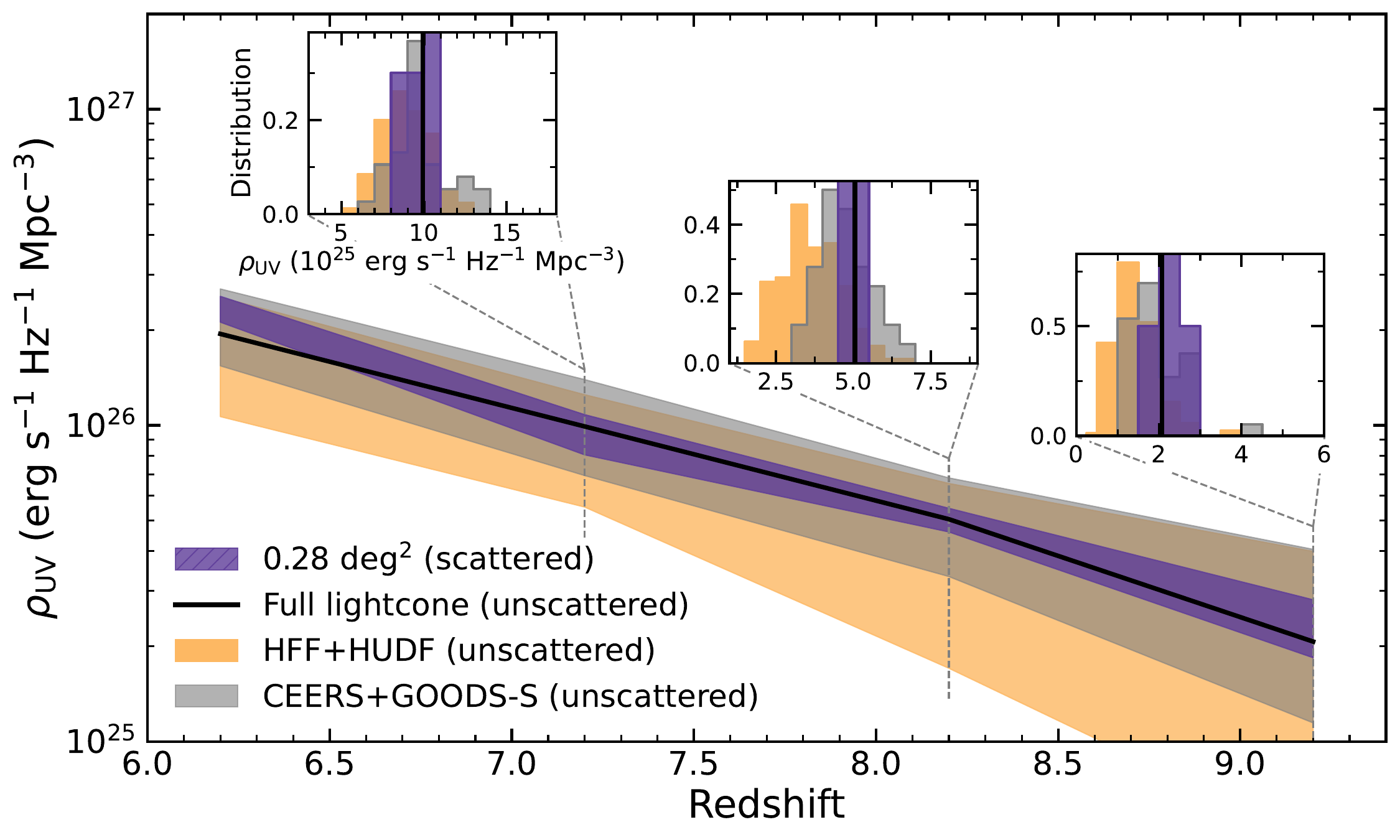}{0.49\textwidth}{}
          \fig{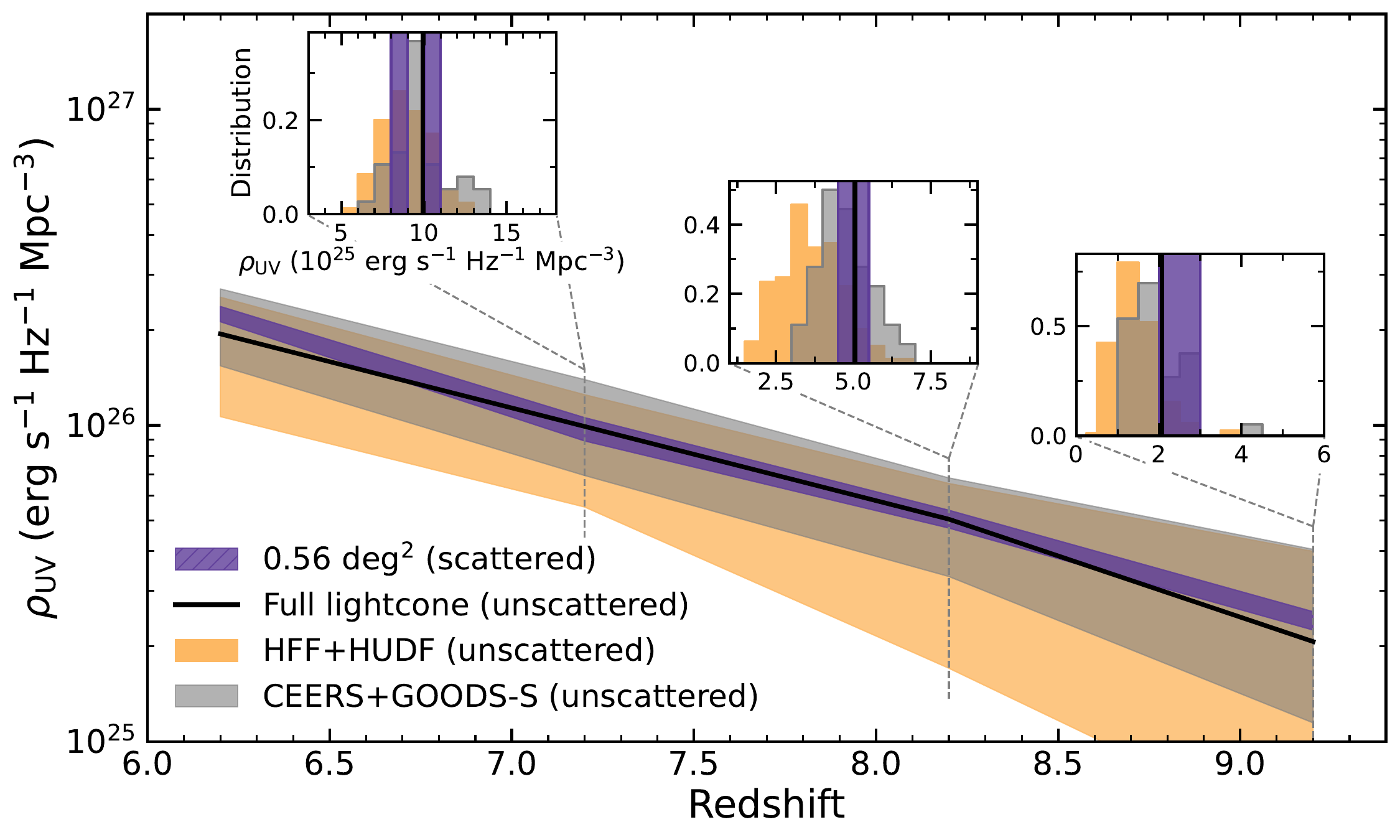}{0.49\textwidth}{}}
\vspace{-8mm}
\gridline{\fig{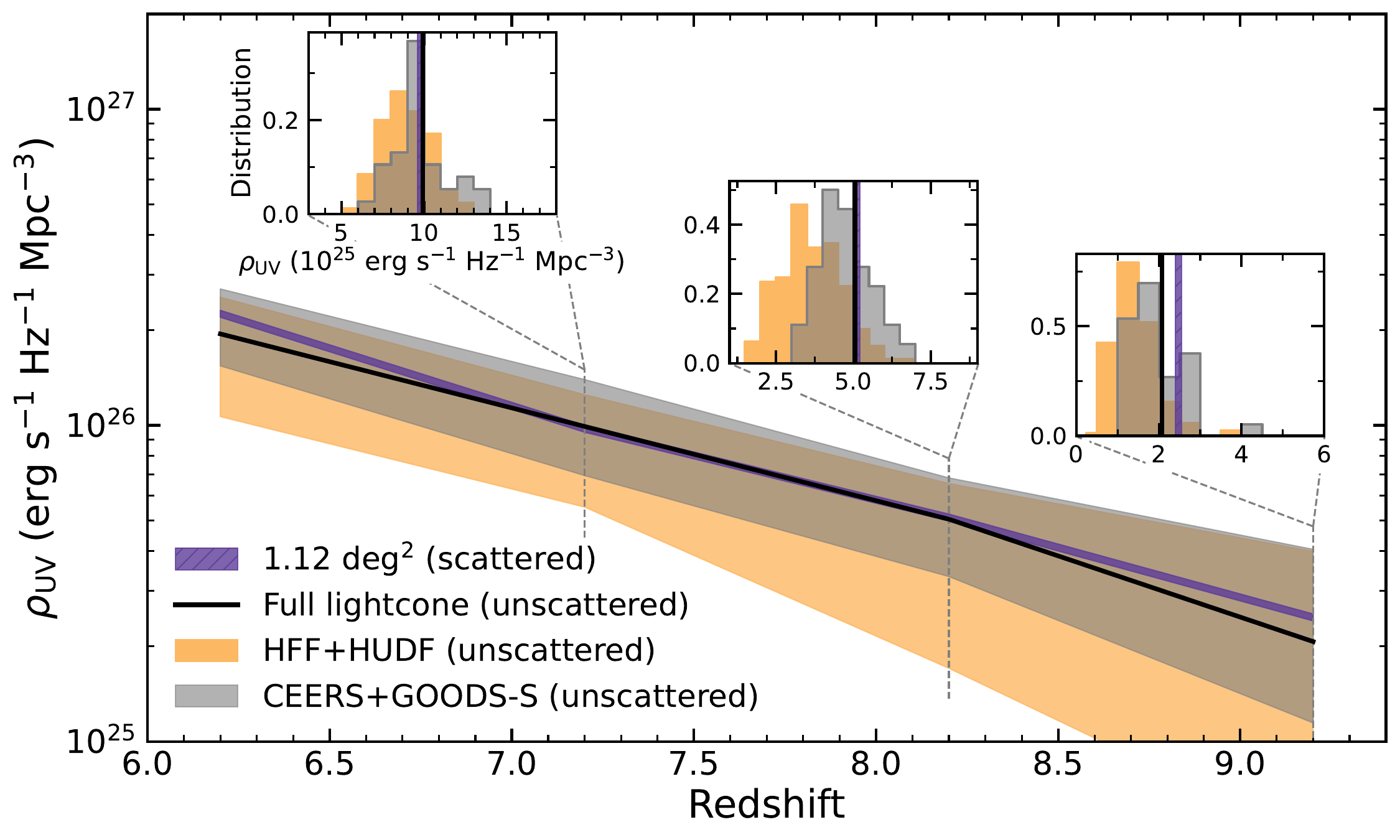}{0.49\textwidth}{}
          \fig{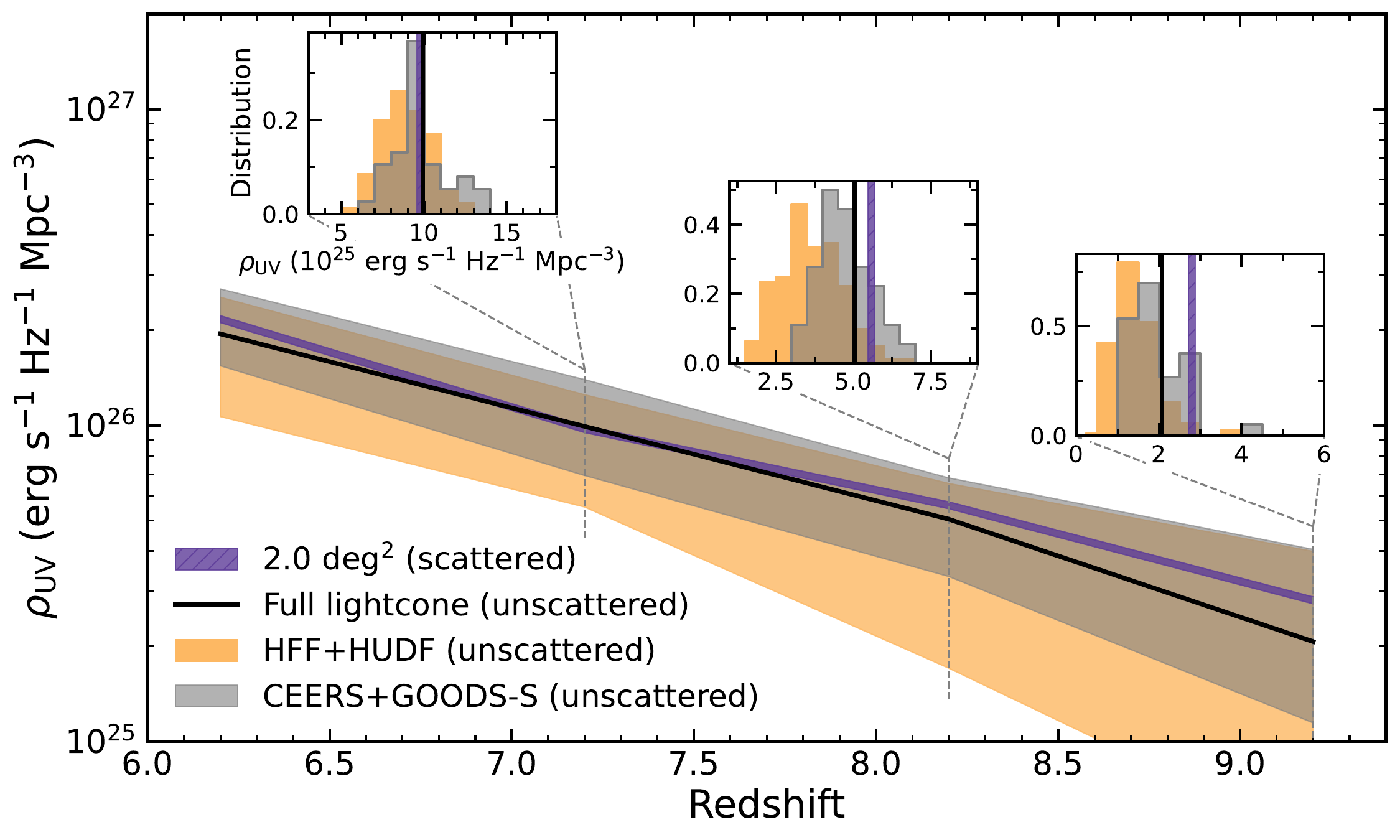}{0.49\textwidth}{}}
\caption{The evolution of the integrated non-ionizing UV luminosity density for a given filter set. Each set of four panels shows results for one filter set for survey areas covering 0.28 (upper left), 0.56 (upper right), 1.12 (lower left) and 2.0 degree$^2$ (lower right). 
In each panel, we show $\rho_{\mathrm{UV}}$ as measured in the ``observed'' \rst\ surveys in purple. For surveys covering 0.28 and 0.56 degrees$^2$, we show the range of $\rho_{\mathrm{UV}}$ values calculated from the multiple survey realizations. The ``true'' $\rho_{\mathrm{UV}}$ calculated from the full, unscattered light cone is plotted as a thick black curve. The grey regions show the range of $\rho_{\mathrm{UV}}$ that are calculated in 35 survey areas that represent a combination of \jwst/NIRCam observations from Cycles 1+2, including CEERS and the deep imaging in GOODS-S from JADES and NGDEEP. We show the range of $\rho_{\mathrm{UV}}$ from 150 HFF+ pointings in orange. The insets show the $\rho_{\mathrm{UV}}$ distributions displayed on a linear scale at $z=7.2$, 8.2 and 9.2. 
The \rst\ surveys closely recover the true $\rho_{\mathrm{UV}}$ at $z\sim7$ and 8, where sample completeness is high and contamination low. This recovery gets worse as the survey area increases and the depth in each filter decreases, as $\rho_{\mathrm{UV}}$ is highly sensitive to the faint end slope of the UV luminosity function The grey and orange regions, representing results from idealized versions of premiere \hst- and \jwst-like surveys, cover ranges of $\rho_{\mathrm{UV}}$ that are $\sim2-7\times$ larger than those measured in the photometrically-scattered \rst\ surveys. 
Results for all surveys are included in the online version as a figure set. We show the results for the surveys using all six filters (Base+\rband+\fband) as an example in the PDF version. (The remaining figures are presented in the Appendix as Figure~\ref{fig:rhouv_temp} for the arXiv version.)
\label{fig:rhouv}}
\end{figure*}

We show the results in Figure~\ref{fig:rhouv} as purple hatched regions. Each of 
the four main panels shows $\rho_{\mathrm{UV}}$ on a log scale as a function of
redshift. The three inset plots show the distribution of $\rho_{\mathrm{UV}}$ 
values at $z=7.2$, 8.2 and 9.2 displayed on a linear scale. Rather than presenting 
the four filter sets for a given survey area as in Figures~\ref{fig:contam_recov} 
and \ref{fig:lfs}, the four panels here show results for the four survey areas 
for a given filter set. In this way, we can see how an increase in area (and a
corresponding decrease in depth in each filter) affects the recovery of 
$\rho_{\mathrm{UV}}$. The ``true'' value, found by integrating the luminosity 
function fit to the full, unscattered light cone to $M_{1500} = -17$ is plotted 
as a black curve (black vertical line in the insets).

The measured $\rho_{\mathrm{UV}}$ is closest to the true values at $z\sim7-8$, and deviates at lower and higher redshifts. This behavior is related to the sample recovery and completeness discussed in Sections~\ref{sec:recov} and \ref{sec:veff}. As seen in Figure ~\ref{fig:contam_recov}, the completeness is highest in the redshift range $6.7 < z < 7.7$ in all filter sets, staying close to 100\% across the redshift range down to the limiting magnitude of the survey. The recovery is low at all magnitudes for much of the $z\sim6$ sample bin, and decreases at faint magnitudes for the $z\sim9$ sample. A lower sample recovery corresponds to lower sample completeness and therefore a smaller effective volume and higher measured volume densities. For example, the lower completeness at faint magnitudes at $z\sim9$ contributes to a steepening of the observed $z\sim9$ luminosity functions (Figure~\ref{fig:lfs}). At a minimum, lower sample completeness at faint magnitudes will result in larger uncertainties on the faint end slope $\alpha$ of the luminosity function. 

Since $\rho_{\mathrm{UV}}$ is calculated by integrating the luminosity function, it is very sensitive to $\alpha$.  
The amplitude of the $\rho_{\mathrm{UV}}$ overestimation therefore increases with increasing survey area and decreasing survey depth. When measured in a survey covering 2.0 degree$^2$, $\rho_{\mathrm{UV}}$ is higher than the true value even at $z\sim8$. 
Indeed, the $5\sigma$ limiting magnitudes of the Base7 surveys are a full magnitude brighter than those of the Base1 surveys (see Table~\ref{tab:grid}). These results indicate that survey depth in all filters is more crucial to measurements of $\rho_{\mathrm{UV}}$ than survey area. In 500 hours, the time is better spent going as deep as possible over at most two \rst\ pointings.

To this figure, we have also added the $\rho_{\mathrm{UV}}$ that is calculated 
in the ``HFF+'' pointings (one field representing the HUDF with $m_{\mathrm{F160W}}=29.5$ and eight pointings reaching 0.5 magnitudes shallower representing the HUDF and HFF parallels). The orange shaded region in Figure~\ref{fig:rhouv} shows the range of $\rho_{\mathrm{UV}}$ calculated in 150 realizations of HFF+ pointings. The \hst\ results cover a range that is $3-7\times$ larger than that of the \rst\ results. Importantly, these results are unscattered, showing that even in the ideal survey that is fully complete and contains no contamination, the premiere \hst\ observations result in 
$\geq3\times$ more variation in $\rho_{\mathrm{UV}}$ than the \rst\ deep fields we consider here.

We also include results representing a handful of Cycle 1+2 \jwst\ programs, with NIRCam imaging combining the area and depth of the Cosmic Evolution Early Release Science Survey \citep[CEERS; ERS 1345; PI: Finkelstein][]{finkelstein2017}, and deep imaging in the GOODS-S region from the Next Generation Deep Extragalactic Exploratory Public Survey \citep[NGDEEP; GO 2079; PIs: Finkelstein, Papovich, Pirzkal;][]{bagley2024b} and the \jwst\ Advanced Deep Extragalactic Survey \citep[JADES; GTO 1180, 1210, PIs: Eisenstein, Luetzgendorf, respectively;][]{eisenstein2023a}.
We note that the \jwst/NIRCam filters are not ideal for selecting $z\sim6-7$ galaxies, and so we use the true, unscattered values of the mock catalog sources for this analysis. For CEERS, we extract regions from the light cone that are $21\farcm6 \times 6\farcm5$ and assume a $5\sigma$ depth of 29.0 for \hband, which is the median F150W depth of all ten CEERS pointings as reported in \citet{finkelstein2023b}. For the deep field surveys, we extract a region $5\farcm0 \times 7\farcm2$ down to $m_{\mathrm{H}158}=29.9$, within which $4\farcm1\times 2\farcm2$ extends down to $m_{\mathrm{H}158}=30.1$ \citep[representing the JADES Deep observations from GTO 1180;][]{eisenstein2023a} with one NIRCam-sized pointing ($2\farcm2 \times 5\farcm0$) off to one side down to $m_{\mathrm{H}158}=30.1$ \citep[representing the JADES Deep observations from GTO 1210 and the JADES Origins Field, GTO 3215, PI: Eisenstein;][]{eisenstein2023a,eisenstein2023b}. One additional NIRCam-sized pointing is extracted on the opposite side down to $m_{\mathrm{H}158}=30.2$ to represent NGDEEP \citep{leung2023}. The grey regions in Figure~\ref{fig:rhouv} show the results from 35 of these CEERS+GOODS-S combined pointings, again using unscattered sources from the mock catalog.

Even as idealized measurements, the $\rho_{\mathrm{UV}}$ measured in the combination of these deep+wide \jwst\ pointings still covers a range $\sim2-4\times$ 
larger than that from a single \rst\ deep field. A one- or two-pointing \rst\ deep field can achieve limiting magnitudes deeper than all but the deepest existing \jwst\ imaging (which currently covers $\sim35$ arcmin$^2$ to $m_{\mathrm{F150W}} > 30$)
while simultaneously covering $\sim30-60$ times the area. The depth is required for well-constrained measurements of $\rho_{\mathrm{UV}}$, and the area is needed to derive accurate constraints on the density of UV-bright galaxies. Additionally, the depth and filter coverage from existing \hst\ and \jwst\ imaging varies from field to field. Constructing a luminosity function that is well-constrained at both the bright and faint ends requires selecting galaxy samples from heterogeneous observations. However, a sufficiently deep \rst\ survey allows for the measurement of well-constrained luminosity functions and the evolution of $\rho_{\mathrm{UV}}$ with redshift.
This comparison highlights the complementary power of a \rst\ ultra deep field in the age of \hst\ and \jwst.

\section{Discussion}\label{sec:discussion}

\subsection{Cosmic Variance} \label{sec:cv}

\begin{figure*}
\epsscale{1.1}
\plotone{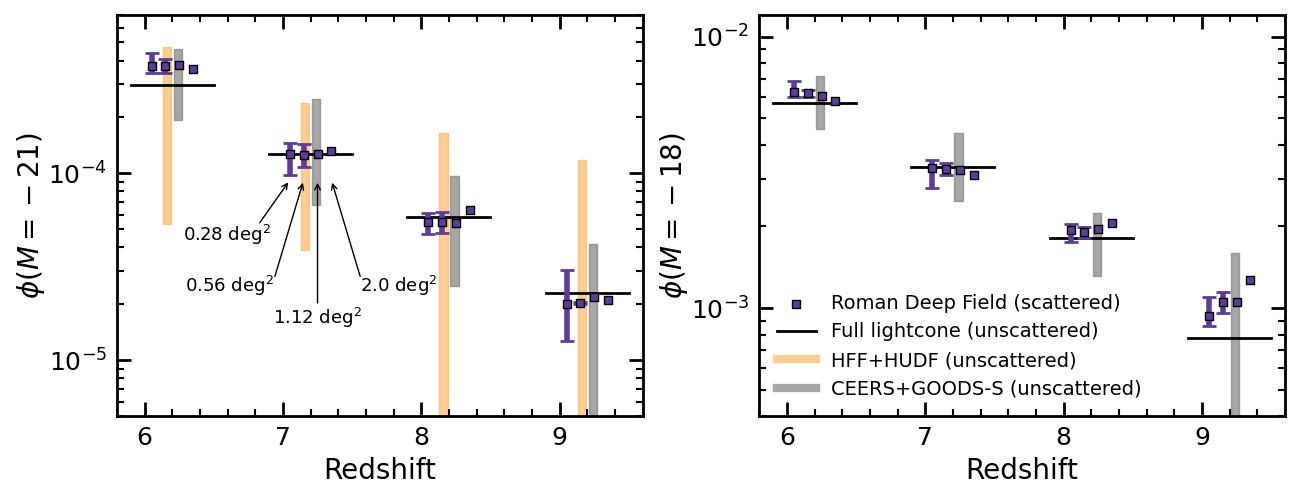}
\caption{The volume number density measured in the Base+\rband+\fband\ survey at $M_{1500}=-21$ (left panel) and $M_{1500}=-18$ (right panel) as a function of redshift. The purple squares show the \rst-calculated densities using the best-fit Schechter function parameters determined in Section~\ref{sec:lfs}, with error bars indicating the range of measurements in the multiple realizations of the 0.28 and 0.56 deg$^2$ surveys. These ranges, albeit based on a very small number of realizations, demonstrate how the increased \rst\ survey area results in measurements that are less susceptible to cosmic variance than those of \hst- (orange bars) and \jwst-like (gray bars) surveys. At fainter magnitudes (i.e., $-18$, right panel), which are crucial for constraining $\rho_{\mathrm{UV}}$, the deepest \jwst\ measurements cover a range that is 7--16 times larger than the \rst\ surveys (and the \hst\ observations are not deep enough). Additionally, the \hst- and \jwst-like measurements plotted here are unscattered (i.e., 100\% complete with 0\% sample contamination) while the \rst\ measurements are from photometrically scattered surveys with full sample selection.
\label{fig:densities}}
\end{figure*}

Cosmic variance uncertainty in an observed sample of galaxies arises from variations in how matter is distributed relative to the average density of the universe, as captured within the boundaries of the survey region. As illustrated in Figure~\ref{fig:sam}, a sample of galaxies observed in a single, small survey may under- or over-count the average population based on whether the survey happens to be observing over or under dense regions. 
At faint magnitudes, where the predicted number of galaxies depends sensitively on the estimate of the UV luminosity function's faint end slope $\alpha$, uncertainties due to cosmic variance can have a large impact on resulting estimates of $\rho_{UV}$. Covering large areas (e.g., using \rst's large field of view)
or multiple independent lines of sight (such as with the HUDF HFF parallel fields)
are both ways to overcome the effects of cosmic variance.  
Here we estimate the improvement in the effect of cosmic variance that can be achieved with \rst\ survey areas compared with those of \hst\ and \jwst. 

We use the cosmic variance calculator presented in \citet{bhowmick2020}, which is based on the \textsc{bluetides} high-resolution hydrodynamic simulation \citep{feng2016}. We consider seven surveys: the four \rst\ surveys covering 0.28, 0.56, 1.12, and 2.0 deg$^2$; a single \hst/WFC3 pointing covering 4.6 arcmin$^2$ (approximating the HUDF); the ``HFF+'' survey that covers nine independent \hst/WFC3 pointings for a total of $\sim$41 arcmin$^2$; and the ``CEERS+GOODS-S'' survey described in Section~\ref{sec:rhouv} that collectively covers $\sim$198 arcmin$^2$. 
We estimate the cosmic variance uncertainty at $z=9.2$ (the midpoint of our highest redshift bin), and calculate it for a survey depth of $m=30.0$.

The expected fractional uncertainty due to cosmic variance drops from 50.3\% for a single \hst\ pointing to 16.8\% and 13.5\% for the HFF+ and CEERS+GOODS-S surveys, respectively. The four \rst\ surveys have uncertainties of 9.1\% (0.28 degree$^2$), 7.2\% (0.56 degree$^2$), 5.8\% (1.12 degree$^2$), and 4.8\% (2 degree$^2$). Even a single pointing with \rst\ reduces the effects of cosmic variance as compared with leading extragalactic \hst\ and \jwst\ surveys. We also note that while the fractional uncertainty estimated here only improves $\sim5-10\%$ from the CEERS+GOODS-S to the \rst\ surveys, we assumed a depth of $m=30$ in all pointings. Covering all $\sim$200 arcmin$^2$ of the CEERS+GOODS-S footprint to that depth would represent a considerable investment of \jwst\ observing time.

Figure~\ref{fig:densities} provides another way to visualize the effect of cosmic variance on the galaxy number counts resulting from a deep \rst\ survey. We show the density $\phi$ as measured in the Base+\rband+\fband\ survey from the best-fit Schechter function parameters as described in Section~\ref{sec:lffits} calculated at two different magnitudes, $M_{1500}=-21$ (left panel) and $M_{1500}=-18$ (right panel). The \rst\ measurements are plotted in purple, with error bars indicating the range of measurements in the multiple realizations of the 0.28 and 0.56 deg$^2$ surveys. We show the range of measurements in the HFF+ (orange) and CEERS+GOODS-S survey realizations (grey). The \rst\ realizations cover a range that is 9--13 times smaller than that of the CEERS+GOODS-S realizations at $M_{1500}=-21$, and 7--16 times smaller at $M_{1500}=-18$, though we note that the small number of \rst\ realizations complicates this comparison. On the other hand, the \hst- and \jwst-like surveys are unscattered (i.e., 100\% complete with 0\% contamination), while the \rst\ surveys have been photometrically scattered and the results are based on a full sample selection process. When measuring the high-redshift UV luminosity function and volume number density, an ultra-deep survey covering a sufficiently large area at \jwst-like photometric depth is required to minimize the uncertainties arising from cosmic variance. \rst\ provides the opportunity to achieve both.

\subsection{Synergies with \rst\ Core Community Surveys} \label{sec:core_surveys}
In this paper, we are presenting a set of potential ultra deep surveys with \rst\ with the goal of optimizing survey design for the measurement of $\rho_{\mathrm{UV}}$ at high redshift. However, many of the survey parameters for which we advocate are likely to be incorporated in some of the \rst\ Core Community Surveys. These surveys are being designed with community input to maximize the science return of the observatory across a broad range of science goals while still meeting the cosmology and exoplanet mission requirements. A significant fraction of mission time in the first five years will be devoted to the three core surveys: the High Latitude Wide Area survey, the High Latitude Time Domain survey, and the Galactic Bulge Time Domain survey.
At the time of this writing, the Core Community Survey Definition Committees have presented proposed implementation plans for each survey, which the Roman Observations Time Allocation Committee (ROTAC) has synthesized into recommendations for the \rst\ WFI observing program \citep{rotac2025}.  

\begin{figure*}
\plotone{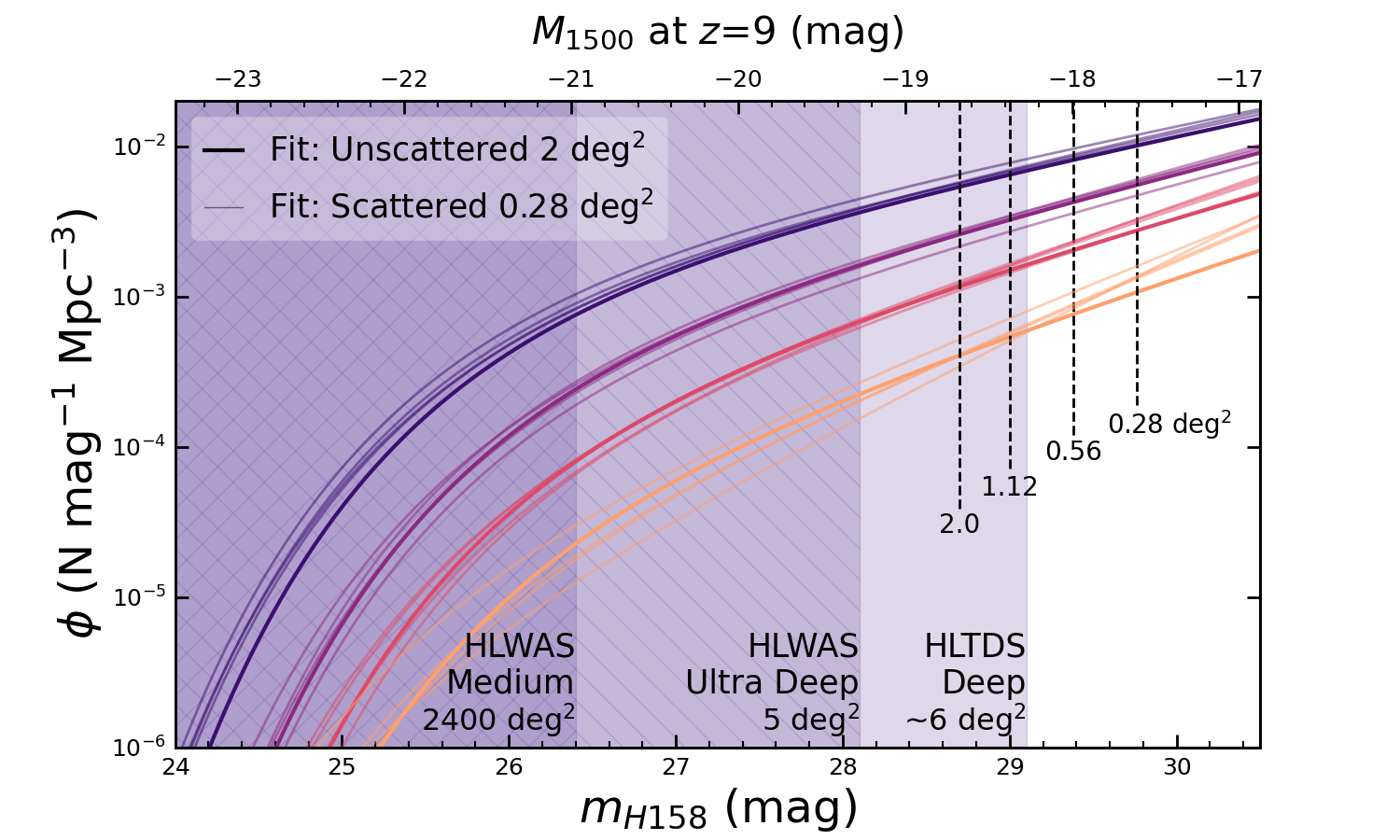}
\caption{The rest-UV luminosity function as measured in the Base1+\rband+\fband\ survey. 
The ``true'' luminosity function, obtained by fitting the full 2 degree$^2$ unscattered light cone down to a magnitude limit of $m_{\mathrm{F160W}} \leq 30.0$, is plotted as a thick curve color-coded by redshift range from $z\sim6$ in purple to $z\sim9$ in yellow (see the Figure~\ref{fig:lfs} legend). The fits to the ``observed'' (photometrically scattered galaxies that are then identified via our selection criteria) pointings are plotted as thin curves. The $5\sigma$ magnitude limits for the Base+\rband+\fband's four survey areas are plotted as vertical dashed lines. 
The purple regions show the $5\sigma$ \hband\ limit for the HLWAS Medium and Ultra Deep tiers and the HLTDS Deep Tier. The HLTDS Deep Imaging Tier will reach approximately the same depth as our Base4+\rband+\fband\ survey, though it will likely not include the \rband\ imaging necessary for improved completeness and contamination in high-redshift galaxy samples. 
\label{fig:core_surveys}
}
\end{figure*}

Specifically, there is significant overlap between the characteristics of the deep fields discussed here and the High Latitude Time Domain survey \citep[HLTDS; see][for details]{rotac2025}). The Core Component of the HLTDS will observe the same regions of the sky repeatedly with a $\sim$5--10-day cadence, slowly building up depth over the course of two years. A Wide Imaging Tier will observe in \rband, \zband, \yband, \jband, and \hband, and a Deep Imaging Tier will use \zband, \yband, \jband, \hband, and \fband. The HLTDS will cover the ELAIS-N1 field in the North (10.68 deg$^2$ in the Wide Imaging Tier and 1.97 deg$^2$ in the Deep) and the Euclid Deep Field South in the South (7.59 deg$^2$ Wide, 4.5 deg$^2$ Deep). Additional visits in each field are planned for the Pilot and Extended Components of the HLTDS, which will be observed in the first year of operations and throughout \rst's five year prime mission, respectively. 
Although the ROTAC report does not provide an estimated total depth for the HLTDS observations, the Core Component is primarily modeled after the Design Reference Survey for supernova cosmology presented in \citet{rose2021}, which estimates a co-added limiting magnitude of $m_{H158}=29.1$ for the Deep Tier. 

The Deep Tier of the HLTDS is therefore expected to cover an area $\sim$3 times larger than that which we consider here and to reach approximately the depth of our Base4 surveys. However, in the current plan, the Deep Tier will not use \rband\ (or $K_{213})$. As discussed in Sections~\ref{sec:recov} and \ref{sec:contam}, galaxy samples selected without \rband\ suffer from relatively low recovery and high contamination. The sample recovery and contamination from lower-redshift galaxies can be significantly improved with the addition of the \rband\ filter. 
This filter is also fundamental in separating the colors of galaxies from those of cool stars and brown dwarfs (Section~\ref{sec:stars}). A deep survey that excludes \rband\ will be limited in how effectively it can select the robust samples of high-redshift galaxies needed to constrain the UV luminosity function and $\rho_{\mathrm{UV}}$ at early times. 
On the other hand, the Wide Tier will include \rband, but exclude \fband, which we have also seen is necessary to increase sample completeness and minimize contamination. Additionally, the shallower depth of the Wide Tier would result in poorer constraints on the faint end of the luminosity function.

A more promising way forward may be to combine the HLTDS with aspects of the deep fields presented here. One option would be to change the HLTDS design to incorporate the \rband\ filter\footnote{As an overguide option, the HLTDS Definition committee suggests adding non-cadenced \rband\ imaging to the Deep Tier to reach $m=29.4$, though the ROTAC does not recommend this option in their report.}. However, as can be seen in Table~\ref{tab:grid}, adding filters to a survey with a fixed total observing time can quickly reduce the depth possible in each filter. Any loss of depth would come at the cost of weaker constraints on the faint end of the luminosity function and therefore on the evolution of $\rho_{\mathrm{UV}}$. Another option would be to use the deep pointing of the HLTDS as the location for a \rst\ ultra deep field, focusing new GO observations on adding depth and imaging in any missing filters. This option leverages the observing time already devoted to the core survey to efficiently build a deep field with full wavelength coverage, potentially adding depth to the base filter set as well.

Finally, although we have not included the $K_{213}$ filter in our analysis here, it would be a welcome inclusion for the search for $z>9$ galaxies. For $9 \lesssim z \lesssim 14$, the $K_{213}$ imaging would provide an additional detection filter to help constrain the SED shape redward of the \lya\ break and minimize sample contamination. It would also allow for a search for very bright $z\gtrsim14$ galaxies, which though rare have already been shown to exist \citep{carniani2024,naidu2026}.

\section{Summary}\label{sec:summary}
We present a trade study of possible ultra deep \rst\ surveys, with the goal
of determining the optimum parameters in the depth-area-filter phase space
for high redshift science. For this work, we use a mock galaxy catalog
generated from a 2 degree$^2$ light cone and the Santa Cruz SAM. The mock
catalog is populated with over $7.6$ million $M_\mathrm{UV} \lesssim -15$ galaxies at $0 < z < 10$, with
realistic clustering properties and relative counts of galaxies in each
redshift slice. The properties of each source, including star formation
rate, star formation efficiency, and feedback mechanisms are forward
modeled to create synthetic photometry measured in many filters, including
those form \hst/WFCS and \rst/WFI. We extract idealized \rst\ pointings
from the light cone, treating the \rst\ field of view as a simplified
$0\fdg75 \times 0\fdg375$ rectangle, covering the appropriate area excluding
chip gaps.

We begin by demonstrating the power of a \rst\ deep field in overcoming the
effects of cosmic variance in high redshift samples, by comparing a single
\rst\ observation with \hst-like observations.
For this comparison, we consider fully idealized surveys, selecting sources
from the
mock light cone using their simulated redshift and absolute magnitude at
1500\AA. In this case, the luminosity functions we measure are essentially
calculated for 100\% completeness and 0\% contamination. There is significant
variation in the Schechter function parameter values and the number density 
of bright ($M<-21$) galaxies as measured in pointings mimicking the
combined HUDF and HFF parallel field observations.  
On the other hand, the \rst\ observations result in luminosity function parameter distributions with $\sim8-20$ times less variation and
closely recover the ``true'' values
obtained from the unscattered, full light cone.

Next, we consider a grid of possible deep field surveys, covering a range of 
area and filter coverages while totalling 500 hours. 
We consider four areas covering 0.28, 0.56, 1.12, and 2 degree$^2$ (1, 2, 4 
or $\sim$7 \rst\ pointings, respectively) along with four filter sets, a base 
set (\zband, \yband, \jband\ and \hband) to which we add \rband, \fband, or
both. We photometrically scatter the sources to create 16 realizations of the
ligh cone mock catalog, one for each of the potential deep fields in the
trade study. In each realization, we perform sample selection using \eazy\
and a combination of criteria related to the integrated redshift PDFs as well
as the detection significance in individual bands. We create samples in four
$\Delta z=1$ redshift bins centered at $z=6.2$, 7.2, 8.2 and 9.2, with bin 
centers chosen such that the low-redshift edge of the $z\sim6$ bins falls
outside of the bluester (\rband) filter.

We evaluate each survey's recovery, and contamination from lower-redshift
galaxies and from stellar sources. We measure the rest-UV luminosity 
function in all four redshift bins and the integrated non-ionizing UV 
luminosity density, $\rho_{\mathrm{UV}}$, as a function of redshift, 
comparing both quantities against the ``truth'' obtained from the unscattered,
fully complete mock catalog. From these checks and comparisons, we come to the 
following conclusions regarding an \rst\ ultra deep field survey:
\begin{itemize}
\item Any study aiming to explore galaxies at $z \sim 5-6$ will require the \rband\ filter. Without this filter probing blueward of the \lya-break, sample contamination can approach 100\% at $z<6.7$.
\item The recovery of galaxies at $z\lesssim6.5$ improves from 0\% at all magnitudes to almost 100\% with the inclusion of the \rband\ filter.
\item The $z>9$ galaxy recovery begins to suffer from incompleteness $0.5-1$ mag deeper with 
the inclusion of the \fband\ filter, as it provides a second detection band redward of the break for constraining galaxy colors.
\item Cleanly identifying stellar contaminants in the galaxy samples requires both the \rband\ and the \fband, especially at the highest redshifts where galaxies may not all be resolved in \rst\ imaging.
\item Rest-UV luminosity functions measured without \rband\ tend to underpredict the bright end at $z\sim6$. 
\item As the depth decreases with widening area, the faint end of the measured $z\sim9$ luminosity function steepens relative to the true slope.  
\item As a result, the measured $\rho$ is over-estimated at $z\sim6$ and $z\gtrsim9$, indicating the need for \rband\ and increased depth in all filters.
\end{itemize}

With these results in mind, we make the following suggestions for a \rst\ ultra deep survey. 
We recommend an ultra deep field target at least two \rst\ pointings ($0.56$ deg$^2$) and include all six filters: \rband, \zband, \yband, \jband, \hband, and \fband\ to optimize survey completeness and minimize contamination from lower-redshift galaxies and stars. Ideally, such a program would build on the deep portion of the existing HLTDS to add imaging in the missing \rband\ to at least $m_{5\sigma}=29.9$ (to match our Base2+\rband+\fband\ survey). Achieving this depth in \rband\ across two \rst\ pointings would collectively require only $\sim$37 hours (see Table~\ref{tab:grid}). Any additional program time could then be devoted to increasing depth in all filters for improving the measurement of the faint-end slope, or to adding $K_{213}$ imaging for improved redshifts at $z>9$. The depth and area achieved by such a survey would provide improved constraints on the faint-end slope of the rest-UV luminosity function, reducing the uncertainty measured on the rest-UV luminosity density by factors of $>2-4$ compared to the deepest existing \jwst\ programs.

\begin{acknowledgements}
We acknowledge that a significant part of our work is
done on stolen land, and we support the efforts of Land
Back and true stewardship to those same peoples whose
land is occupied. 
A majority of this work took place at the University of Texas at Austin, 
which sits on part of Turtle Island.
The Tonkawa lived in central Texas, and the
Comanche and Apache moved through this area. We pay our
respects to all the American Indian and Indigenous Peoples and
communities who have been or have become a part of these
lands and territories in Texas, on this piece of Turtle Island.

The authors
acknowledge the Texas Advanced Computing Center (TACC)
at The University of Texas at Austin for providing HPC and
visualization resources that have contributed to the research
results reported within this pape
The material is based upon work
supported by NASA under award No. 80GSFC21M0002, and
via the WFIRST Science Investigation Team contract
NNG16PJ33C, ``Studying Cosmic Dawn with WFIRST.''
\end{acknowledgements}

\software{Astropy \citep{astropy},
          SciPy \citep{scipy},
          Scikit-learn \citep{scikit-learn},
          \eazy\ \citep{brammer2008}
          }

\facilities{\rst\ (WFI), HST (ACS, WFC3), JWST (NIRCam)}

\bibliography{roman_lfs}

\appendix

\begin{figure*}
\gridline{\fig{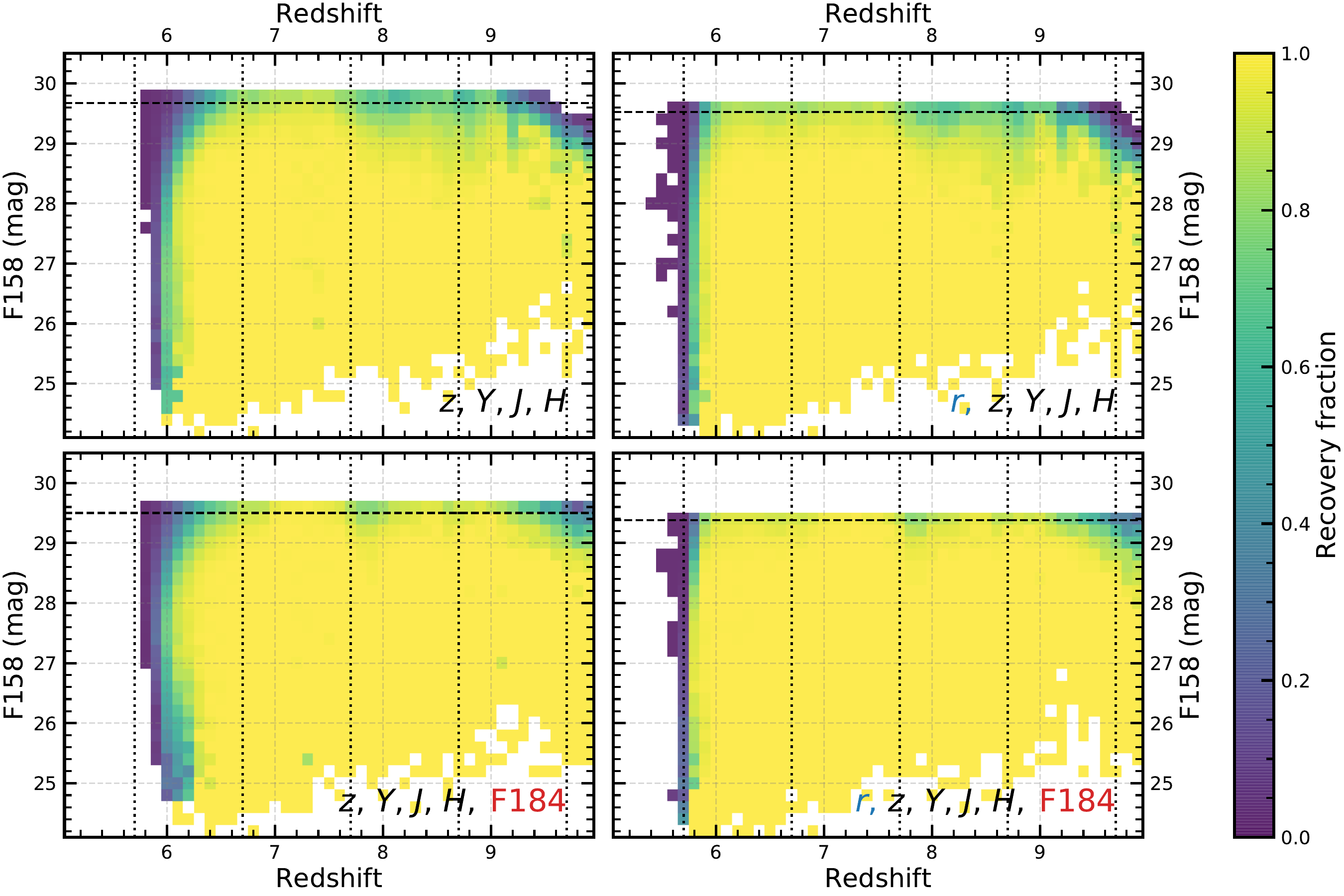}{0.48\textwidth}{}
          \fig{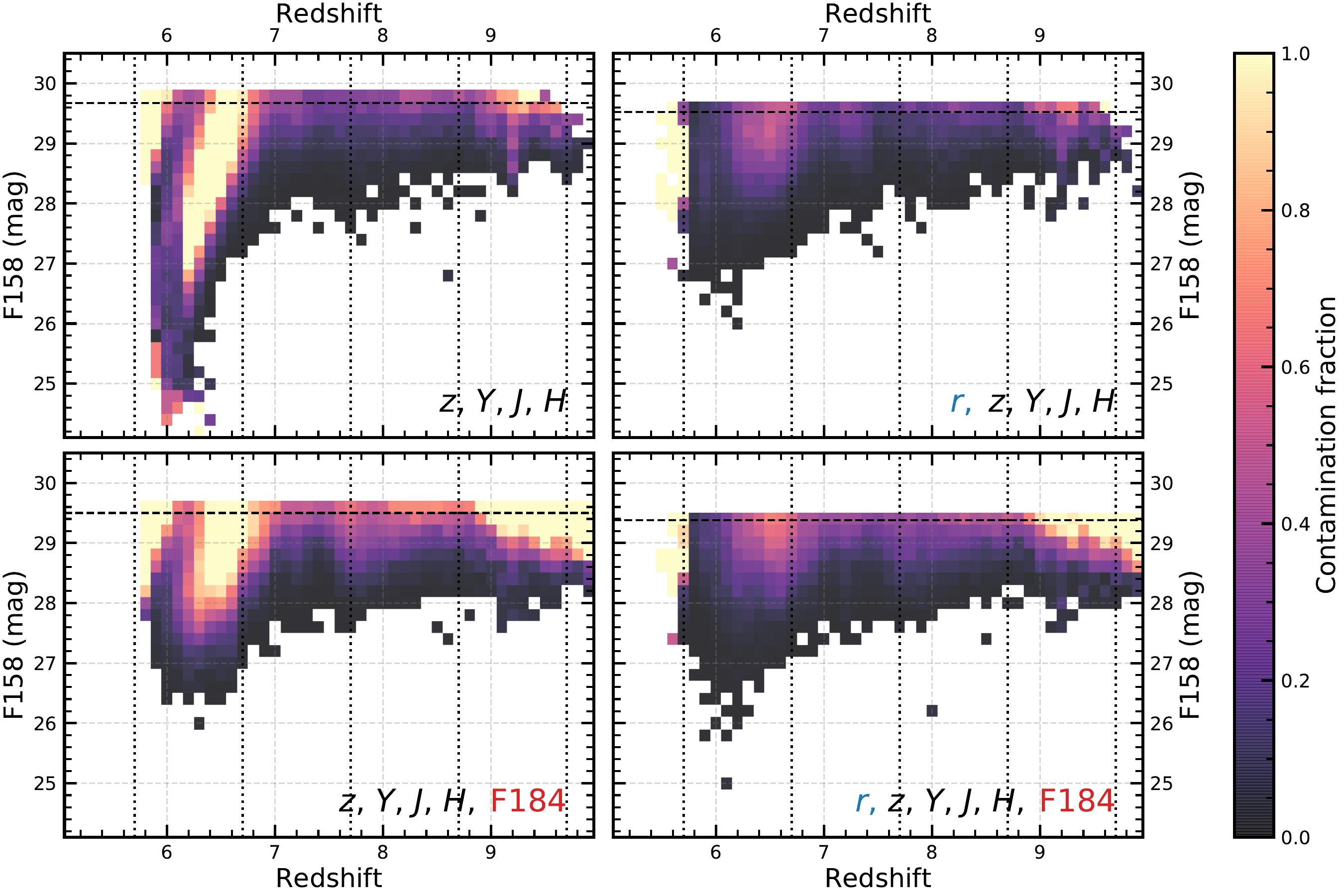}{0.48\textwidth}{}}
\vspace{-6mm}
\gridline{\fig{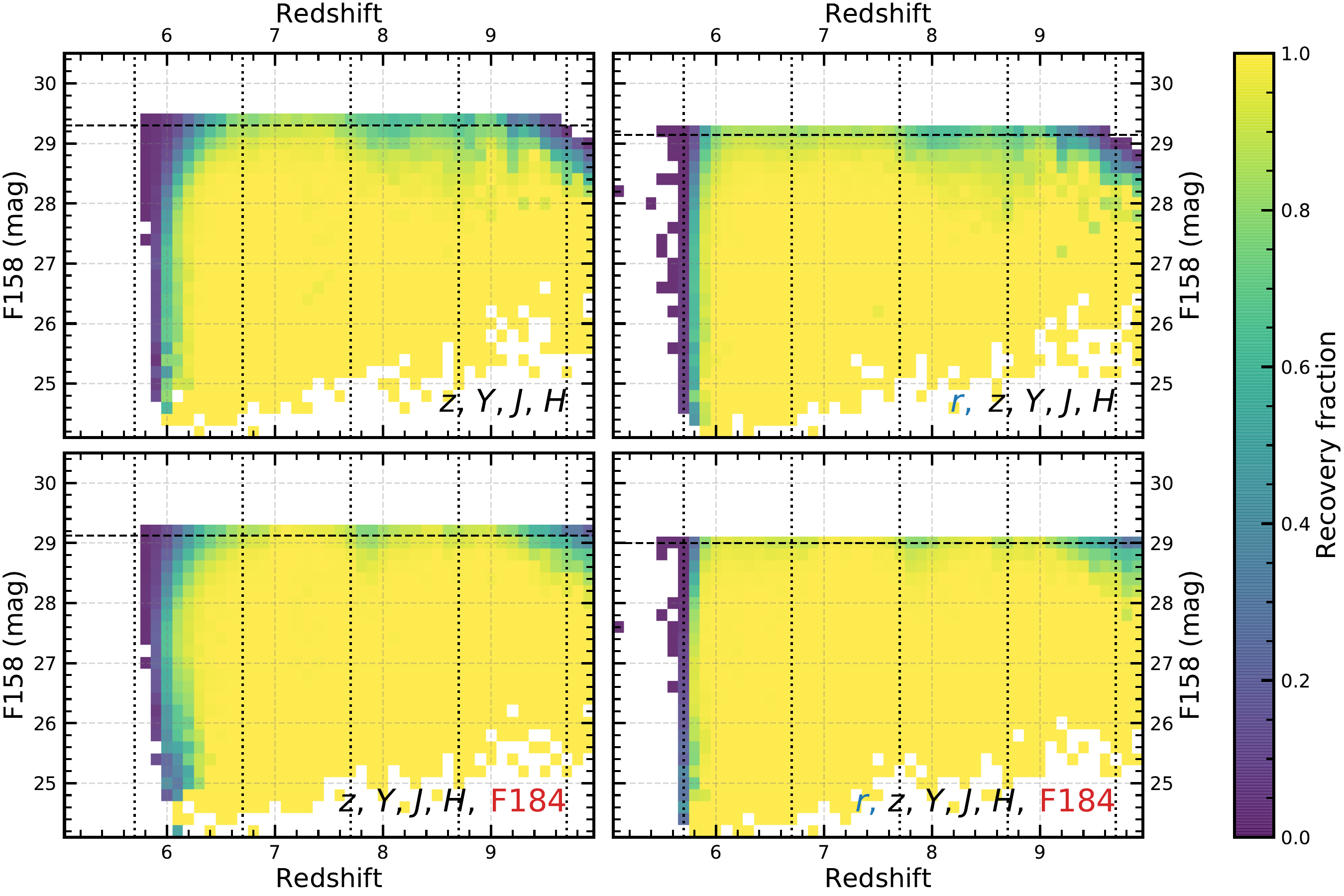}{0.48\textwidth}{}
          \fig{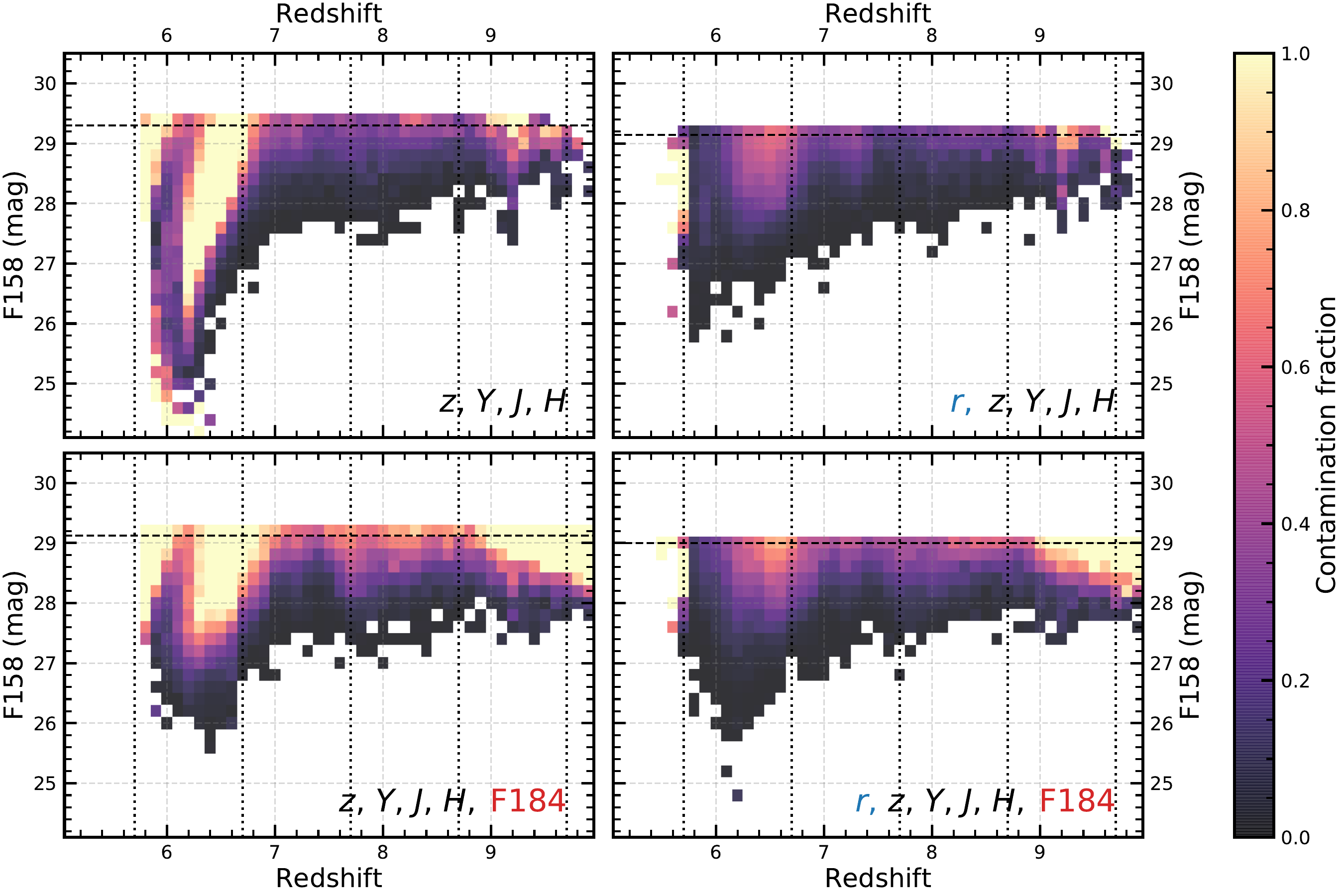}{0.48\textwidth}{}}
\vspace{-6mm}
\gridline{\fig{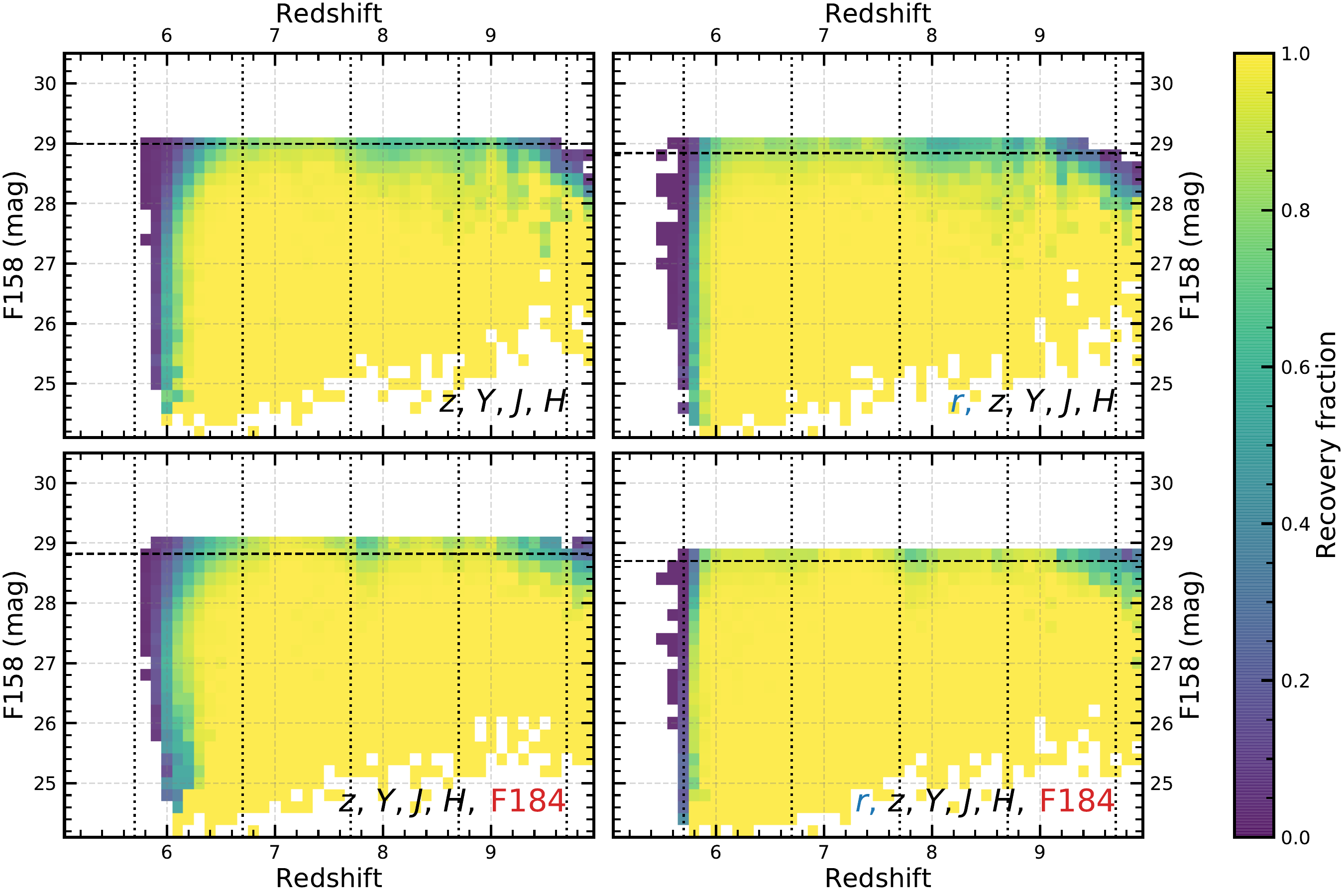}{0.48\textwidth}{}
          \fig{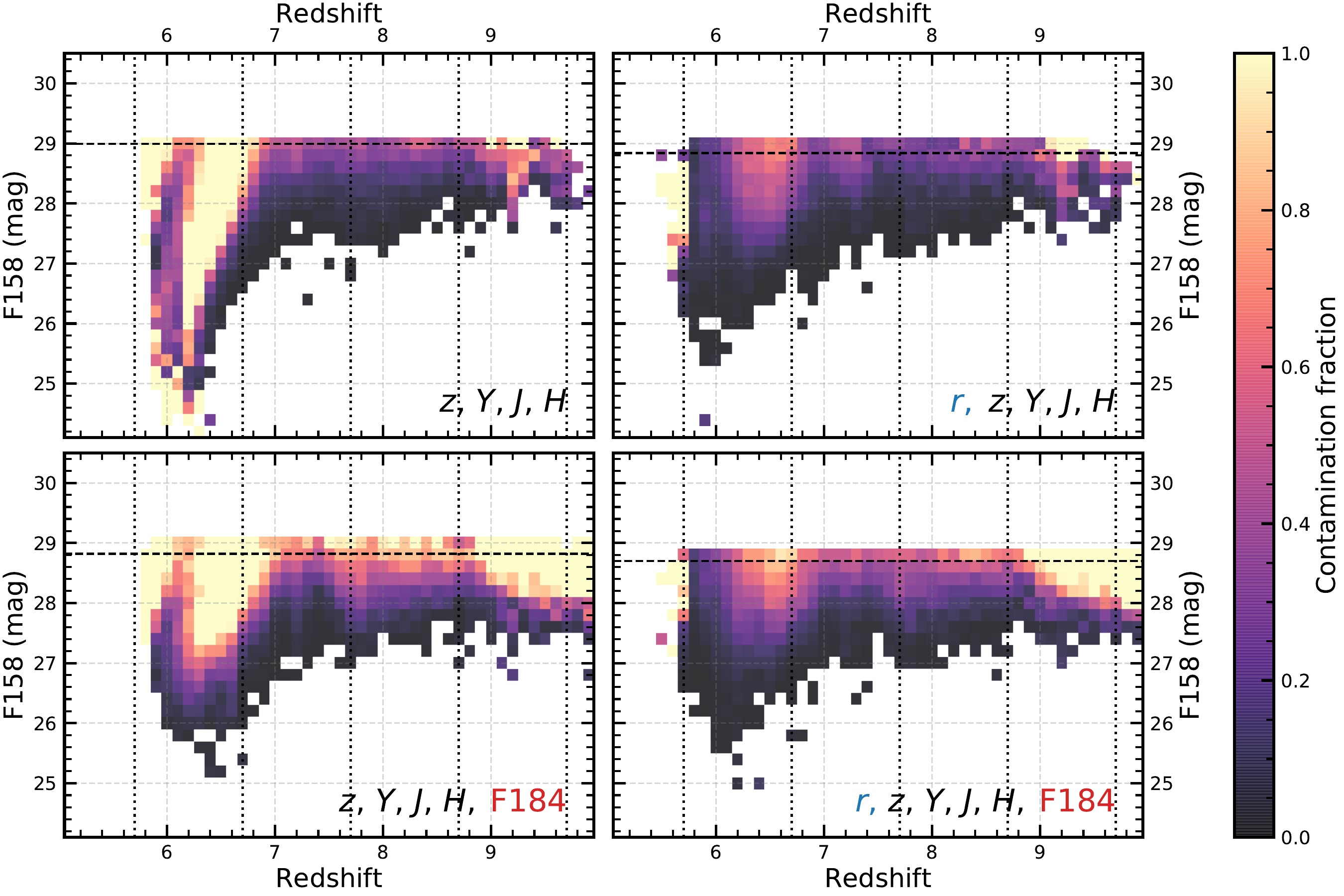}{0.48\textwidth}{}}
\caption{The recovery and contamination of the surveys covering 
0.56 degree$^2$ (2 \rst\ pointings, top row), 1.12 degree$^2$ (4 pointings,
middle row), and 2.0 degree$^2$ ($\sim$7 pointings, bottom row). See 
Figure~\ref{fig:contam_recov} caption for details.
These figures are included in Figure Set~\ref{fig:contam_recov} and are shown here for reference in the preprint version.
\label{fig:contam_recov_temp}}
\end{figure*}

\begin{figure*}
\gridline{\fig{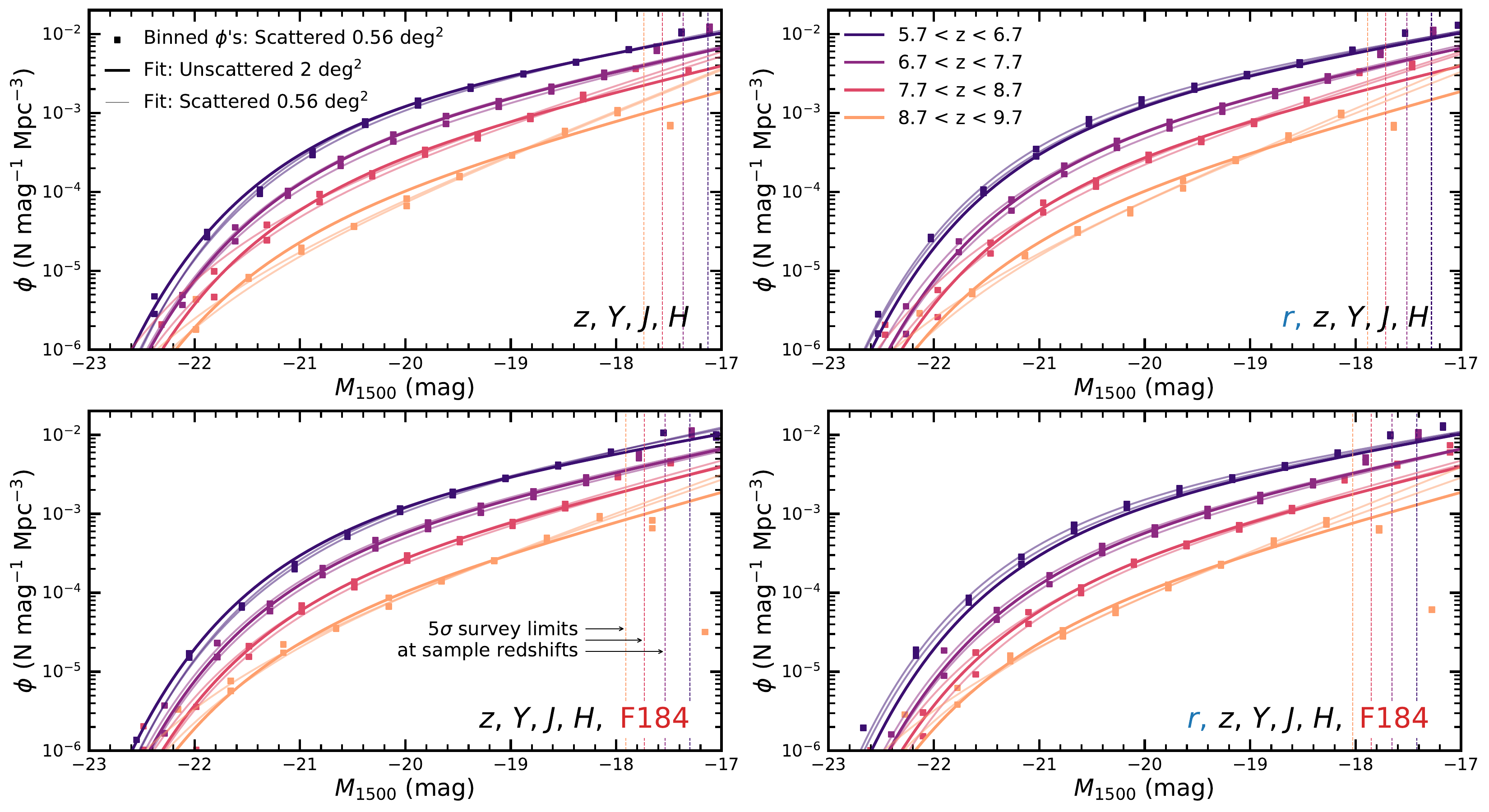}{0.95\textwidth}{}}
\vspace{-2mm}
\gridline{\fig{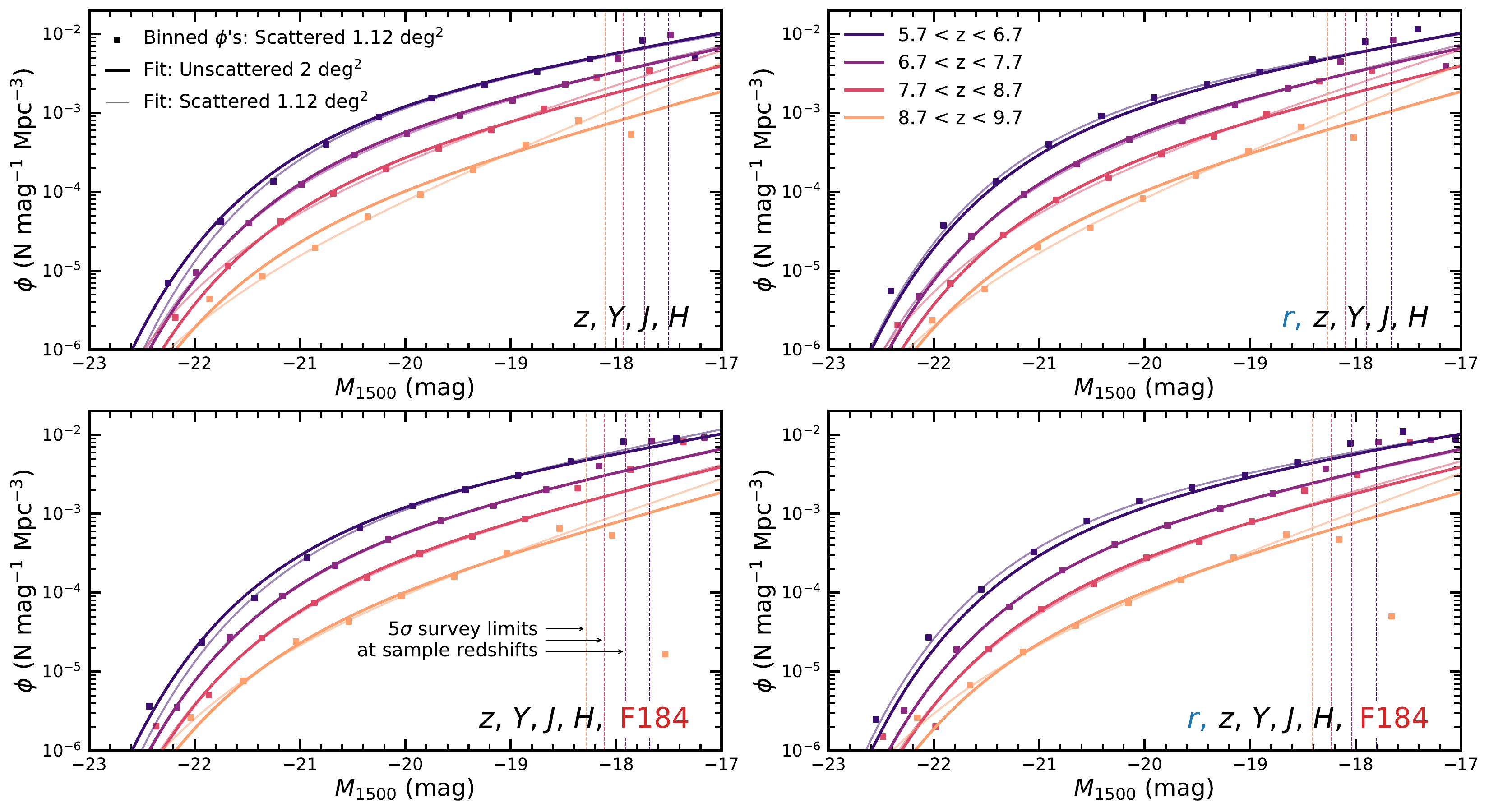}{0.95\textwidth}{}}
\caption{The rest-UV luminosity functions measured in surveys covering 0.56 degree$^2$ (2 \rst\ pointings, top four panels) and 1.12 degree$^2$ (4 pointings, bottom four panels). See Figure~\ref{fig:lfs} caption for details. 
These figures are included in Figure Set~\ref{fig:lfs} and are shown here for reference in the preprint version.
\label{fig:lfs_temp}}
\end{figure*}

\setcounter{figure}{13}
\begin{figure*}
\epsscale{1.12}
\plotone{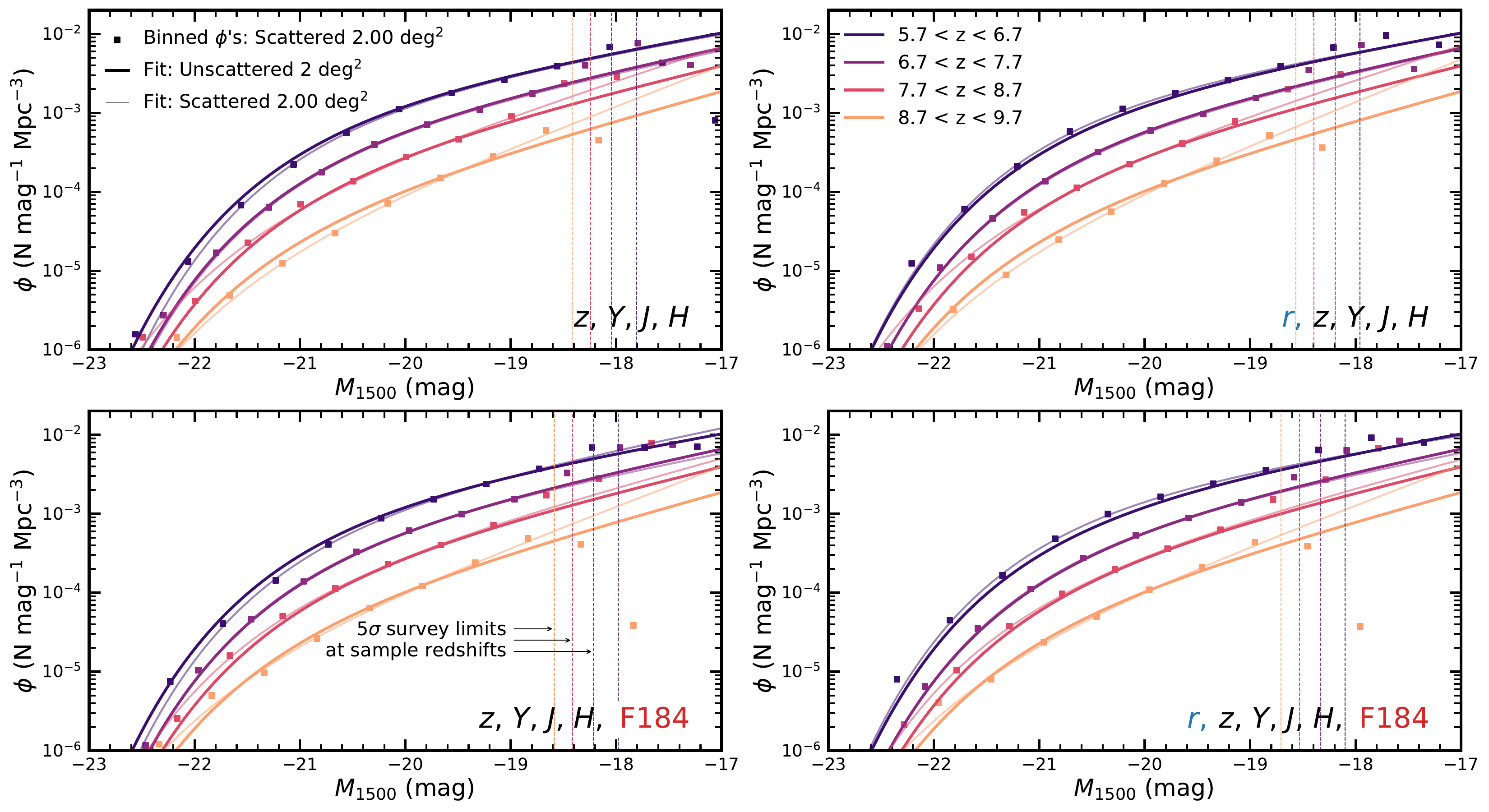}
\caption{\textbf{(continued)} The rest-UV luminosity functions measured in surveys covering 2.0 degree$^2$ ($\sim$7 pointings). See Figure~\ref{fig:lfs} caption for details. These figures are included in Figure Set~\ref{fig:lfs} and are shown here for reference in the preprint version.
\label{fig:lfs_temp_continued}}
\end{figure*}

\begin{figure*}
\gridline{\fig{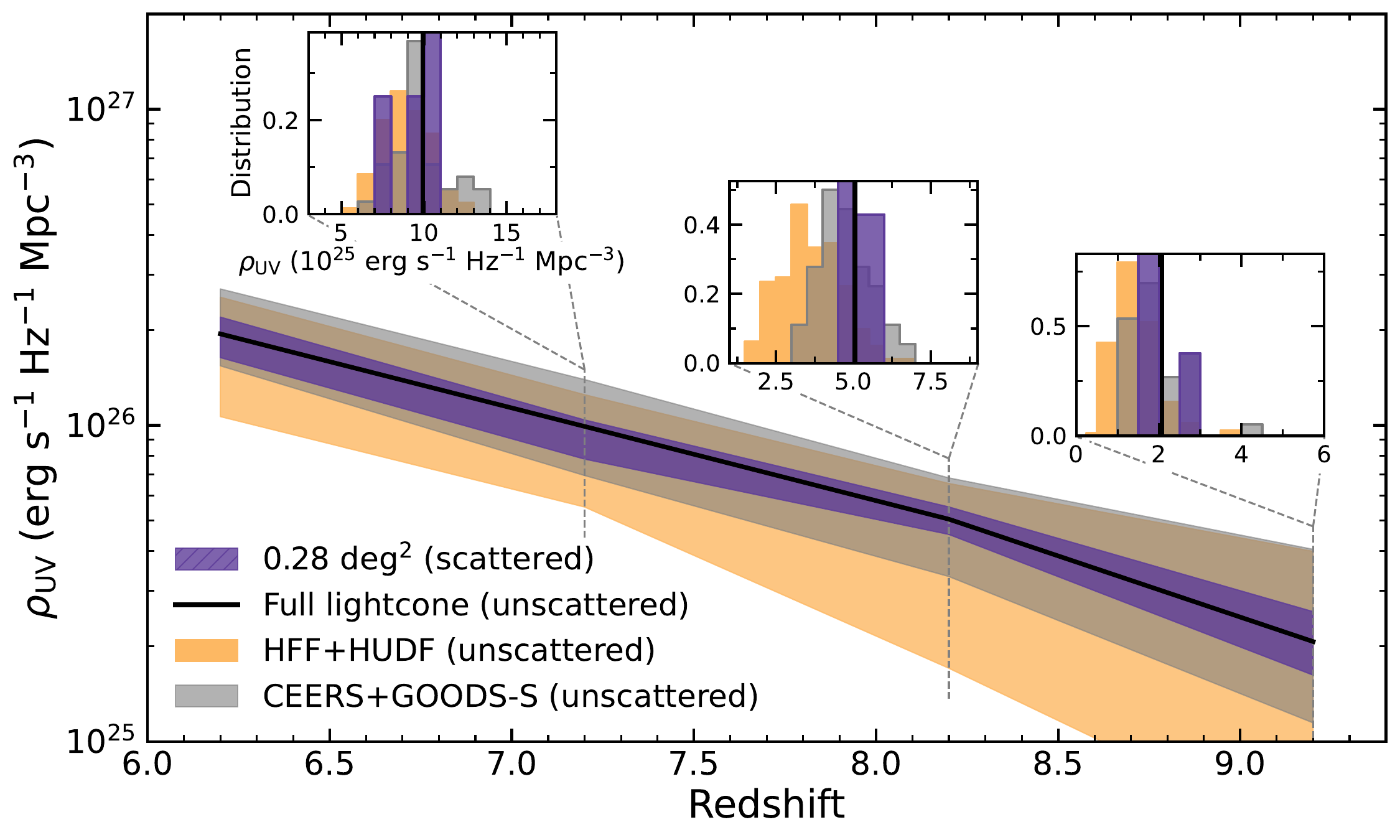}{0.48\textwidth}{}
          \fig{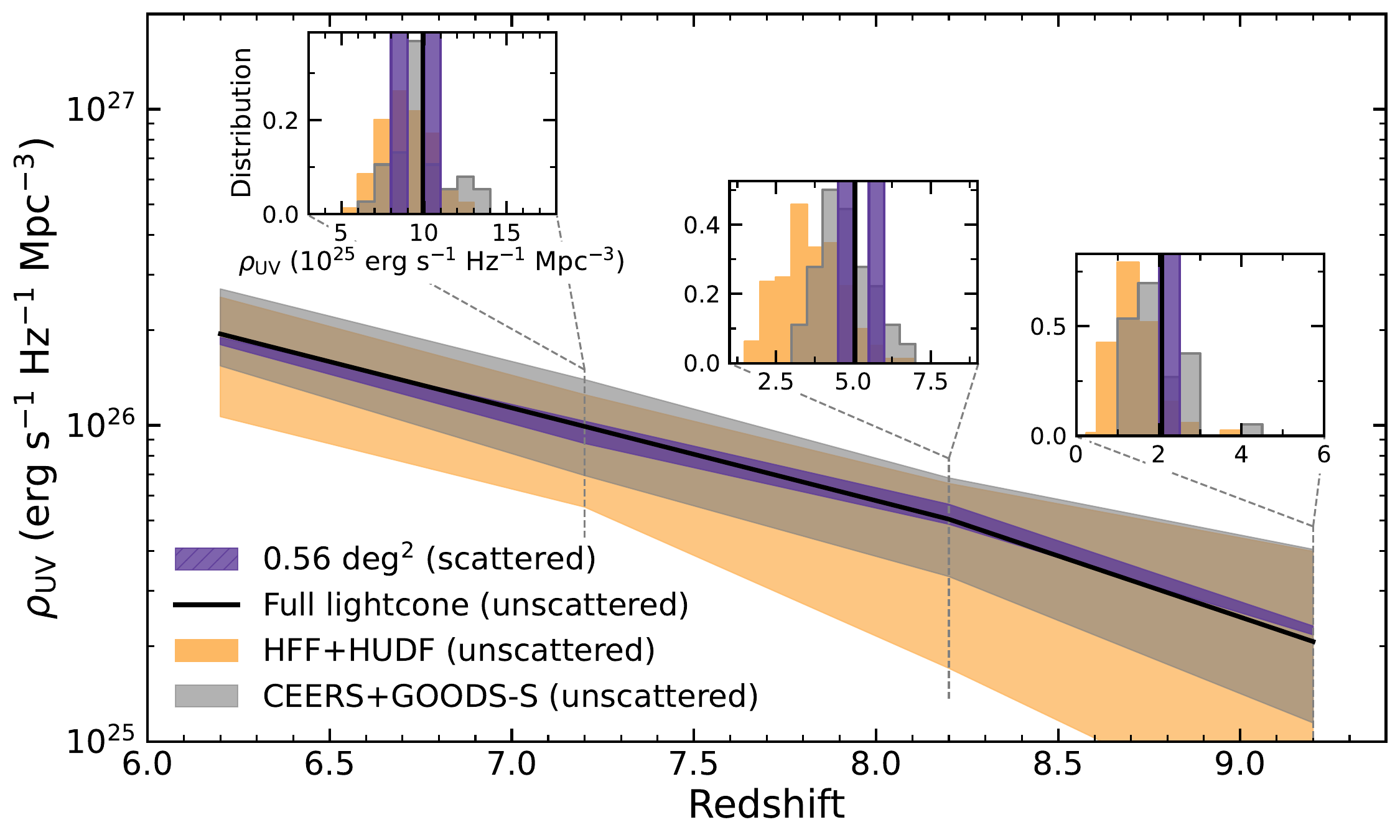}{0.48\textwidth}{}}
\vspace{-8mm}
\gridline{\fig{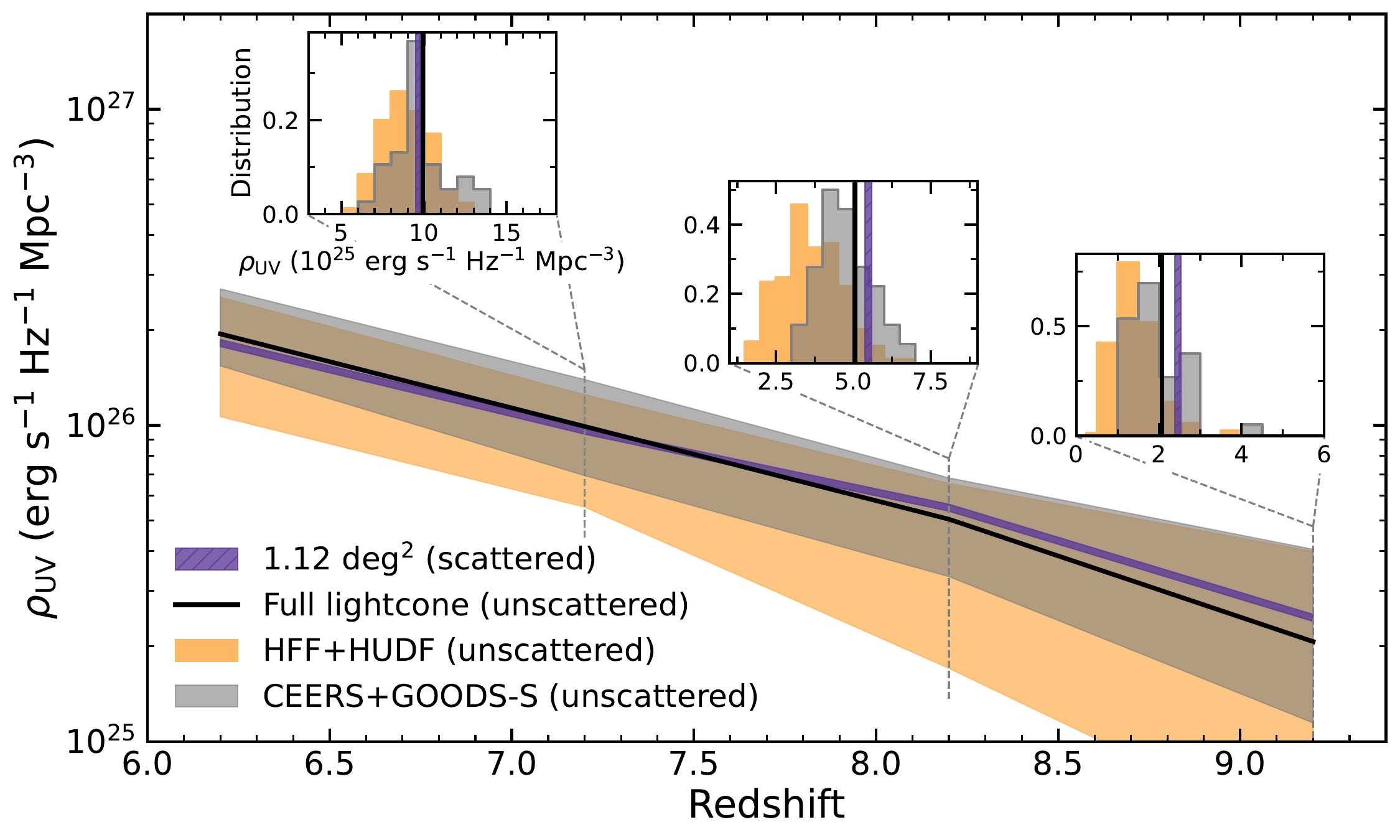}{0.48\textwidth}{}
          \fig{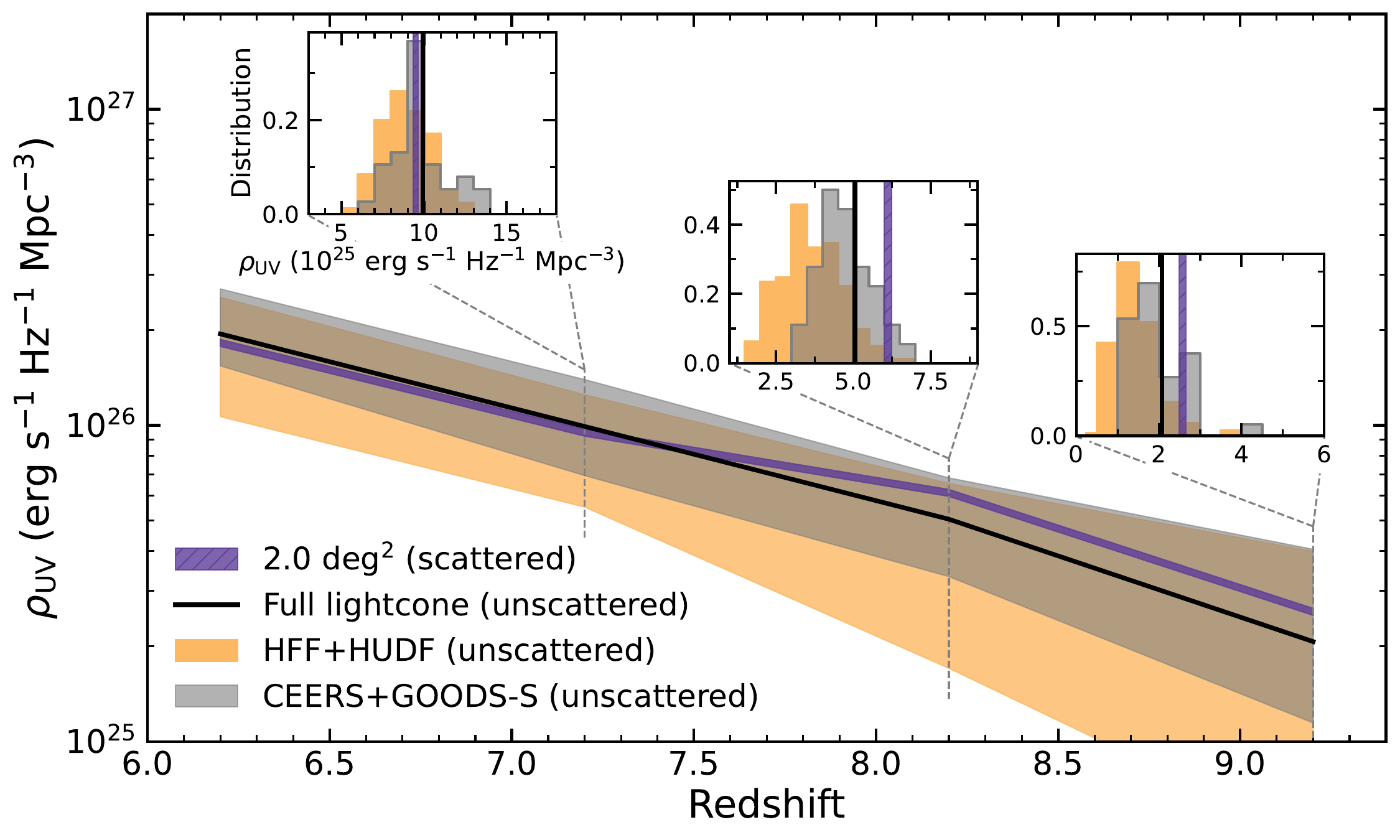}{0.48\textwidth}{}}
\vspace{-4mm}
\gridline{\fig{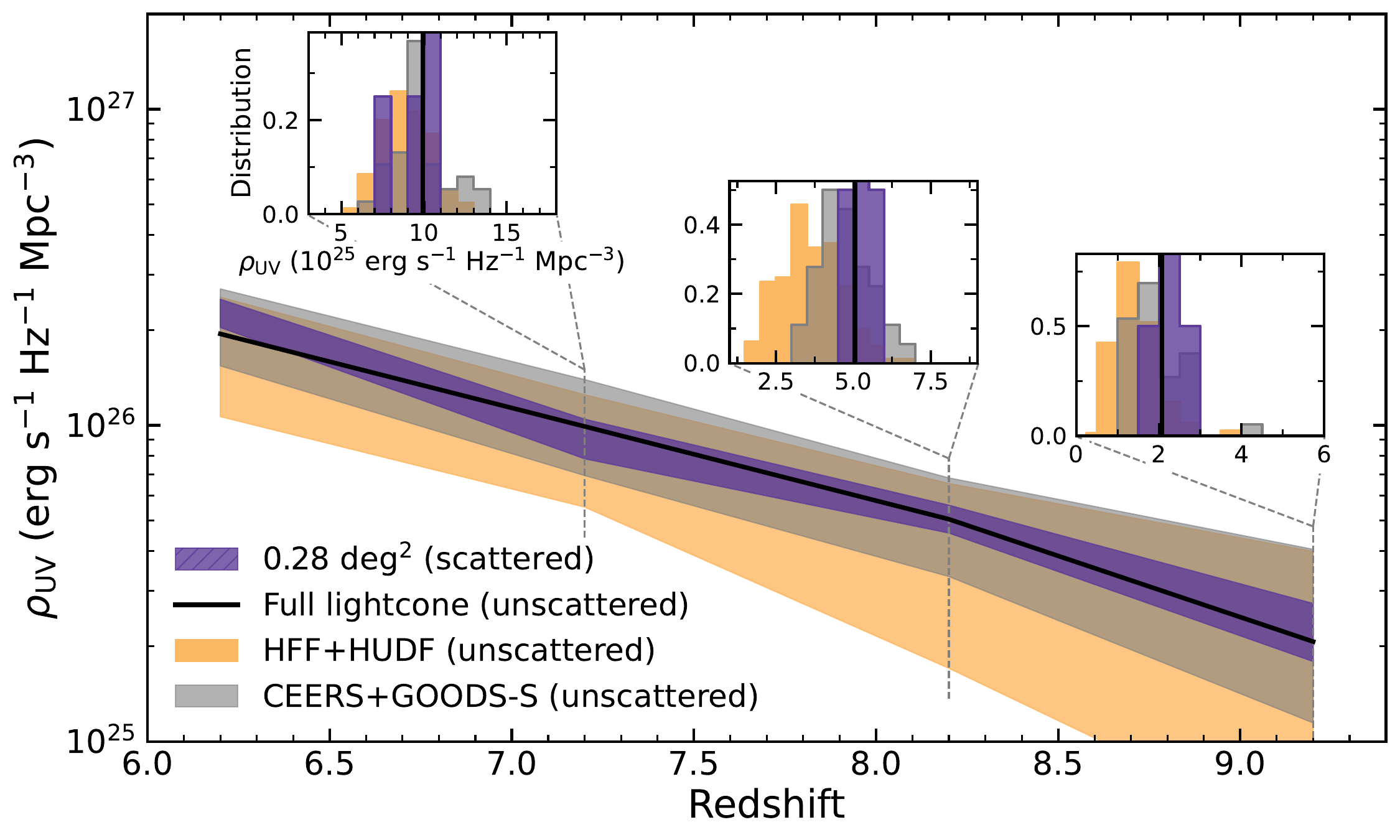}{0.49\textwidth}{}
          \fig{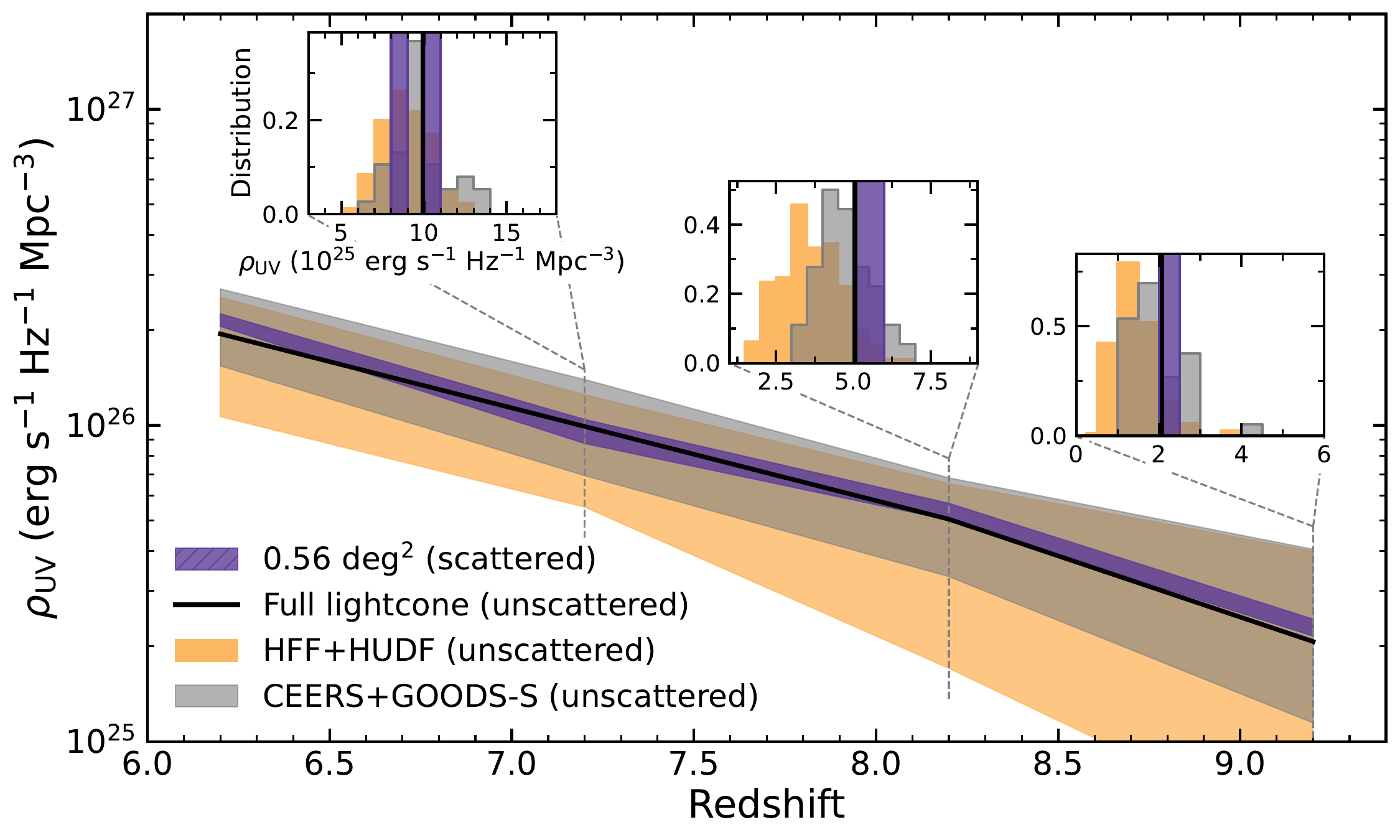}{0.49\textwidth}{}}
\vspace{-8mm}
\gridline{\fig{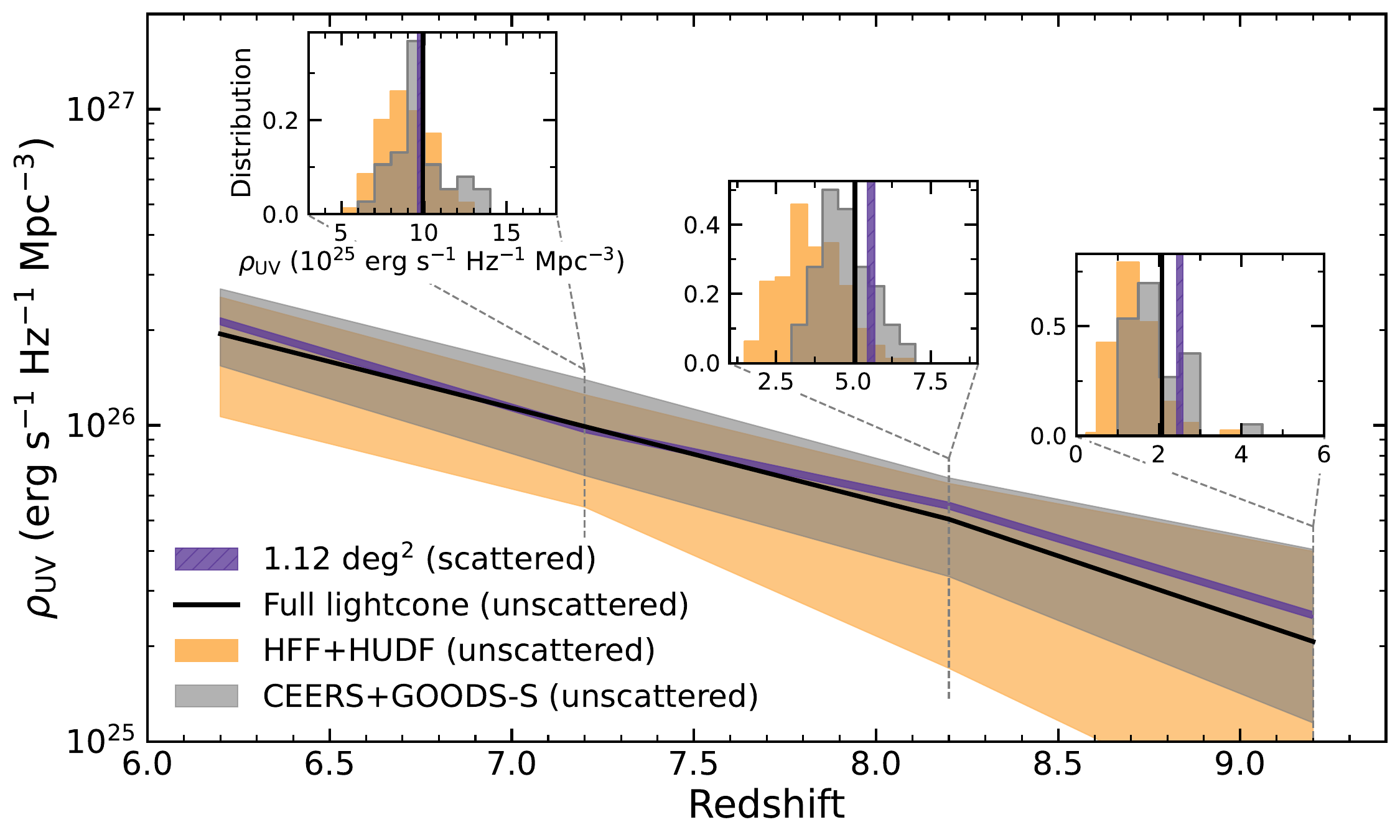}{0.49\textwidth}{}
          \fig{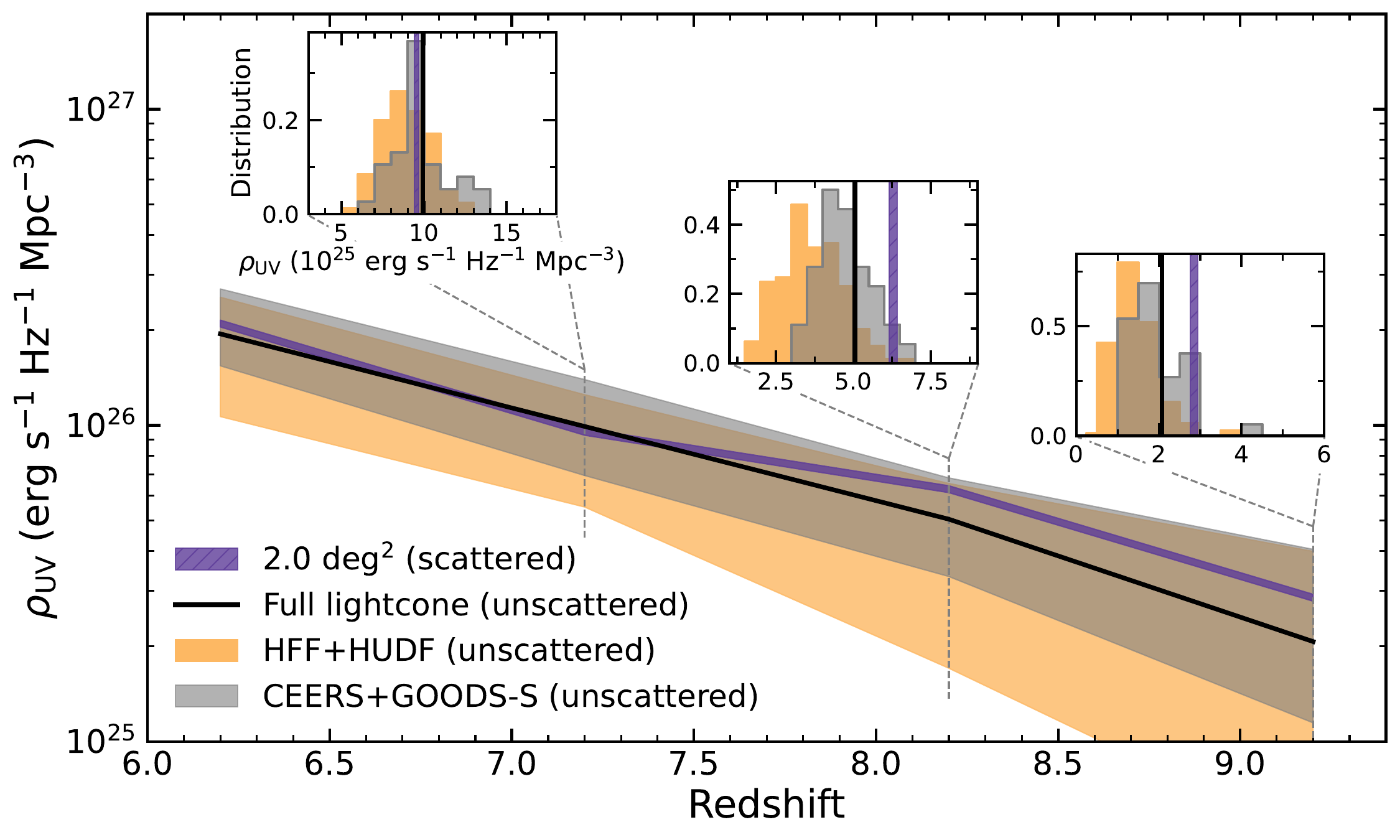}{0.49\textwidth}{}}
\vspace{-5mm}
\caption{The evolution of the integrated, non-ionizing rest-UV luminosity density in surveys using the Base filter set (top four panels) and the Base+\rband\ set (bottom four panels). See Figure~\ref{fig:rhouv} caption for details. 
These figures are included in Figure Set~\ref{fig:rhouv} and are shown here for reference in the preprint version.
\label{fig:rhouv_temp}}
\end{figure*}

\setcounter{figure}{14}
\begin{figure*}
\gridline{\fig{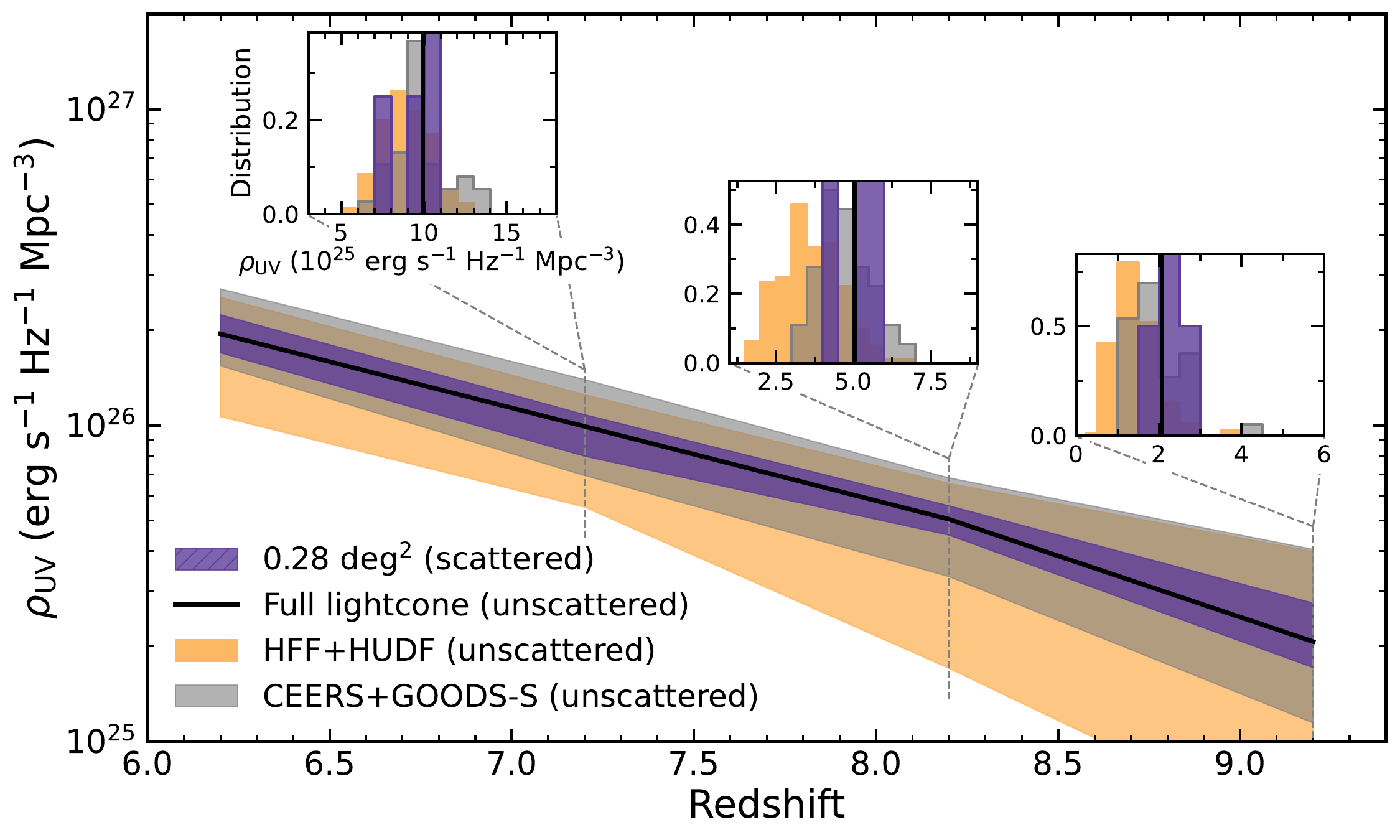}{0.49\textwidth}{}
          \fig{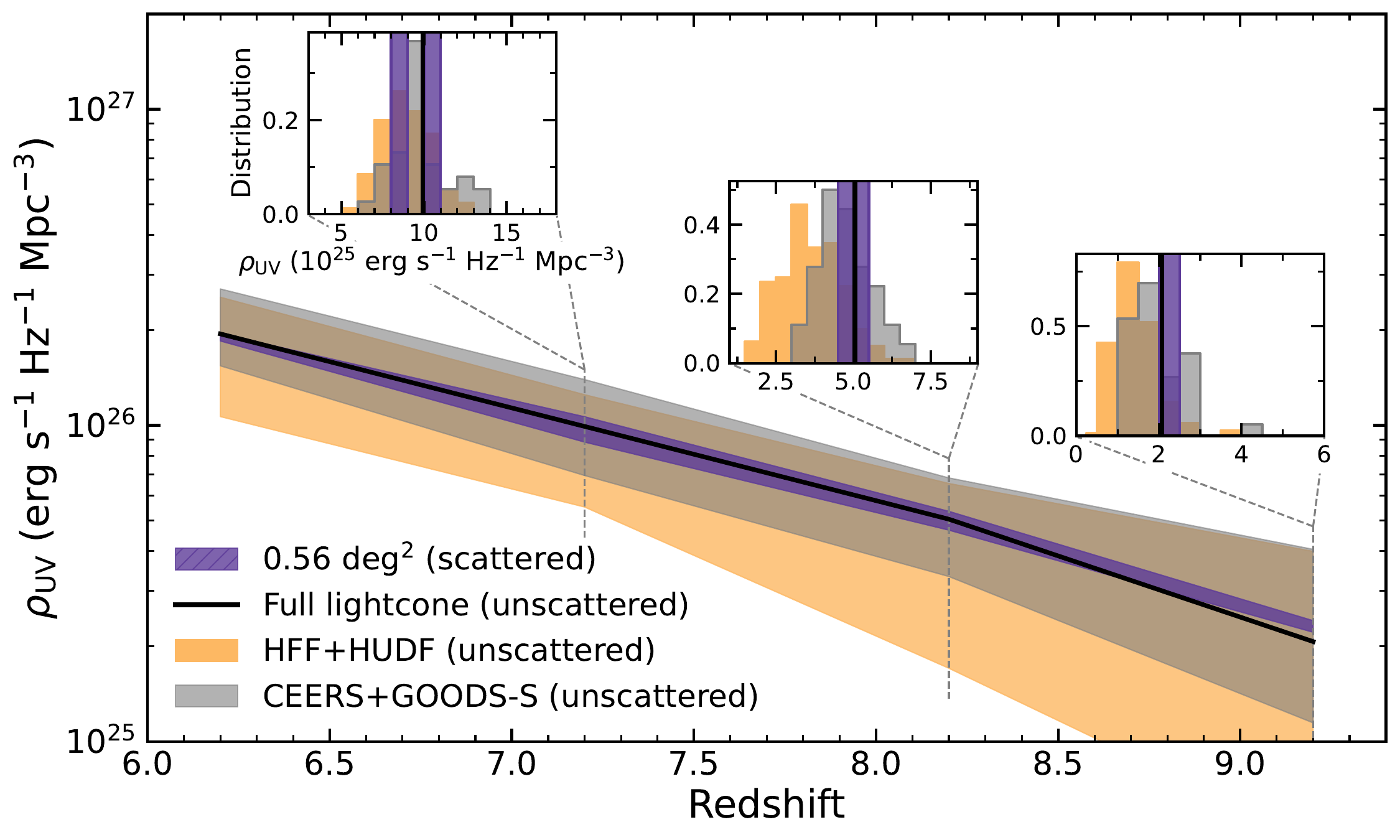}{0.49\textwidth}{}}
\vspace{-8mm}
\gridline{\fig{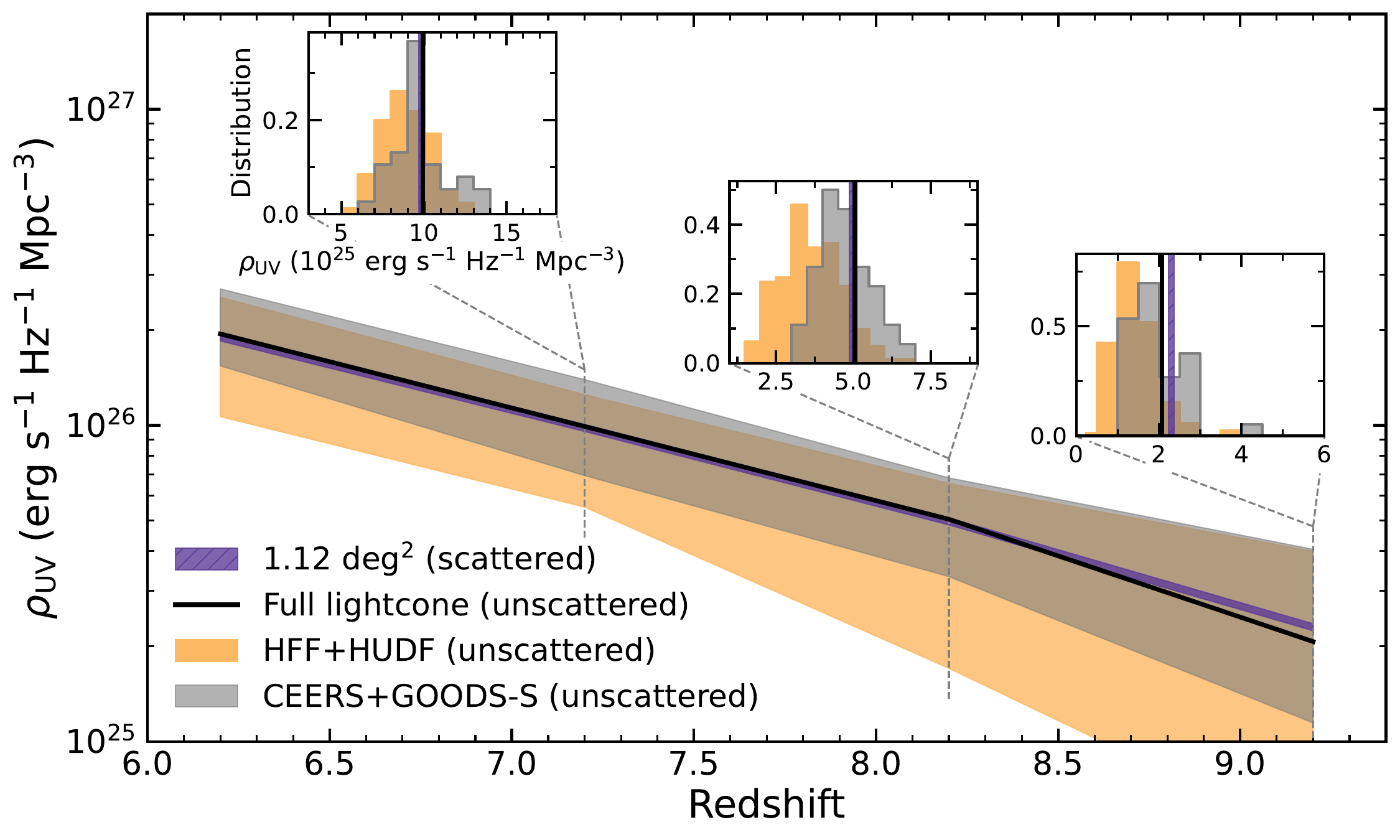}{0.49\textwidth}{}
          \fig{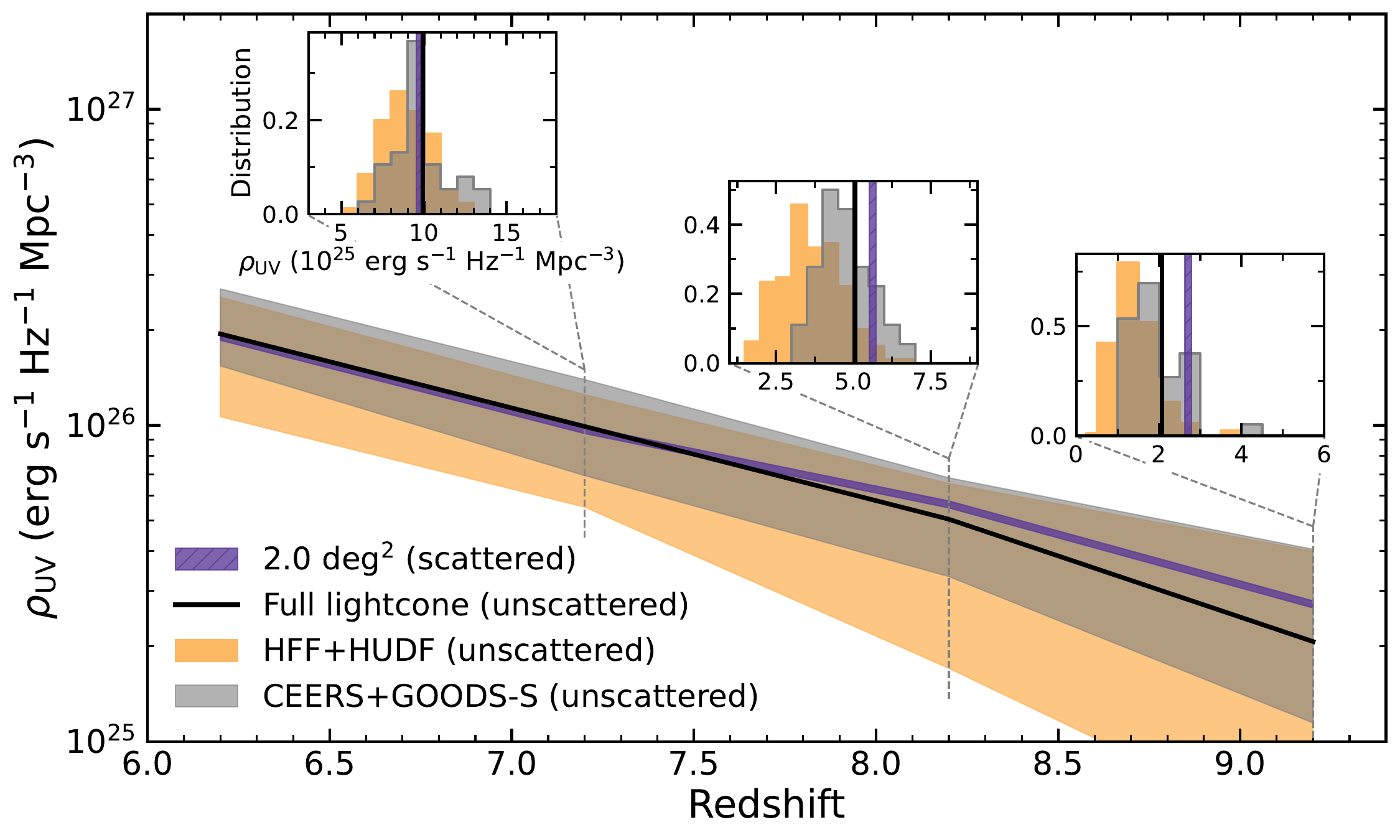}{0.49\textwidth}{}}
\caption{\textbf{(continued)} The evolution of the integrated, non-ionizing rest-UV luminosity density in surveys using the Base+\fband\ filter set. See Figure~\ref{fig:rhouv} caption for details. 
These figures are included in Figure Set~\ref{fig:rhouv} and are shown here for reference in the preprint version.
\label{fig:rhouv_temp_continued}}
\end{figure*}

\end{document}